\documentstyle[12pt,agums]{article}
\lefthead{LIPATOV, COOPER, Sittler, Hartle, Simpson}
\received{}
\revised{}
\journalid{}{JOUR.DATE}
\authoraddr{A.\ S.\ Lipatov, J.\ F.\ Cooper, W.\ R.\ Paterson,  
E.\ C.\ Sittler Jr., R.\ E.\ Hartle, D.\ G.\ Simpson,
NASA GSFC, Greenbelt, MD 
}
\slugcomment{Submitted to Planetary and Space Science, 2011}
\begin{document}

%\setkeys{Gin}{draft=false}

\title{Jovian plasma torus interaction with Europa. Plasma wake structure
and effect of inductive magnetic field: 
3D Hybrid kinetic simulation}  
\author{A.S. Lipatov$^{a,d,*}$, J.F. Cooper$^{b}$, W.R. Paterson$^{b}$, 
E.C. Sittler Jr.$^{b}$, R.E. Hartle$^{b}$, and D.G. Simpson$^{b}$}
%\email{alexander.lipatov-1@nasa.gov; john.f.cooper@nasa.gov} 
\affil{
$^{a}$ Goddard Planetary Heliophysics Institute, 
UMBC/NASA GSFC, Greenbelt, MD 20771, USA\\
$^{b}$ NASA Goddard Space Flight Center, 
Greenbelt, MD 20771, USA\\
$^{d}$ Department of Problems of Physics and Power Engineering,
Moscow Institute of Physics and Technology, Russia}

$^*$ Corresponding author. NASA GSFC, Code 673, Bld. 21, Rm. 247, 
8800 Greenbelt Rd., Greenbelt, MD 20771, USA. 
Tel.: +1 301 286 0906; fax: +1 301 286 1648.

E-mail address: Alexander.Lipatov-1@nasa.gov, alipatov@umbc.edu 
(A.S. Lipatov), John.F.Cooper@nasa.gov (J.F. Cooper),
William.R.Paterson@nasa.gov (W.R. Paterson),
Edward.C.Sittler@nasa.gov (E.C. Sittler Jr.), 
Richard.E.Hartle@nasa.gov (R.E. Hartle),
David.G.Simpson@nasa.gov (D.G. Simpson)

 \begin{abstract}
The hybrid kinetic model supports comprehensive simulation
of the interaction between different spatial and energetic elements
of the Europa moon-magnetosphere system with respect a to variable
upstream magnetic field and flux or density distributions of plasma
and energetic ions, electrons, and neutral atoms.
This capability is critical for improving the interpretation of the 
existing Europa flyby measurements from the Galileo Orbiter mission, 
and for planning flyby and orbital measurements (including the
surface and atmospheric compositions) for future missions.
The simulations are based on  recent models of the atmosphere of
Europa (Cassidy et al., 2007; Shematovich et al., 2005). 
In contrast to previous approaches with MHD simulations, the hybrid model
allows us to fully take into account the finite gyroradius effect and 
electron pressure, and to correctly estimate the ion 
velocity distribution and
the fluxes along the magnetic field (assuming an initial  
Maxwellian velocity distribution for upstream background ions).
Photoionization, electron-impact ionization, charge exchange and 
collisions between the ions and neutrals are also included in our model.
We consider  the models with $O^{++}$ and $S^{++}$ background plasma, 
and various betas
for background ions and electrons, and pickup electrons. 
The majority of $O_2$ atmosphere is thermal with an extended non-thermal 
population (Cassidy et al., 2007). 
In this paper we discuss two tasks: (1) the plasma wake structure
dependence 
on the parameters of the upstream plasma and Europa's atmosphere
(model I, cases (a) and (b) with a homogeneous Jovian magnetosphere field, 
an inductive magnetic dipole and high
oceanic shell conductivity); 
and (2) estimation of the possible effect of an induced magnetic field 
arising from oceanic shell conductivity. This effect was estimated   
based on the difference between the observed and modeled magnetic fields  
(model II, case (c) with an inhomogeneous Jovian magnetosphere field, 
an inductive magnetic dipole and low
oceanic shell conductivity).

\end{abstract}

\section*{1.\ \ Introduction}\vspace*{-3mm} 
 
The interaction of the Jovian plasma torus with Europa and other 
moons is a fundamental problem in magnetospheric physics (see 
e.g., Goertz, 1980; Southwood et al., 1980; Southwood et al., 
1984; Wolf-Gladrow et al., 1987; Ip, 1990; Schreier et al., 1993; 
Lellouch, 1996). The plasma environment near Europa was studied by 
flyby observations during the Galileo prime mission and the extended 
Galileo Europa mission 
(Kivelson et al., 1997; Khurana et al., 1998; Kivelson et al., 
1999, Paterson et al., 1999).

Europa, one of the icy moons of Jupiter, was encountered by the 
Galileo satellite three times during its primary mission, seven times 
during its Galileo Europa Mission (GEM), and once during Galileo
Millennium Mission (GMM). Europa is located at a radial distance 
of 9.4 $R_J$ (Jovian radii, 71,492\,km) from Jupiter, and has
a radius of 1560\,km (1 $R_E$).

The interaction of Europa with the magnetized plasma of the Jovian
plasma sheet gives rise to a so-called Alfv\'en wing, which has
been extensively studied in the case of Io (e.g., Neubauer, 1980; 
Southwood et al., 1980; Herbert, 1985; Lipatov and Combi, 2006).
Neubauer (1998; 1999) has shown theoretically how an Alfv\'en wing is
modified by an induced magnetic field, such as that found at
Europa (Kivelson et al., 2000).
Observations by Kivelson et al. (1992) show the generation of
ultra-low frequency electromagnetic waves in Europa's wake. 
These waves have frequencies near and below the gyrofrequencies of the 
ion species in the plasma torus (e.g., ionized sulfur, oxygen, 
and protons).
Ion cyclotron waves grow when ion distribution functions are
sufficiently anisotropic, as occurs when ion pickup creates a ring 
distribution of ions (in velocity space). 
The analysis of these waves has been done  by
Huddleston et al. (1997) (Io), Volwerk et al. (2001)
and Kivelson, Khurana and Volwerk (2009) (Europa).
They found intensive wave power at low frequencies (near and below
the cyclotron frequencies of heavy ions) in Europa's wake 
during the E11 and E15 flybys.
However, our current 3D hybrid modeling cannot yet produce these waves due 
to insufficient spatial grid resolution. 

The most general and accurate theoretical approach to this 
problem would require the solution of
a nonlinear coupled set of integro-MHD/kinetic-Boltzmann 
equations which describe the dynamics of Jupiter's corotating  
magnetospheric plasma, pickup ions, and ionosphere, 
together with the neutrals from  
Europa's atmosphere. To first order, 
the plasma and neutral atoms and molecules 
are coupled by charge exchange and ionization. 
The characteristic scale of the ionized components is usually 
determined by the typical ion gyroradius, which for Europa is 
much less than characteristic global magnetospheric scales of 
interest, but which may be comparable to the  
thickness of the plasma structures near Europa. 
Kinetic approaches, such as Direct Simulation Monte Carlo, have been
applied to the understanding of global aspects of the neutral atmosphere 
(Marconi et al., 1996; Austin and Goldstein, 2000). 
Plasma kinetic modeling is, however, much more complicated, 
and even at the current stage of computational technology require 
some approximations and compromises to make some initial progress. 
Several approaches have been formulated for including the neutral 
component and pickup 
ions self-consistently in models that describe the interaction of the
plasma torus with Europa. 

There have been recent efforts to improve and extend the 
pre-Galileo models for Europa, Io and Ganymede, both in terms 
of the MHD (Kabin et al., 1999; Combi et al., 1998; 
Linker et al., 1998; Kabin et al., 2001; Jia et al., 2008),  
the electrodynamic (Saur et al., 1998; Saur et al., 1999; 
Schilling et al., 2008), 
and hybrid kinetic (Lipatov and Combi, 2006; Lipatov et al., 2010) 
approaches. 
These approaches are distinguished
by the physical assumptions that they include.
MHD and hybrid kinetic models cannot, at least yet, include the  
charge separation effects 
which are likely to be important very close to the moon where the 
neutral densities are large and the electric potential can 
introduce non-symmetric flow around the body.
MHD models for Io either include constant artificial conductivity 
(Linker et., 1998) or assume perfect conductivity (Combi et al., 
1998). Comparisons of the sets of published results do not 
indicate that this choice has any important consequences.
The MHD model of Europa developed by Kabin et al. (1999) includes
an exospheric mass loading, ion-neutral charge exchange, and 
recombination. Further development of this model by Liu et 
al. (2000) already includes a possible intrinsic dipole magnetic 
field of Europa. 
Schilling, Neubauer and Saur (2007; 2008) 
found that a conductivity of Europa's ocean 
of 500 mS/m or largecombined with an ocean thickness of 100\,km or
smaller is most suitable for explaining the magnetic flyby data.
They also found that the influence of the fields induced by the time 
variable plasma interaction is small compared to the induction caused
by the time-varying background field.

Hybrid kinetic models can include the finite ion 
gyroradius effects, non-Maxwellian velocity distribution for ions,
and correct flux of pickup ions along the magnetic field.
Hybrid modeling of Io has demonstrated several features. 
The kinetic behavior of ion dynamics reproduces the inverse structure
of the magnetic field (due to drift current) which cannot be explained
by standard MHD or electrodynamic modeling which do not
account for anisotropic ion pressure.
The diamagnetic effect of non-isotropic gyrating pickup ions
broadens the B-field perturbation and produces increased temperatures
in the flanks of the wake, as observed by the Galileo spacecraft,
but had not been explained by previous models. 
The temperatures of the electrons which are created and cooled
by collisions with neutrals in the exosphere and inside
the ionosphere may strongly affect the pickup ion dynamics along 
the magnetic field and consequently the pickup distribution
across the wake. 
The physical chemistry in Io's corona was considered in the paper by
Dols et al. (2008). 
They couple a model of the plasma flow around Io with a multi-species 
chemistry model and compare the model results to 
the Galileo observation in Io's wake.

Galileo flyby measurements 
E4, E6 (plasma only), E11, E12, E14, 
E15, E19, and E26 demonstrate several features in the plasma 
environment: Alfv\'en wing formation and an induced magnetosphere, 
possible existence of the dipole-type induced magnetic field, and
variation of the magnetic field in the plasma wake due to  
diamagnetic currents. The measurements also demonstrate mass 
loading of the plasma torus plasma by pickup ions and the 
interaction of the ions with the surface of Europa.
For an interpretation of these data we need to use a kinetic 
model because of effects of the finite ion gyroradius.

Hybrid models have been shown to be very useful in studying the  
complex plasma wave processes of space, astrophysical, and 
laboratory plasmas. These models provide a kinetic description of 
plasmas in local regions, together with the possibility of performing 
global modeling of the whole plasma system.
Revolutionary advances in computational speed and memory are 
making hybrid modeling of various space plasma problems a 
much more effective general tool.

In this paper, we apply a time-dependent Boltzmann equation 
(a ``particle in cell" approach) together with a hybrid kinetic 
plasma (ion kinetic) model in three spatial dimensions (see, e.g.
Lipatov and Combi, 2006; Lipatov et al., 2010), 
using a prescribed but adjustable 
neutral atmosphere model for Europa.  A Boltzmann simulation is 
applied to model charge exchange between incoming and pickup
ions and the immobile atmospheric neutrals.
In this paper we discuss the results of the hybrid kinetic modeling
of Europa's environment - namely, the global plasma
structures (formation of the magnetic barrier, Alfv\'en
wing, pickup ion tail etc.).
The results of these kinetic modeling are
compared with the Galileo E4 flyby observational data.
Currently, we are working on the hybrid model of the E12 flyby. 
The remarkable aspect of this flyby is a strong variation in the upstream
plasma density profile approximately from 400\,cm$^{-3}$ to 80\,cm$^{-3}$.
The results of this modeling will be discussed
in future publications.

The paper is organized as follows: in Section 2 we present the
computational model and a formulation of the problem.
In Section 3 we present the results of the modeling of the plasma
environment near Europa and the comparison with observational 
data.
Finally, in Section 4 we summarize our results and discuss the
future development of our computational model.

\section*{2.\ \ Formulation of the Problem 
and Mathematical Model} \vspace*{-3mm}    

To study the interaction of the plasma torus 
with the ionized and neutral
components of Europa's environment, we use a quasineutral hybrid 
model for ions and electrons. The model includes ionization
(which in the Europa environment is dominated 
by electron impact ionization, 
not photoionization) and 
charge exchange.
The atmosphere is considered to be an immobile component
in this paper. 

In our hybrid modeling, the dynamics of upstream ions and implanted 
ions are described in a kinetic approach, while the
dynamics of the electrons are described in a hydrodynamical 
approximation. The details of this plasma-neutral approach were developed
early for the study of the Io-Jovian plasma interaction 
(Lipatov and Combi, 2006). 

The single ion particle motion is described by the equations (see, e.g.
Eqs. (1) and (14) from Mankofsky, Sudan and Denavit (1987)):
\begin{equation}
{{d{\bf r}_{s,l}}\over{dt}}={\bf v}_{s,l}; \quad
{{d{\bf v}_{s,l}} \over dt}= \frac{e}{M_{\mathrm{i}}}
\left( {\bf E}+ \frac{{\bf v}_{s,l} \times {\bf B}}{c} \right)
-\frac{m_e\nu_{ie}}{M_i} ({\bf v}_{s,l}-{\bf U}_i)
 -\frac{m_e \nu_{ie}}{M_i e n_i} {\bf J}-\nu_{io} {\bf v}_{s,l}.
\label{eq:1}
\end{equation}
Here we assume that the charge state is $Z_i=1$. 
${\bf U}_i$, and ${\bf J}$ denote the charge-averaged
velocity
of all (incoming and pickup) ions and the total current, Eq. (\ref{eq:5}).
The subscript $s$ denotes the ion population ($s=1,2$ for incoming ions
and $s=3,4$ for pickup ions) and the index $l$ is the particle index.
$\nu_{ie}$ and $\nu_{io}$ are collision
frequencies between ions
and electrons, and ions and neutrals that
may include
Coulomb collisions and collisions due to particle-wave interaction.

For a plasma, the thermal velocity, $v_{\alpha}'$ ($\alpha=i,e$),
is assumed greater than the drift velocity, so
we take
\begin{equation}
\nu_{\alpha, o}=n_{o}\sigma^{o,\alpha} v_{\mathrm{\alpha}}', 
\label{eq:2}
\end{equation}
where the cross section
$\sigma^{o, \alpha}$ is typically about $5\times 10^{-15}$\,cm$^2$ 
(see, e.g., Eq. (17) from Mankofsky, Sudan and Denavit (1987)).

For massless electrons
the equation of motion of
the electron fluid
takes the form
of the standard generalized Ohm's law
(e.g. Braginskii, 1965):
\begin{equation}
{\bf E}=\frac{1}{en_{\mathrm{e}}c}({\bf J}_{\mathrm{e}} \times {\bf B})
-\frac{1}{en_{\mathrm{e}}}\nabla p_{\mathrm{e}}
-\frac{m_{\mathrm{e}}}{e}\left[\sum_s\nu_{e,s} [({\bf U}_i-{\bf U}_s)
-\frac{{\bf J}}{ne}]
+\nu_{a,eo} {\bf U}_e\right], 
\label{eq:3}
\end{equation}
where $p_{\mathrm{e}}=nm_{\mathrm{e}}\langle 
v_{\mathrm{e}}'^2\rangle/3=n_{\mathrm{e}} k_B T_{\mathrm{e}}$, and
$v_{\mathrm{e}}'$
are the scalar
electron pressure and the thermal velocity of electrons, and the electron
current is estimated from Eq. (~\ref{eq:5}).

The induction equation (Faraday's law) has a form
\begin{equation}
\frac{1}{c}\frac{\partial {\bf B}}{\partial t}+
\nabla \times {\bf E} =0.
\label{eq:4}
\end{equation}

The total current is given by
\begin{equation}
{\bf J} = {\bf J}_{\mathrm{e}}+ {\bf J}_{\mathrm{i}}; \quad
{\bf J}_{\mathrm{i}}= \sum_{s=1}^2 en_s {\bf U}_s=e n_i {\bf U}_i,
\label{eq:5}
\end{equation}
where ${\bf U}_s$ is the bulk velocity of ions of the type $s$.

Since we suppose that electron heating due to collisions with ions is
very small, the electron fluid is
considered adiabatic. For simplicity we assume that the total
electron pressure may be represented as a sum of partial pressures
of all electron
populations:
\begin{equation}
p_{\mathrm{e}} \propto \frac{(\beta_{\mathrm{e}} n_{\mathrm{i, up}}^{5/3}+
\beta_{\mathrm{e,PI}} n_{\mathrm{i,PI}}^{5/3}
)}{\beta_{\mathrm{e}}},
\label{eq:6}
\end{equation}
where $\beta_{\mathrm{e}}$ and $\beta_{\mathrm{e,PI}}$
denote electron upwind, and pickup betas. 
Note that $\beta_{e,k}=p_{e,k}/(B^2/8\pi)$, 
where $k$ is a population of electrons. 
We also assume here that $n_{\mathrm{e,up}}=n_{\mathrm{i, up}}$,
$n_{\mathrm{e,PI}}=n_{\mathrm{i, PI}}$.

The neutral atmosphere of Europa serves as a source of new ions, 
mainly by electron impact ionization from corotating (or nearly 
corotating) plasma and also by photoionization. 
The neutral atmospheric molecules also serve as collisional
targets for charge exchange by corotating ions. 
The impacting ions consist of both upstream torus ions and  
newly implanted ions which are picked
up by the motional electric field.

In the current model we assume that the background plasma 
contains only ions with molecular mass/charge of 8 and 16
corresponding to $O^{++}$ and $S^{++}$, respectively.

We assume that Europa has a radius $R_{E}=1560$\,km.  
We have also adopted a two-species description
for the neutral $O_2$ exosphere of exponential form 
(Shematovich et al., 2005)
\begin{equation}
n_{\mathrm{neutral,k}} \approx n_{\mathrm{atmos,k}} 
\exp{[-(r-r_{\mathrm{exobase,k}})/
h_{atmos,k}]}, \label{eq:7}
\end{equation}
where $n_{\mathrm{atmos,k}}$ denotes the maximum value of the 
neutral density extrapolated to the exobase 
($n_{\mathrm{atmos,1}}=3\times 10^{4}$cm$^{-3}$;  
$n_{\mathrm{atmos,2}}=8.5\times 10^{7}$cm$^{-3}$; 
$r_{\mathrm{exobase},1}\approx  1700$\,km;
$r_{\mathrm{exobase},2}\approx 1560$\,km), and 
index $k$ denotes either non-thermal ($k=1$) or thermal ($k=2$) species.
Here the scale heights $h_{\mathrm{atmos,1}}=200$\,km and 
$h_{\mathrm{atmos,2}}=30$\,km. 

The production rate of new ions from the exosphere near Europa 
corresponds to
\begin{equation}
G_{\mathrm{exo,k}} \propto \nu_{i,k} n_{\mathrm{atmos,k}} 
\exp[-(r-r_{\mathrm{exobase},k})/h_{\mathrm{atmos,k}}], 
\label{eq:8}
\end{equation}
where $n_{\mathrm{atmos,k}}$ denotes the value of the neutral 
component density at $r=r_{\mathrm{exobase},k}$ 
and $\nu_{i,k}$ is the effective
ionization rate per atom or molecule of species $k$. 
$\nu_{i,k}$ includes the photoionozation $\nu_{ph}$, and the 
electron impact ionization by the magnetospheric 
electrons $\nu_{e,im}$. 
We assume that our model of the atmosphere mainly consists of $O_2$,
and we use the effective photoionization rate $1.7 \times 10^{-8}$\,s$^{-1}$ 
(Johnson et al., 2009).
We also adopt the effective electron impact ionization rates of
$2.4\times10^{-8}$\,cm$^{3}$/s 
(for 20\,eV electrons) and
$1.1\times10^{-7}$\,cm$^{3}$/s 
(for 250\,eV electrons) (see e.g. Johnson et al., 2009). 
Since the hot electrons represent only 5\% of the total electron density 
(see Voyager 1 plasma science (PLS) measurements
analyzed by Sittler and Strobel (1987) and Bagenal (1994)) 
we use the same composition 
for computing the impact ionization rate.
We assume that the Sun is located in the direction opposite the $x$ axis.

The interaction of ions with neutral particles by charge exchange
(see Eqs. (12) - (15) from Lipatov and Combi, 2006) currently includes
for the following reactions:
$$O^{++}+O_2 \rightarrow O^{+}+O_2^{+}$$
$$S^{++}+O_2 \rightarrow S^{+}+O_2^{+}$$
\begin{equation}
O_2^{+}+O_2 \rightarrow O_2+O_2^{+}
\label{eq:9}
\end{equation}
 The effective cross section for charge exchange 
($\sigma_{c,ex}=2.6\times 10^{-19}$\,m$^{2}$)
was the same as that used in the hybrid modeling of Io's plasma environment 
(see Lipatov and Combi, 2006; and McGrath and Johnson, 1989).
A more complete list of reactions 
will be considered in future modeling. Of course, this also requires
the addition of Monte Carlo computations.
However, this approach is beyond the scope of this paper.

Our code solves equations
(\ref{eq:1}) - (\ref{eq:9}).

We discuss two models of the interaction between the Jovian
magnetosphere and Europa. In Sect. 3.1 we discuss the interaction model
for the cases with different ion and electron betas, 
different pickup ion production rates near the 
surface of Europa, and a homogeneous global Jovian 
magnetic field (model I, cases (a) and (b)).  In
in Sect. 3.2 we consider model II, case (c) with a realistic global
Jovian magnetic field and the internal dipole magnetic field placed 
in the center of Europa.
To study the interaction of the plasma torus with the ionosphere 
of Europa, the following Jovian plasma torus and ionosphere 
parameters were adopted in accordance with the Galileo Europa 
E4 flyby observational data (Paterson, Frank and Ackerson, 1999; 
Khurana et al., 1998; Kivelson et al., 1997; Kivelson et al., 
1998): magnetic field, $B_0=469$\,nT and
${\bf B}=(77.6,-140.7, -441.3)$\,nT;
torus plasma speed relative to Europa 
(Paterson, Frank and Ackerson, 1999),
$U_0=105$\,km/s;
upstream ion densities, $\rho_{O^{++}}=10$\,cm$^{-3}$; 
$\rho_{S^{++}}=10$\,cm$^{-3}$ and ion temperature, 
$T_i=(25-100)$\,eV (Paterson, Frank and Ackerson, 1999);
electron temperature for suprathermal population, $T_e=20$\,eV 
(Sittler and Strobel, 1987);
ratio of specific heats, $\gamma=5/3$;
Alfv\'en  and sonic Mach numbers, $M_A=0.25$; $M_s=3.66$.

\begin{figure}
\vspace*{10.5cm}
\includegraphics{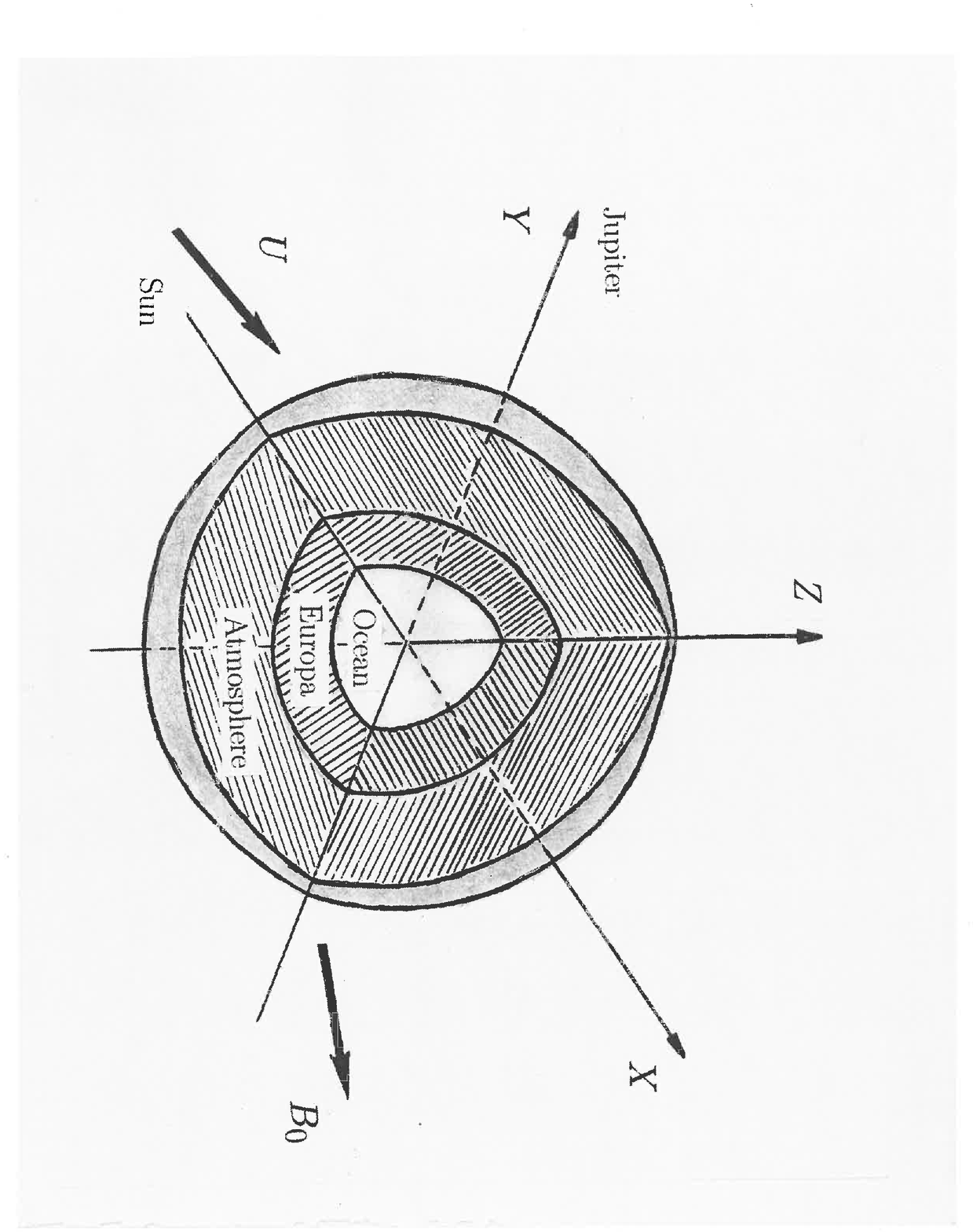}
%\special{psfile=ut1.eps hscale=100.0 vscale=100.0 hoffset=-55
%voffset=90}
\includegraphics{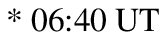}
\includegraphics{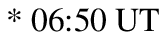}
\includegraphics{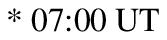}
\includegraphics{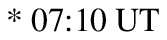}
\caption{Europa's environment and system of coordinates.}
\label{fig:1}
\end{figure}

\underline{Initial Conditions.}
Initially, the computational domain contains only supersonic and 
sub-Alfv\'enic  plasma torus flow with a homogeneous spatial 
distribution and a Maxwellian velocity distribution;
the pickup ions have a weak density and spherical spatial 
distribution.
The magnetic and electric fields are ${\bf B=B}_0$ and
${\bf E} = - {\bf U}_0 \times {\bf B}_0$. Inside Europa the 
electromagnetic fields are ${\bf E}=0$ and ${\bf B=B}_0$, 
and the bulk velocities of ions and electrons are also equal to 
zero. 
Here the $X$ - axis is directed in the corotation direction, the
$Y$ - axis is directed toward Jupiter, and the $Z$ - axis is directed to 
the north, as shown in Fig. \ref{fig:1}.
In model I, cases (a) and (b) we use a homogeneous magnetic field 
for the initial and boundary conditions (see paragraph above). 
In model II, case (c) we use an extrapolation of the magnetic field 
profile along the E4 trajectory (see, Kivelson et al., 1999; 2009) onto 
the computation domain for the initial and boundary conditions. 
The effect of global variation on
the magnetic field in the rest of Europa was not taken into 
account directly in the modeling but it was included in  
the modeling as an internal magnetic
dipole (see, Schilling et al., 2007; 2008).

At $t > 0$ we begin to inject the pickup ions with a spatial 
distribution according to Eq. (\ref{eq:8}). 
Far upstream ($x=-15\,R_{E}$), the background ion flux 
is assumed to have a
Maxwellian velocity distribution.

\underline{Boundary Conditions.}
On the side boundaries ($y=\pm DY/2$ and $z=\pm DZ/2$),
periodic boundary conditions were 
imposed for incoming flow particles.
The pickup ions exit the computational
domain when they intersect the side boundary surfaces  
$y=DY/2-5\times \Delta y$, 
$y=-DY/2+5\times \Delta y$,
$z=DZ/2-\times \Delta z$, 
$z=-DZ/2+5\times \Delta z$. Thus there is no influx of pickup ions 
at the side boundaries.

At the side boundaries we also use a damping boundary condition for
the electromagnetic field (see e.g., Lipatov and Combi, 2006;
Umeda, Omura and Matsumoto, 2001).
This procedure allows us to reduce outcoming electromagnetic 
perturbations, which may be reflected at the boundaries.

Far downstream ($x=12\,R_{E}$), we adopted a
free escape condition for particles and the
``Sommerfeld"  radiation condition for the magnetic field
(see e.g., Tikhonov and Samarskii,
1963) and a free escape condition
for particles with re-entry of a portion of the particles from the
outflow plasma.

At Europa's surface, $r=R_{E}\approx 1560$\,km, 
the particles are absorbed. In model I, there
is no boundary condition at Europa's surface for the electromagnetic field; 
we also use our equations for the electromagnetic field, (Eqs.
(2), (4) and (9) from Lipatov and Combi (2006))
inside Europa  but using the low internal conductivity (Reynolds number,
$Re=0.5$)
and a very small value for the bulk velocity 
that is calculated from the particles.
In model II, we
also  use an inductive magnetic dipole $(0, 0, -72.5)$\,nT\,$R_{E}^3$ for the
boundary condition at Europa's surface
that simulates the effect of a nonstationary Jovian magnetic field 
at the position of Europa.
In this way the jump in the electric field is due to
the variation of the value of the conductivity and bulk velocity 
across Europa's surface. 
(Note that the center of Europa is at $x=0, y =0, z=0$).

The three-dimensional computational domain has dimensions 
$DX=27\,R_{E}$, $DY = 30\,R_{E}$ and $DZ=30\,R_{E}$.  
We used mesh of $301 \times 301 \times 271$
grid points, and  $5 \times 10^8$ and $5 \times 10^8$ particles 
for ions and pickup ions, respectively, for a homogeneous mesh 
computation.
The particle time step $\Delta t_p$ and
the electromagnetic field time step $\Delta t_{EB}$ 
satisfy the following condition:
$v_{max} \Delta t_p \le \min(\Delta x, \Delta y, \Delta z)/8$ and
$v_{max} \Delta t_{EB} \le \min(\Delta x, \Delta y, \Delta z)/256$.

The global physics in Europa's environment is controlled by a set
of dimensionless independent parameters such as $M_{\mathrm{A}}$, 
$\beta_{\mathrm{i}}$, $\beta_{\mathrm{e}}$, 
$M_{\mathrm{i}}/M_{\mathrm{p}}$, ion production and charge
exchange rates, diffusion lengths, and the ion gyroradius 
$\epsilon=\rho_{\mathrm{ci}}/R_{E}$. 
Here $\rho_{\mathrm{ci}}=U_0/(eB/M_{\mathrm{i}}c)=M_{\mathrm{A}} 
c/\omega_{\mathrm{pi}}$ and the ion plasma frequency 
$\omega_{\mathrm{pi}}=\sqrt{4\pi n_0e^2/M_{\mathrm{i}}}$. 
$M_i$ and $M_p$ denote
the ion and proton masses.
For real values of the magnetic field, the value of the ion 
gyroradius is about $80$\,km,  which is calculated
from the local bulk velocity.
The dimensionless ion gyroradius and grid spacing have
the values $\epsilon=0.05$ and $\Delta_x/R_{E}=0.1$. 

In order to study ion kinetic effects (e.g. excitation of
low-frequency oscillations ($\omega << \Omega_{b}$) by mass loading),
we must satisfy the condition
$\Delta \le (10-20)c/\omega_{pb}$, where $\Omega_{b}$ and $\omega_{pb}$
denote the gyrofrequency and 
plasma frequency for upstream ions (Winske et al., 1985).
The above estimation of the plasma parameters shows that we
have good resolution for the low-frequency waves 
(see also Lipatov et al, 2012).

There is another problem - numerical resolution of the gyroradius 
on the spatial grid. This
becomes very important near Europa's surface where the MHD model 
cannot to be used and we have to use a kinetic model to 
study the trajectory 
of heavy ions and their interaction
with the surface of Europa. 
Our current model still does resolve this last effect and
we expect to improve the model by use of a spherical system of coordinates 
in future research.

\section*{3.\ \ Results of Europa's Environment Simulation} \vspace*{-3mm} 

\subsection*{3.1\ \ Effects of plasma betas on the plasma wake structure} 
\vspace*{-3mm} 

In order to study the effect of plasma parameters
on the structure of the plasma wake and the Alfv\'en wing,
we have performed modeling (model I)
for two cases (a) and (b) with different values of the upstream background
ion temperatures, 
pickup electron temperatures, 
and a value of the pickup production rate near the surface of Europa.

The following plasma parameters are chosen the same for both models: 
full magnetosphere corotation speed is
$U_0=105$\,km/s;
upstream densities are
$\rho_{O^{++}}=10$\,cm$^{-3}$,
$\rho_{S^{++}}=10$\,cm$^{-3}$;
magnetic field is
$B_0=469$\,nT;
${\bf B}=(77.6,-140.7, -441.3)$\,nT;
Alfv\'enic Mach number
$M_A=0.25$; magnetosonic Mach number $M_s=3.66$.
The model of $O_2$ atmosphere was taken from
Cassidy et al. (2007), Shematovich et al. (2005) and Smyth and Marconi (2006).
In model I, cases (a) and (b), Europa's interior is represented 
as low conducting
body with Reynolds number $Re=0.5$.

\underline{Model I, case (a)}: 
upstream ion temperatures are 
$T_{O^{++}}=25$\,eV;
$T_{S^{++}}=25$\,eV
and upstream electron temperature is
$T_{e,0}=20$\,eV.
Temperatures of electrons connected with non-thermal and thermal $O_2^+$ 
pickup ions are
$T_{e,non-thermal}=20$\,eV;
$T_{e,thermal}=20$\,eV.

\underline{Model I, case (b)} (reduced density for thermal $O_2$ 
by a factor 60 near surface and higher electron temperatures; 
increased upstream ion temperatures,
$T_{O^{++}}=100$\,eV; $T_{S^{++}}=100$\,eV): the
upstream electron temperature is
$T_{e,0}=20$\,eV;
temperatures of electrons connected with non-thermal 
and thermal $O_2^+$ pickup ions
$T_{e,non-thermal}=200$\,eV;
$T_{e,thermal}=200$\,eV.

We have computed several hybrid models with different ion 
and electron betas, and different production rate for $O_2^{+}$ 
pickup ions, but we discuss here only the models that
fit the observations.

The initial thermal velocities of $O_2^+$ non-thermal and thermal ions
are chosen as the following: $v_{th,non-thermal}=3.0$\,km/s (2\,eV)
and $v_{th,thermal}=0.5$\,km/s (0.05\,eV). The initial bulk velocity of
$O_2^{+}$ pickup ions is about 1\,km/s. 
Eq. \ref{eq:8} gives the following total pickup ion production rate:
$Q_{O_2^{+},thermal}=0.825\times 10^{28}\,s^{-1}$ and
$Q_{O_2^{+},non-thermal}=1.95\times 10^{26}\,s^{-1}$.

\begin{figure}
\vspace*{12.cm}
\includegraphics{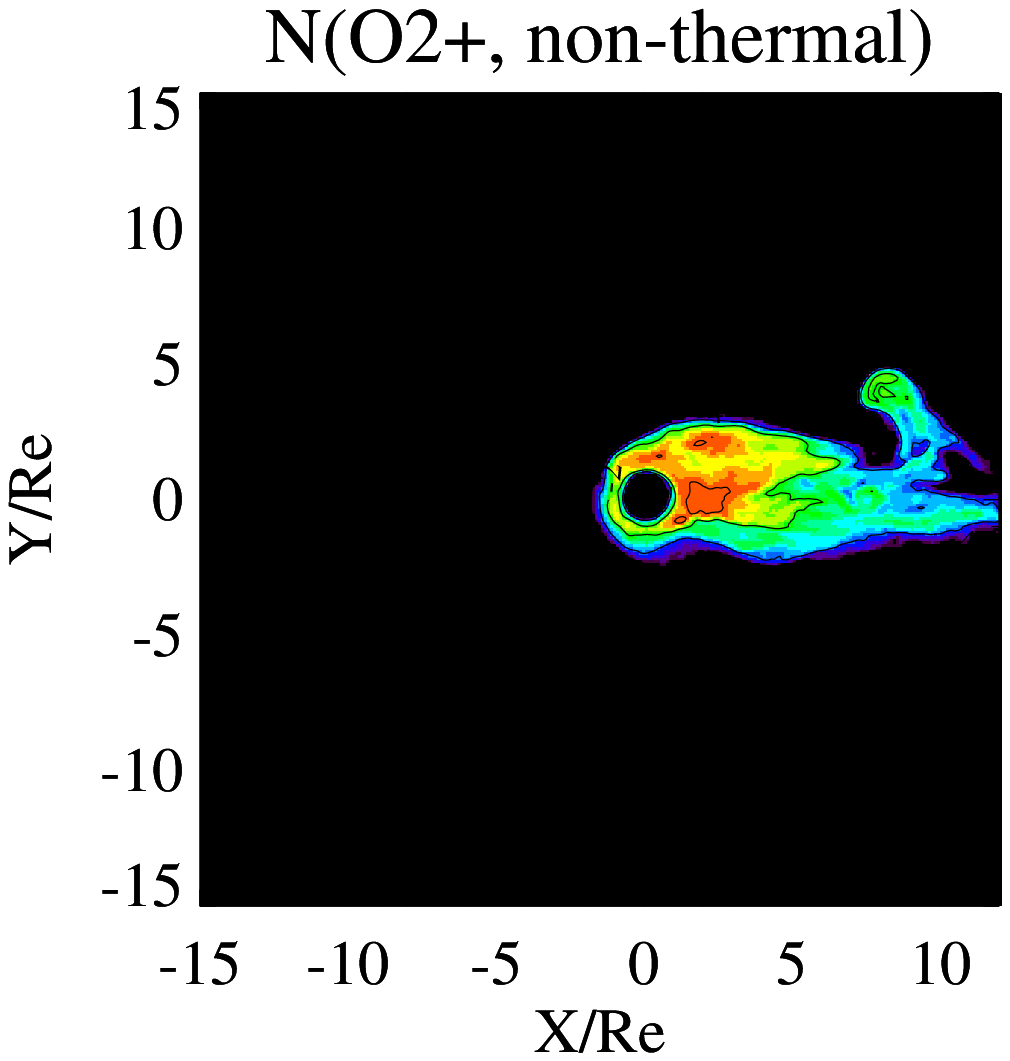}
\includegraphics{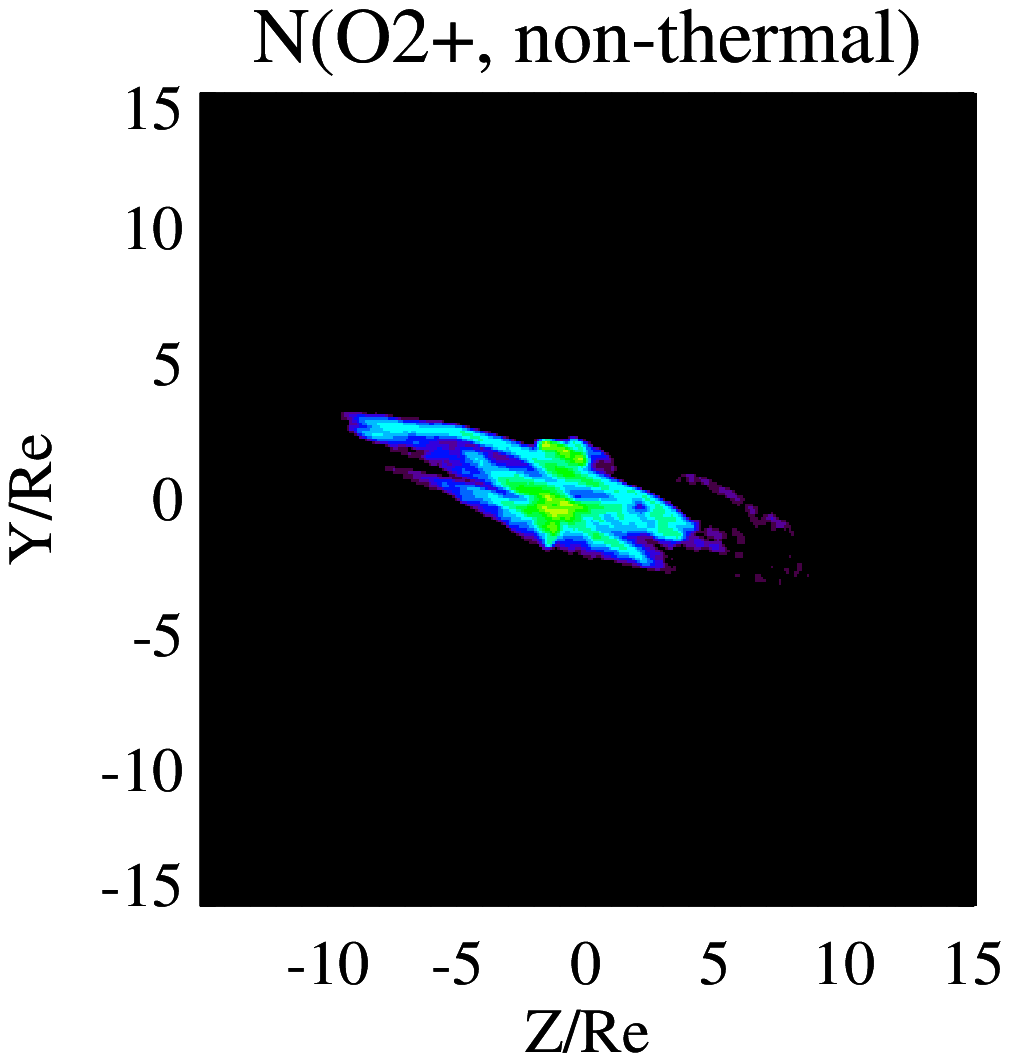}
\includegraphics{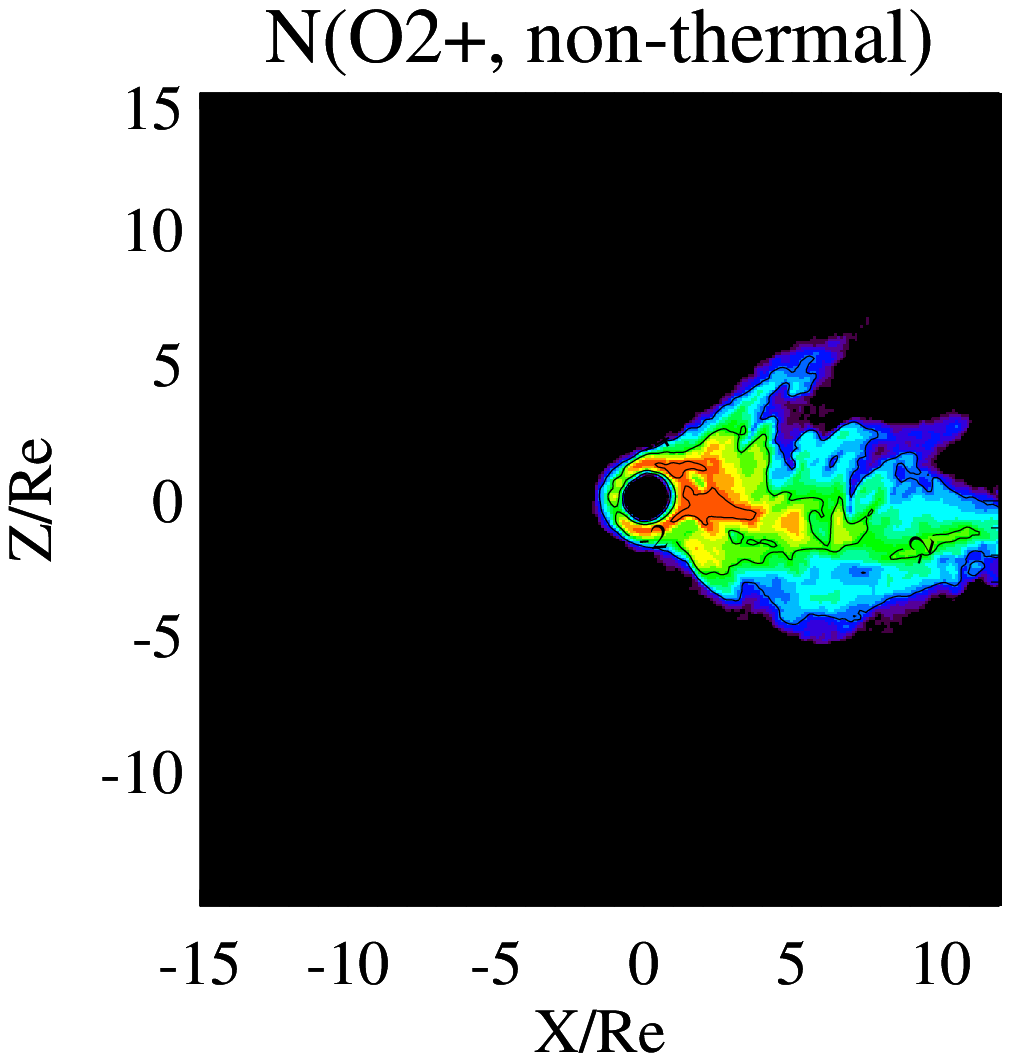}
\includegraphics{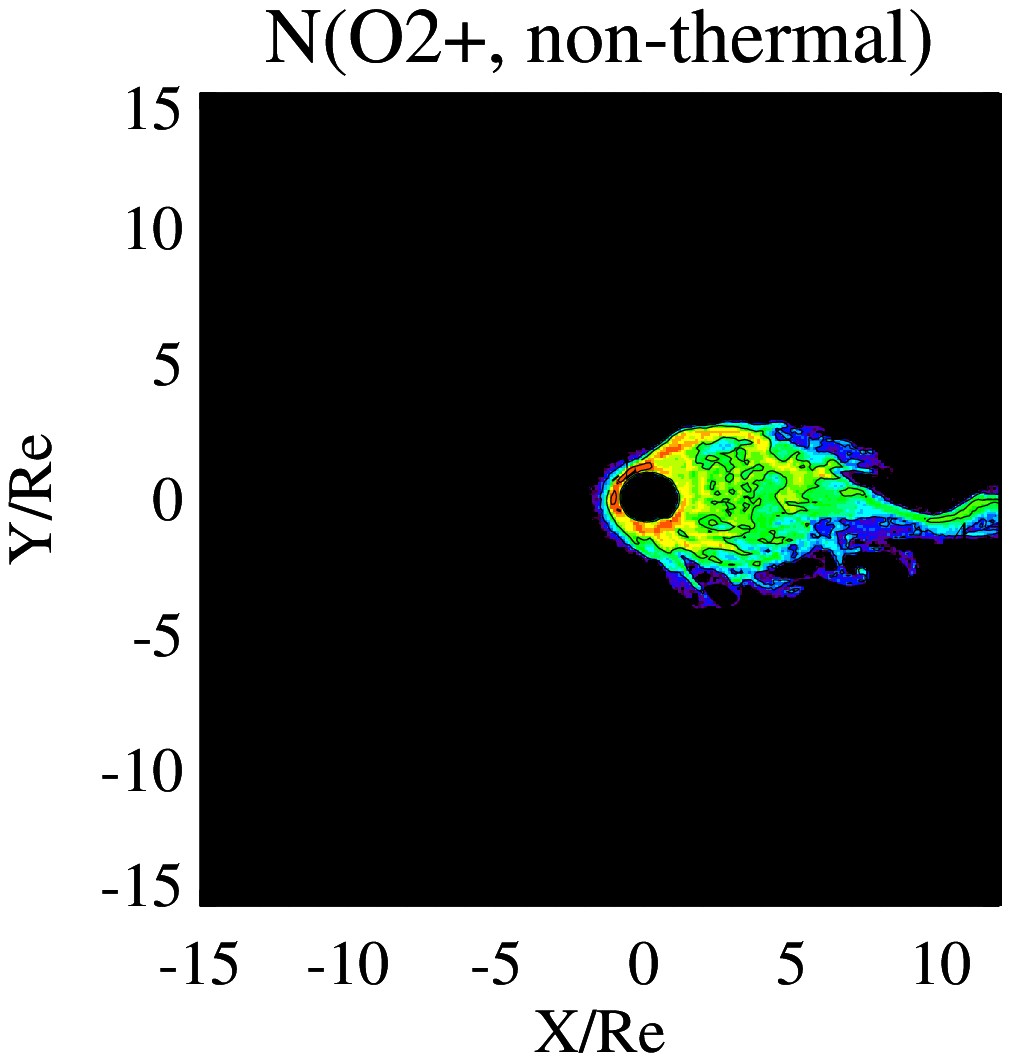}
\includegraphics{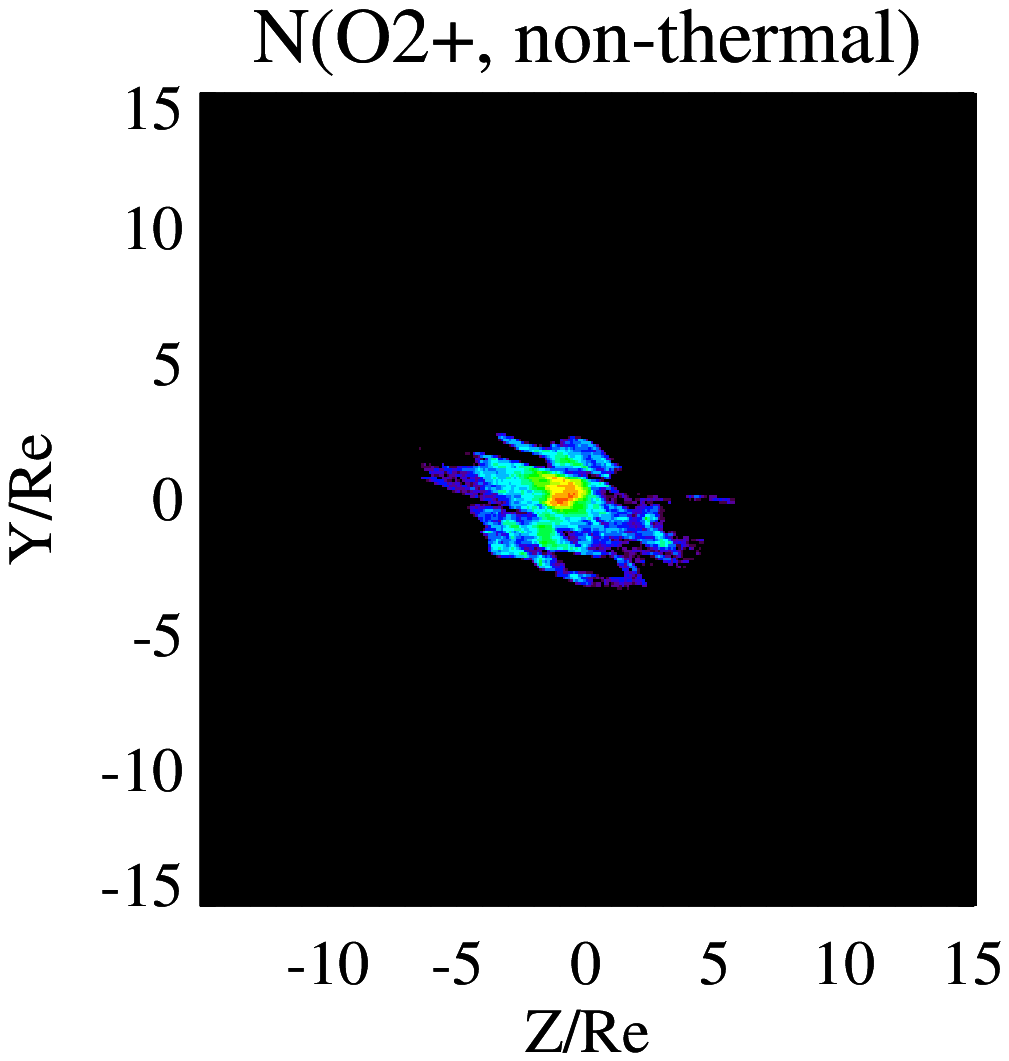}
\includegraphics{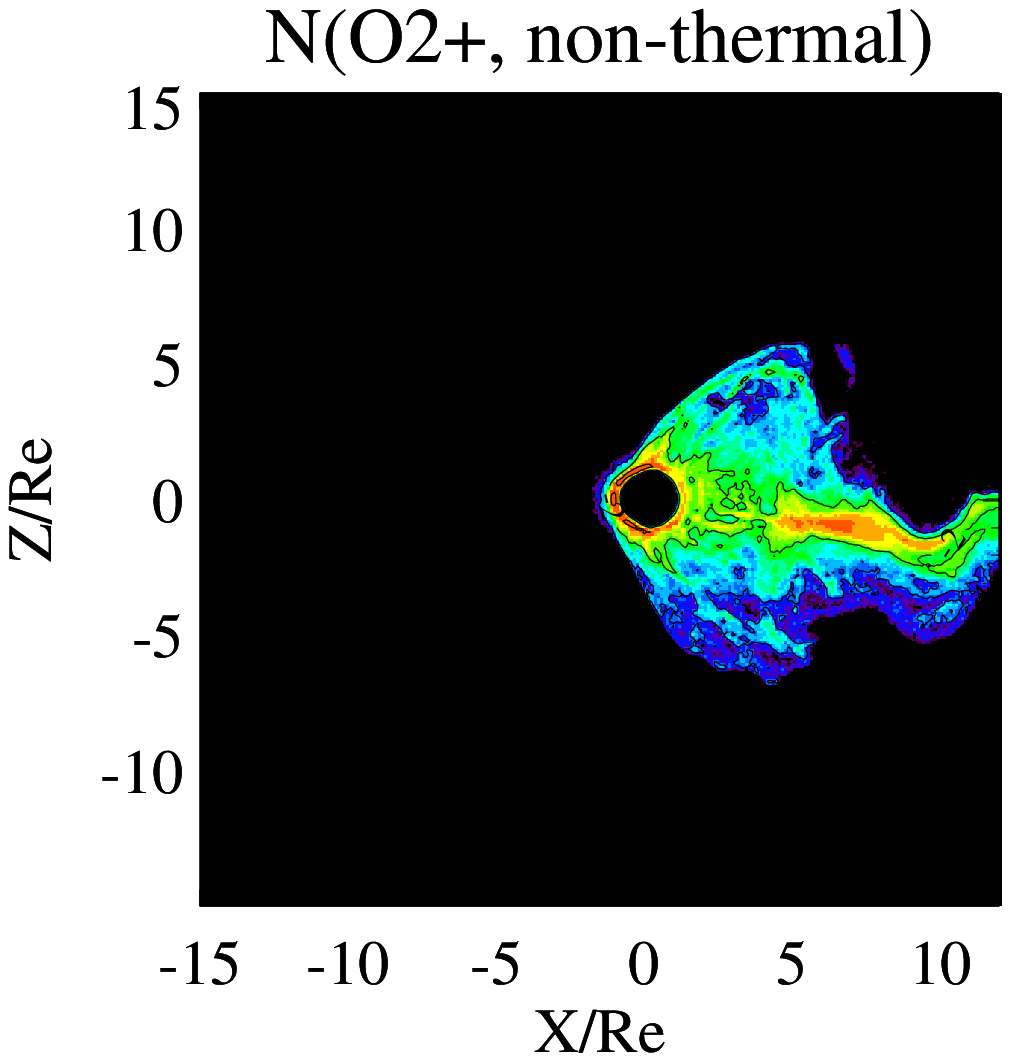}
\includegraphics{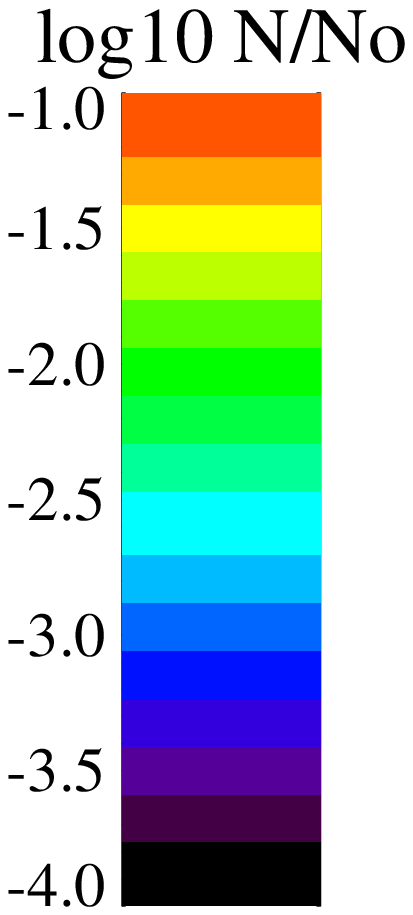}
\caption{2-D cuts of non-thermal pickup ion $O_2^+$ density profile.
Model I, case (a) (top) and case (b) (bottom). 
$x-y$ cuts (left column) are located at $z=0$, 
$y-z$ cuts are located at $x/R_{E}=7$, and $x-z$ cuts (right column) are
located at $y=0$.}
\label{fig:2}
\end{figure}

\begin{figure}
\vspace*{12.cm}
\includegraphics{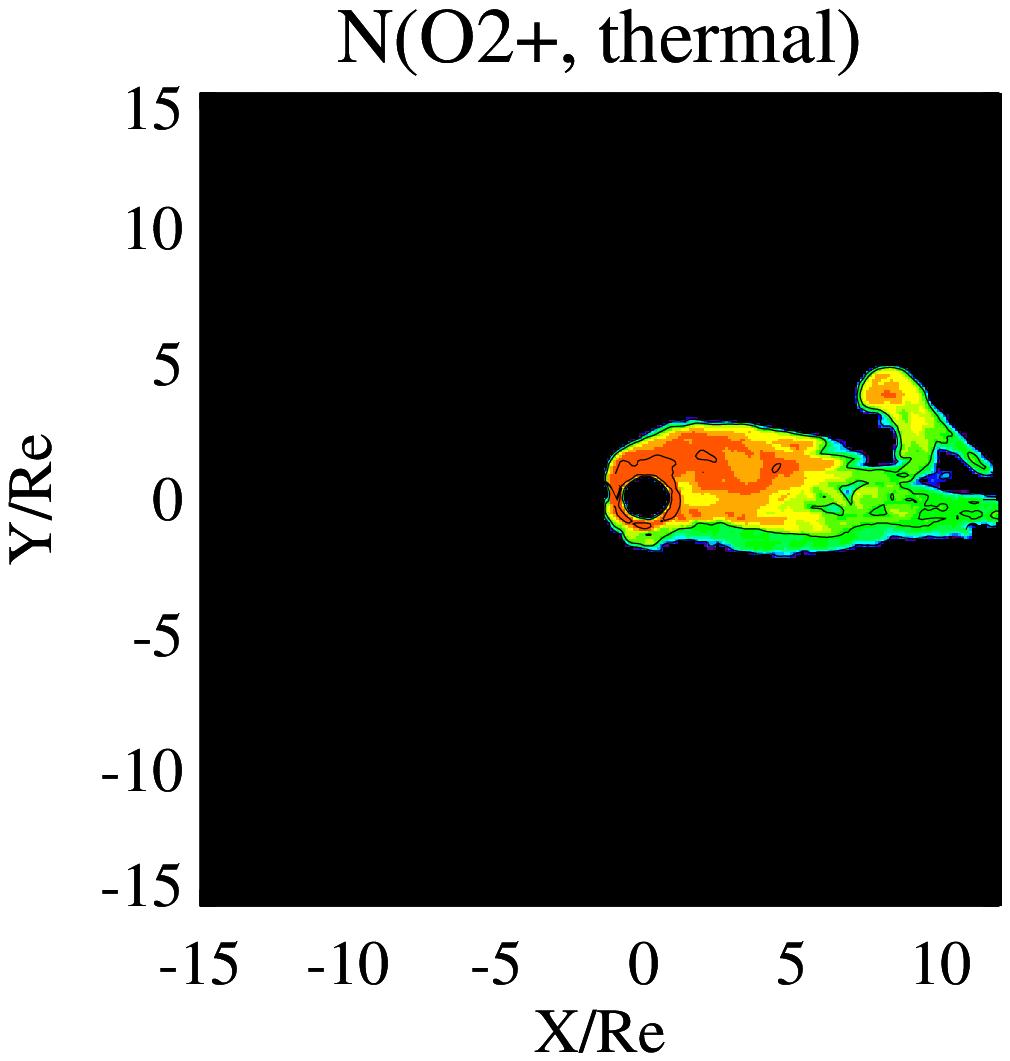}
\includegraphics{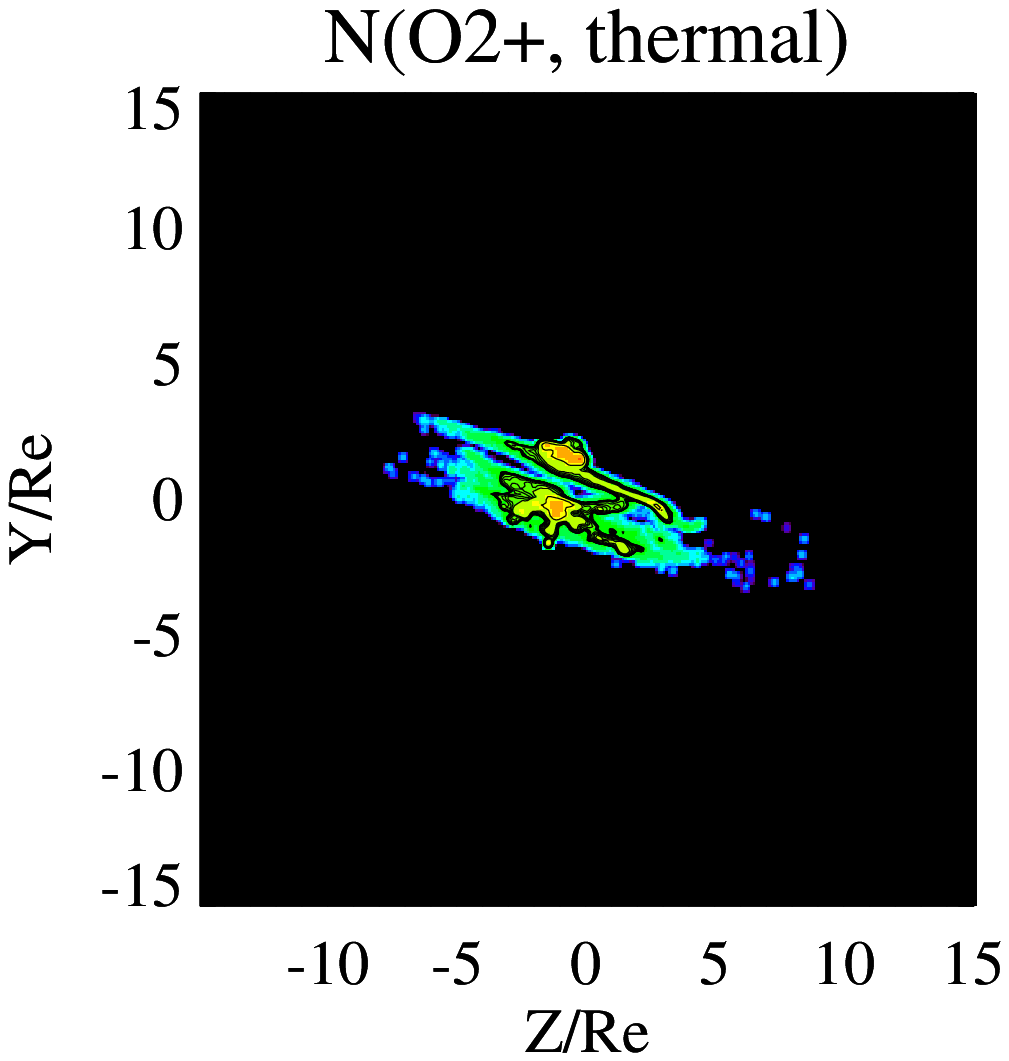}
\includegraphics{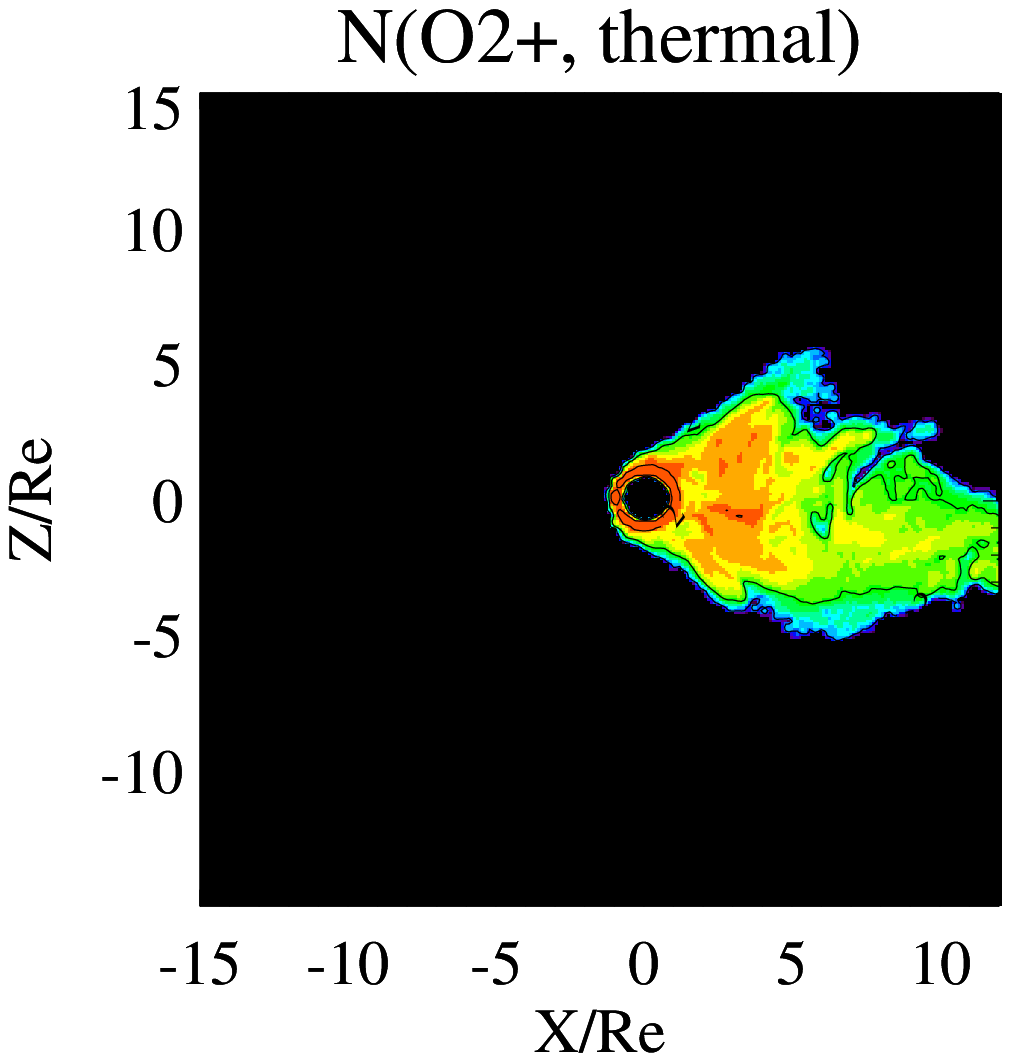}
\includegraphics{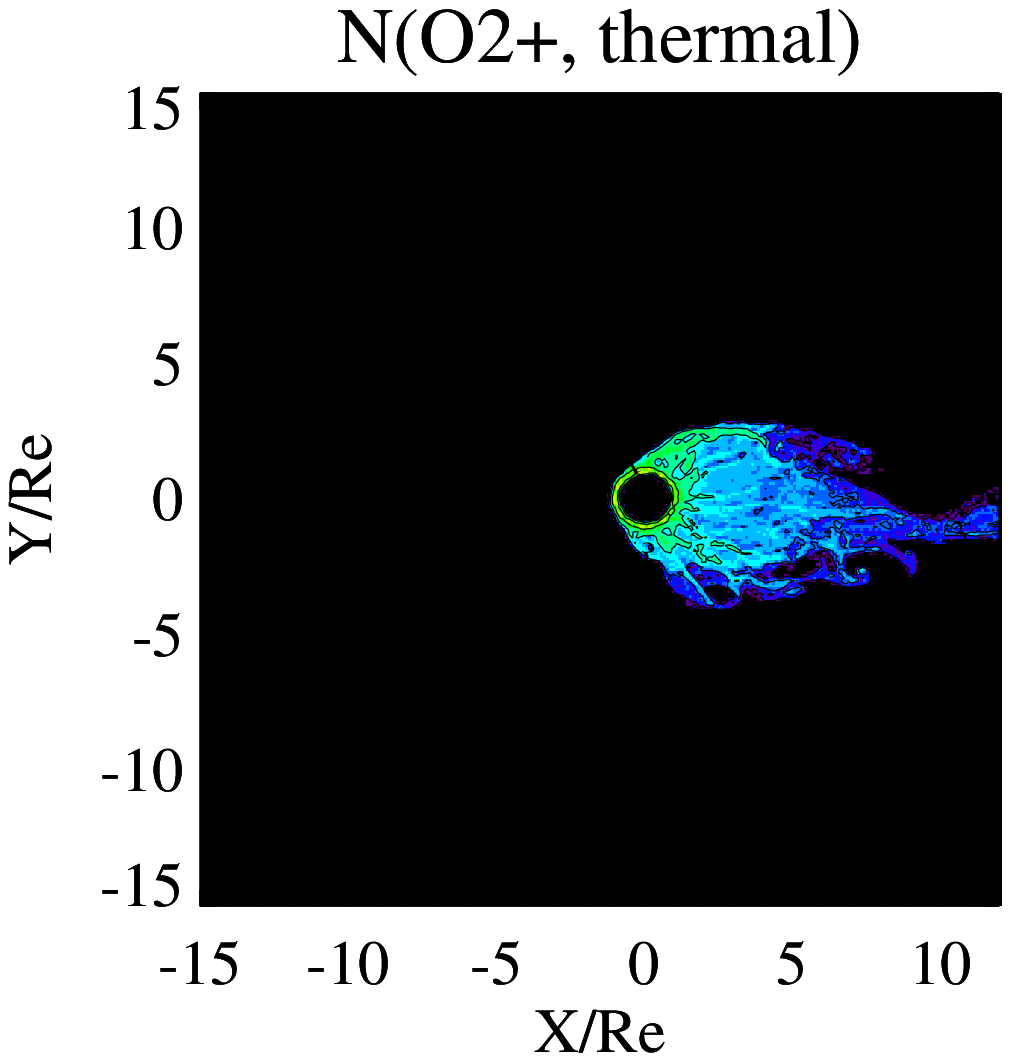}
\includegraphics{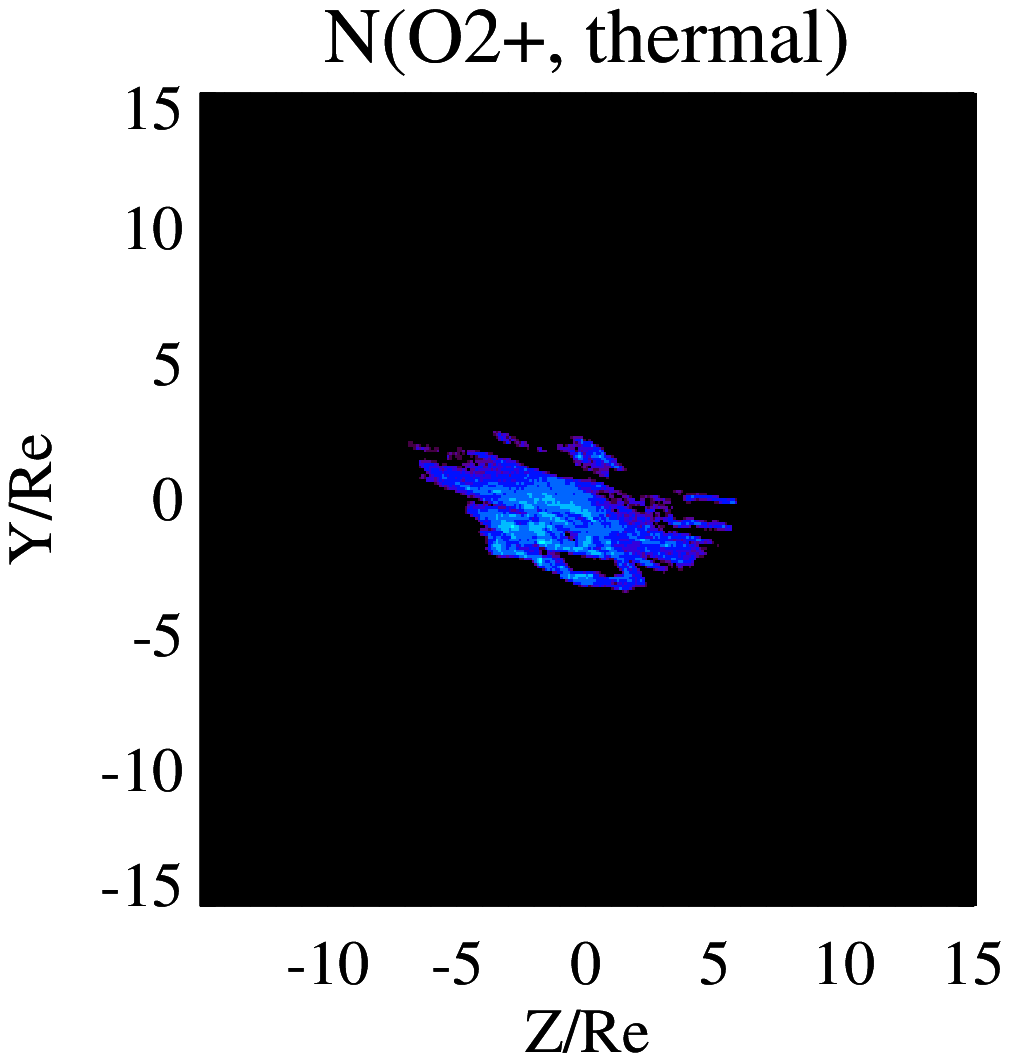}
\includegraphics{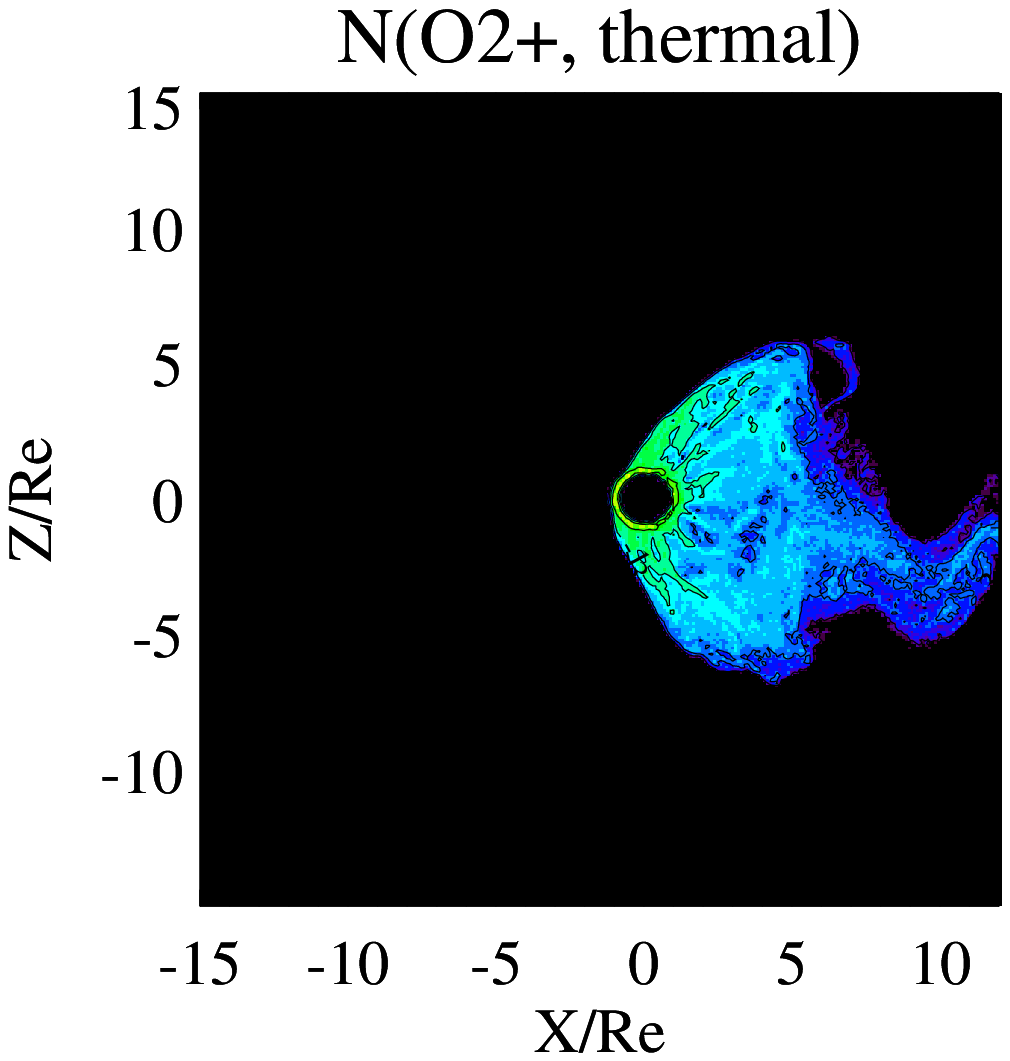}
\includegraphics{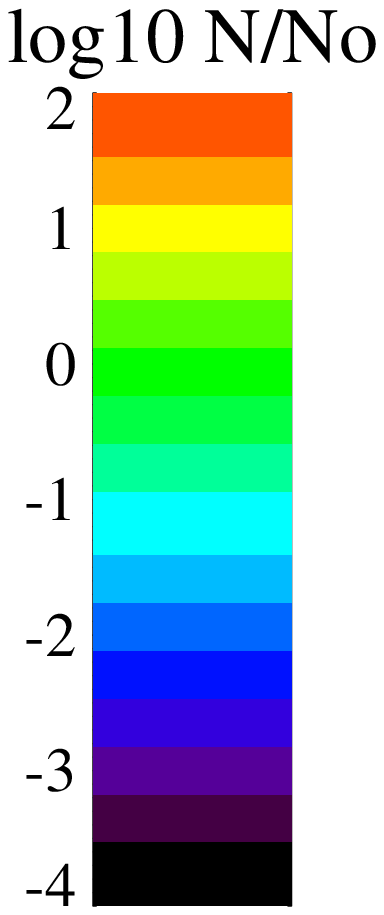}
\caption{2-D cuts of the thermal pickup ion $O_2^+$ density profile.
Model I, case (a) (top) and case (b) (bottom). 
$x-y$ cuts (left column) are located at $z=0$, 
$y-z$ cuts are located at $x/R_{E}=7$, and $x-z$ cuts (right column) are
located at $y=0$.}
\label{fig:3}
\end{figure}

\begin{figure}
\vspace*{12.cm}
\includegraphics{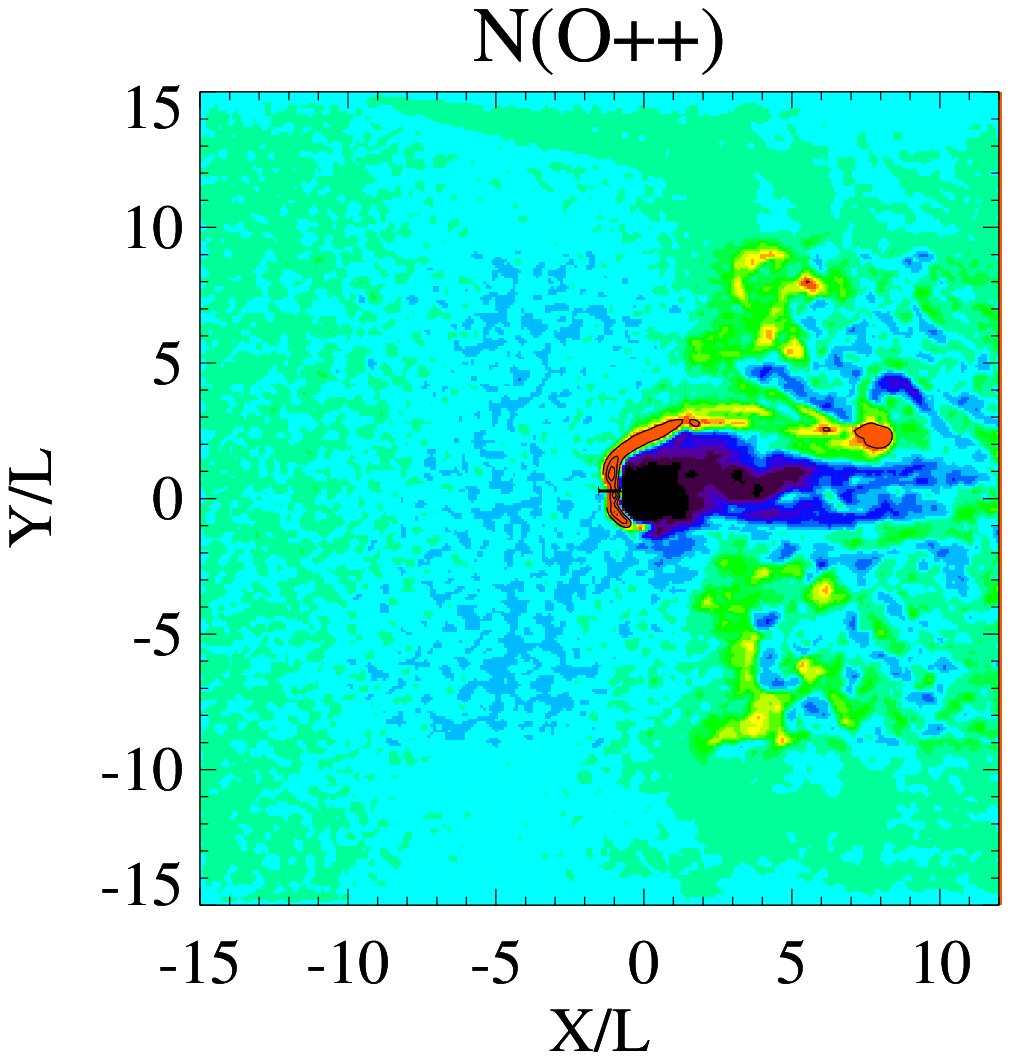}
\includegraphics{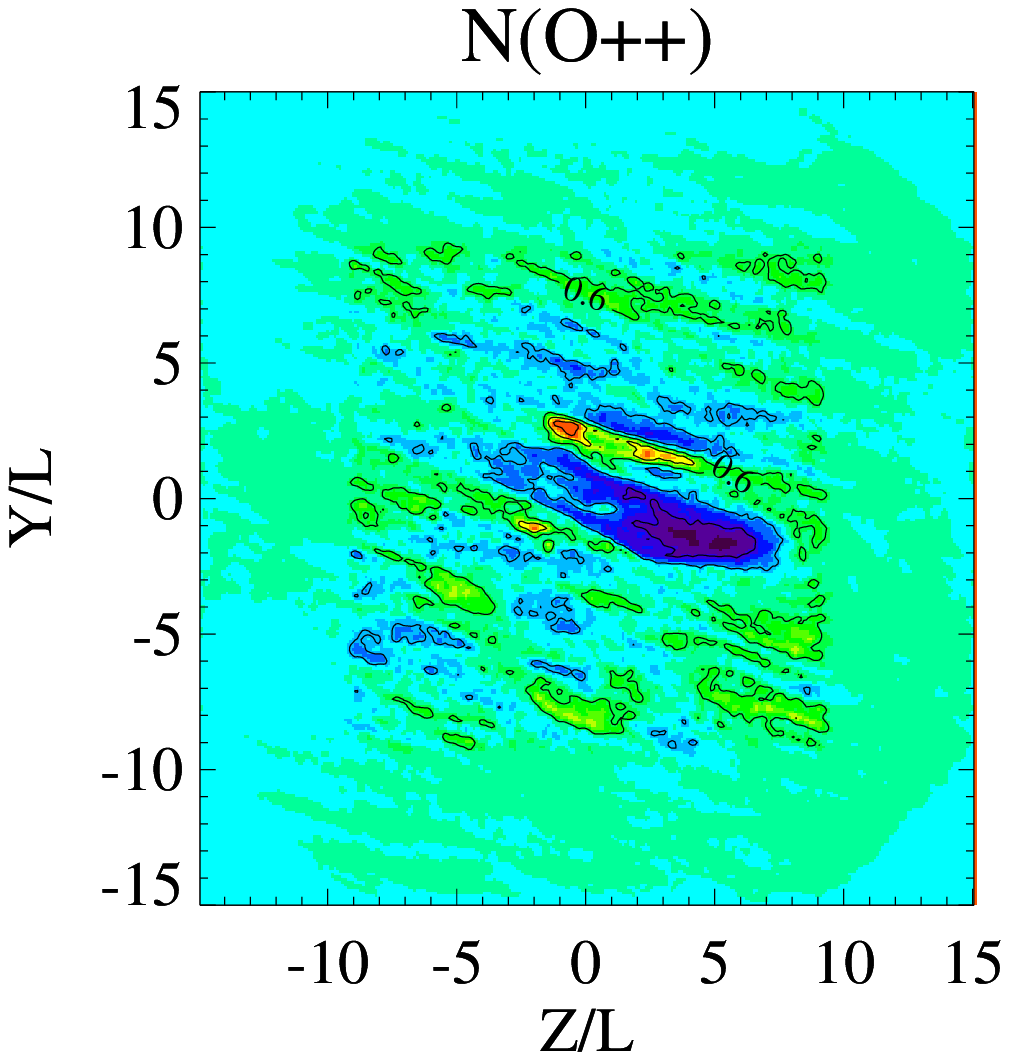}
\includegraphics{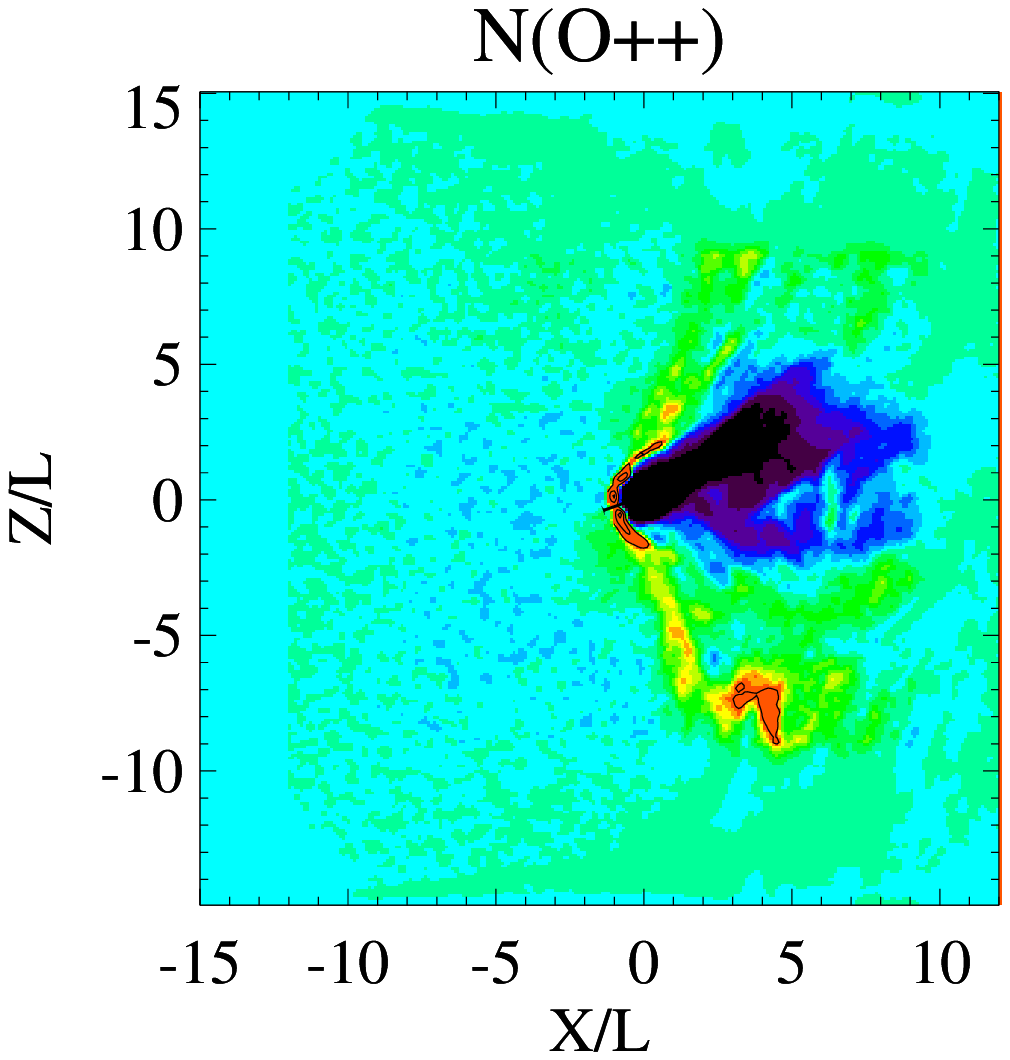}
\includegraphics{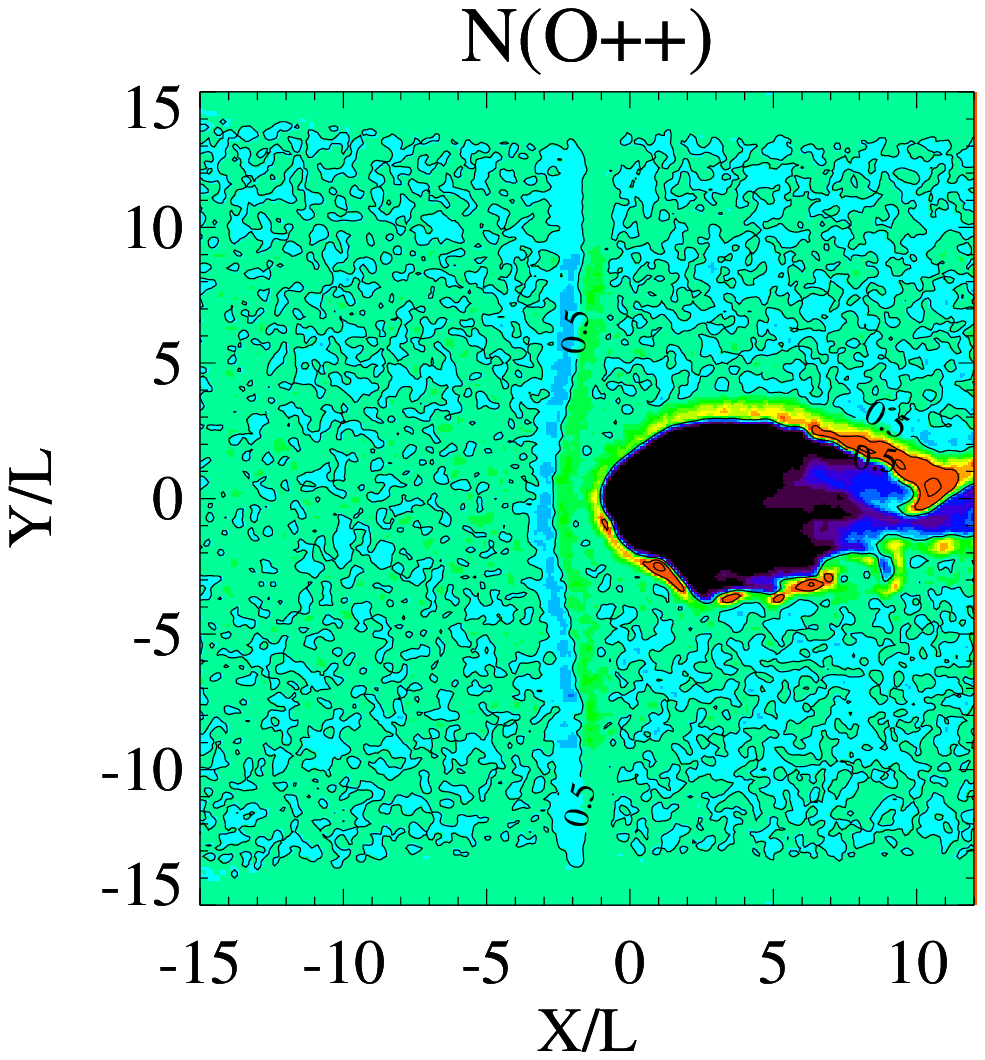}
\includegraphics{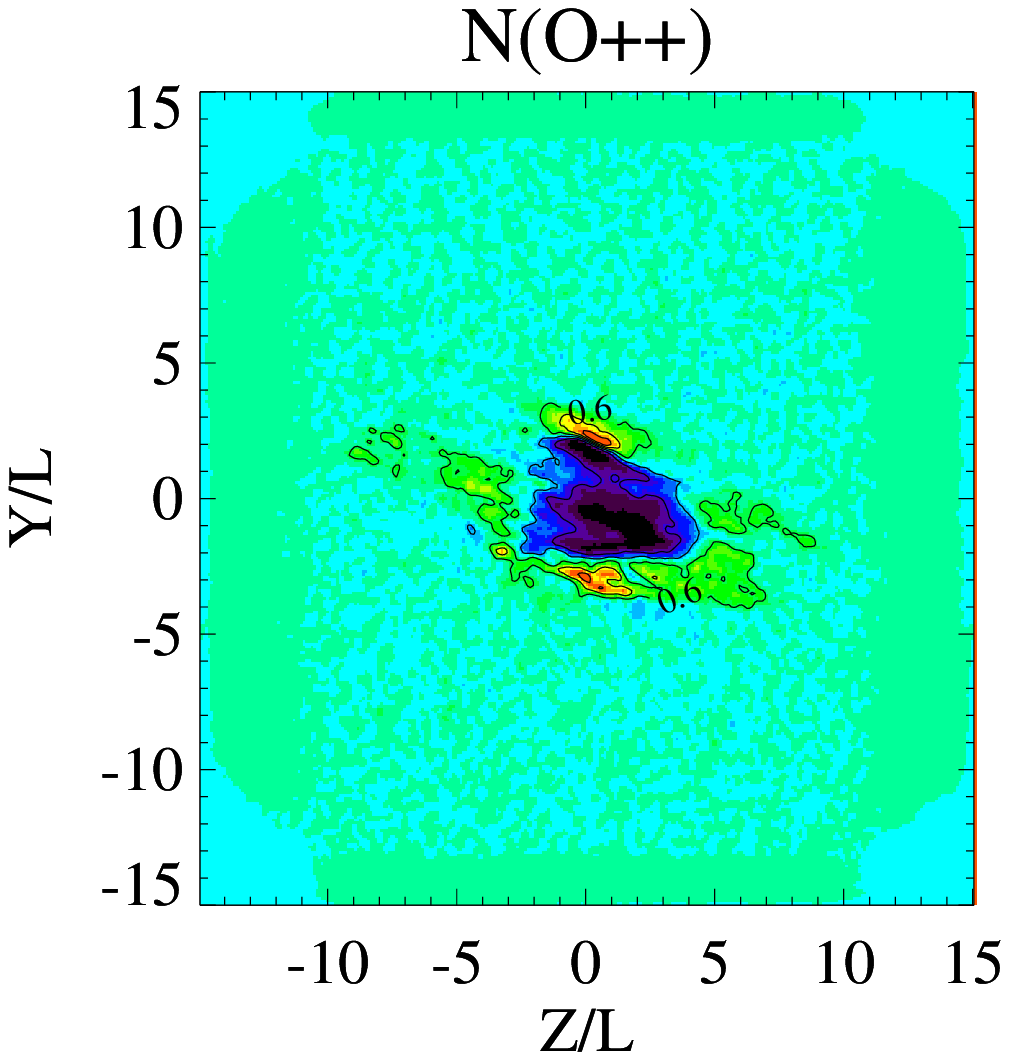}
\includegraphics{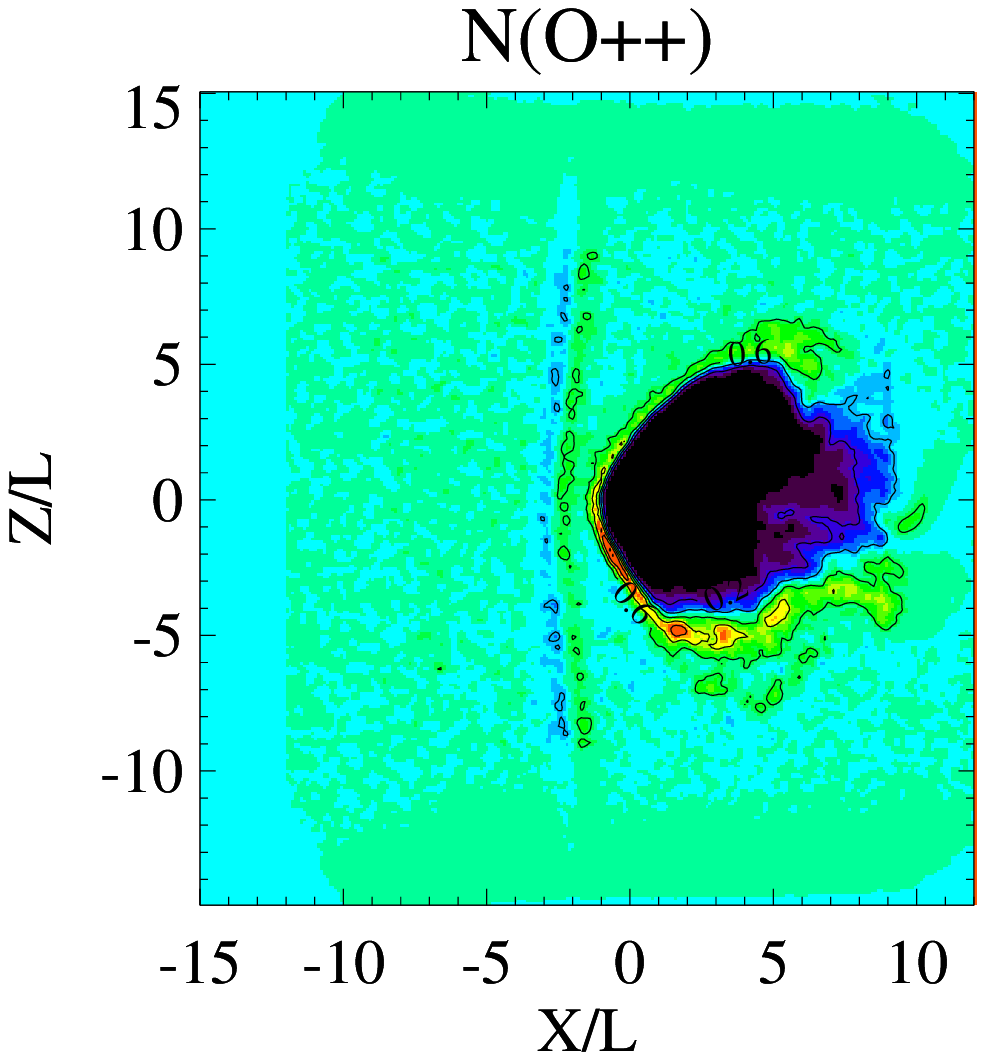}
\includegraphics{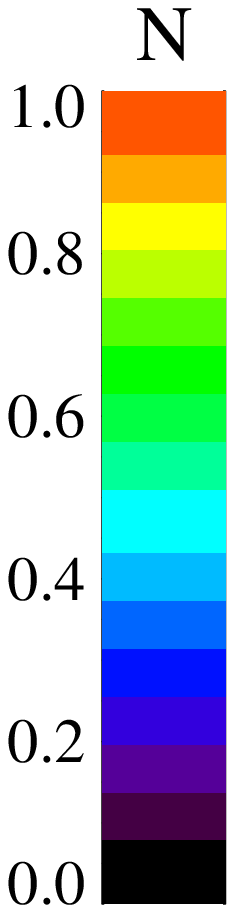}
\caption{2-D cuts of the background $O^{++}$ ion density profiles.
Model I, case (a) (top) and case (b) (bottom). 
$x-y$ cuts (left column) are located at $z=0$, 
$y-z$ cuts are located at $x/R_{E}=7$, and $x-z$ cuts (right column) are
located at $y=0$.}
\label{fig:4}
\end{figure}

Let us consider first the global picture of the
interaction of the plasma torus with Europa.
The results of this modeling are shown in Figs. \ref{fig:2}, \ref{fig:3},
and \ref{fig:4}.
Figures \ref{fig:2} and \ref{fig:3} demonstrate 2D cuts
for non-thermal and thermal $O_2^+$ pickup ion density profiles.
One can observe the asymmetrical 
distribution of the pickup ion density (top, case (a)) and 
(bottom, case (b)) in the $x$-$y$, $y$-$z$\,$(x=5\,R_{E})$ 
and $z$-$x$ planes.  
The pickup ion motion is determined mainly by
the electromagnetic drift. 
The motion along the magnetic field is due to
the thermal velocity and the gradient of the electron pressure.
A more wider density profile of the 
pickup ions was observed in the case (b), Figs.~\ref{fig:2} and 
\ref{fig:3} (bottom).

The figures demonstrate a strong structuring in the non-thermal
and thermal $O_2^+$ ion density profiles.
While case (a) produces a much higher peak in the thermal $O_2^+$ ion
density as was seen in E4 observations, case (b)
produces much better agreement with observation for 
the thermal $O_2^+$ ion density
as shown in Figs. \ref{fig:2} and \ref{fig:3}.

The modeling also demonstrates the asymmetrical 
distribution of the background $O^{++}$ ion density in the 
$x$-$y$, $y$-$z$\,$(x=5\,R_{E})$ and $z$-$x$ planes, 
Fig. \ref{fig:4}. 
The asymmetrical distribution of the background 
ions in the $x$-$y$ plane may be explained by the 
existence of a strong $B_z$ component in the upstream 
magnetic field. One can also see an increase in the plasma
density near Europa due to the formation of a magnetic barrier 
(not shown here).
In case (b) this effect is stronger than in case (a).
The density profiles for $SO^{++}$ background ions are close 
to the density profiles for $O^{++}$ ions.

The inclination of the magnetic field results in an asymmetrical 
boundary condition for ion dynamics (penetration and reflection) 
in Europa's ionosphere and an asymmetrical Alfv\'en wing.

Note that the background ion flow around the effective obstacle 
that is produced by pickup ions and the ionosphere.
The pickup ions flow from the ``corona'' across the magnetic 
field due to electromagnetic drift, 
whereas the motion along the magnetic field is
determined by the thermal velocity of ions 
and the electron pressure.

\begin{figure}
\vspace*{14.cm}
\includegraphics{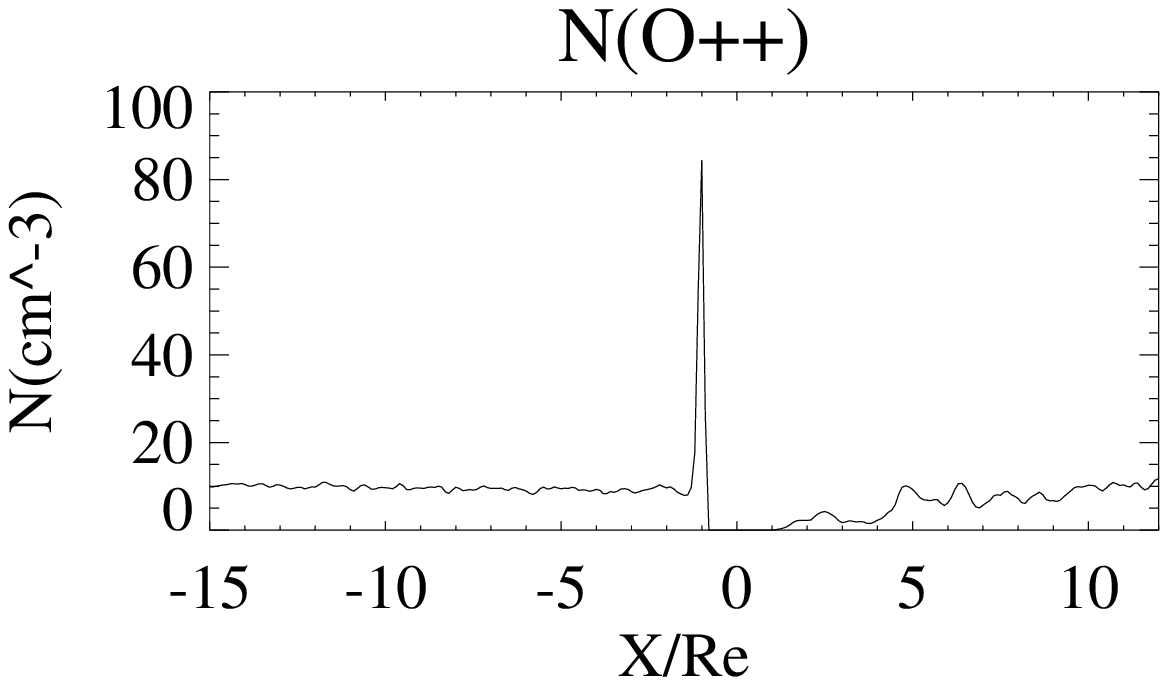}
\includegraphics{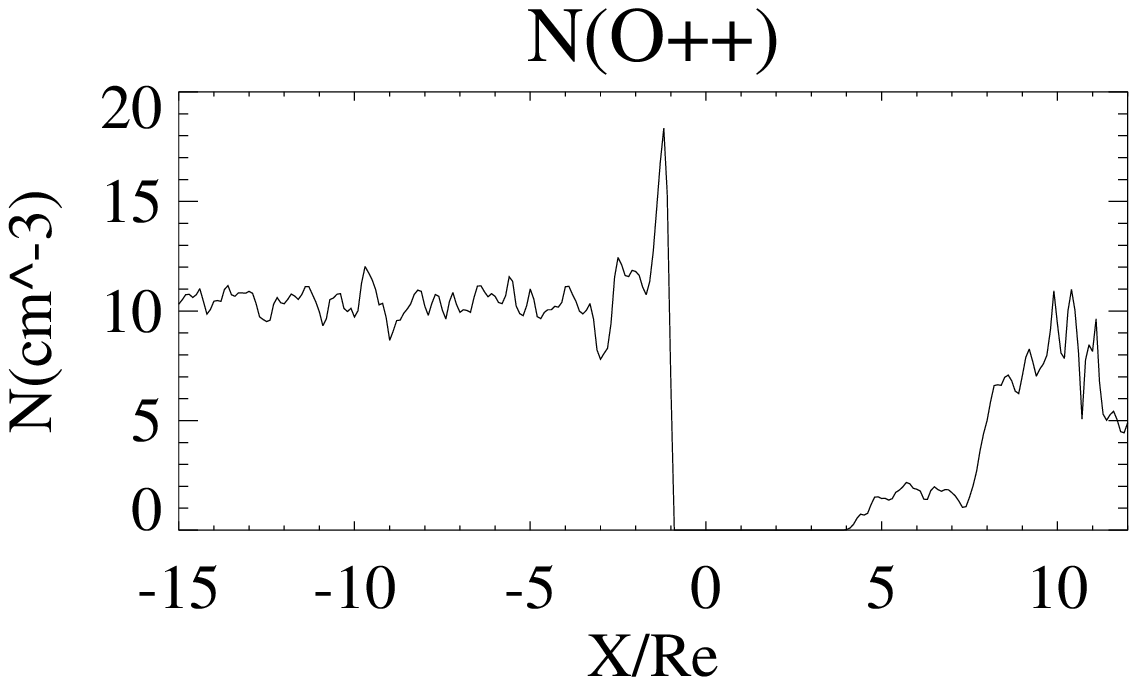}
\caption{1D cuts of the background $O^{++}$ ion density profile. 
The cuts are located at $y=0$, $z=0$. 
Model I, case (a) (top) and case (b) (bottom).} 
\label{fig:5}
\end{figure}

Figure \ref{fig:5} demonstrates the 1D cuts ($y=0$, $z=0$) 
of the background density $O^{++}$ for case (a) (top) 
and case (b) (bottom).
Strong jumps in the plasma density with 
$N_{O^{++}, max}= 80$cm\,$^{-3}$ 
(case (a)) and $N_{O^{++}, max}= 17$cm\,$^{-3}$ (case (b)) are observed 
on the day-side
of the ionosphere, whereas a reduction in the plasma density 
is observed in the plasma wake. 
Note that the jump in the plasma density
profile is stronger in case (a) than it is in case (b). 
Both jumps are located near the surface of Europa.

\begin{figure}
\vspace*{14.cm}
\includegraphics{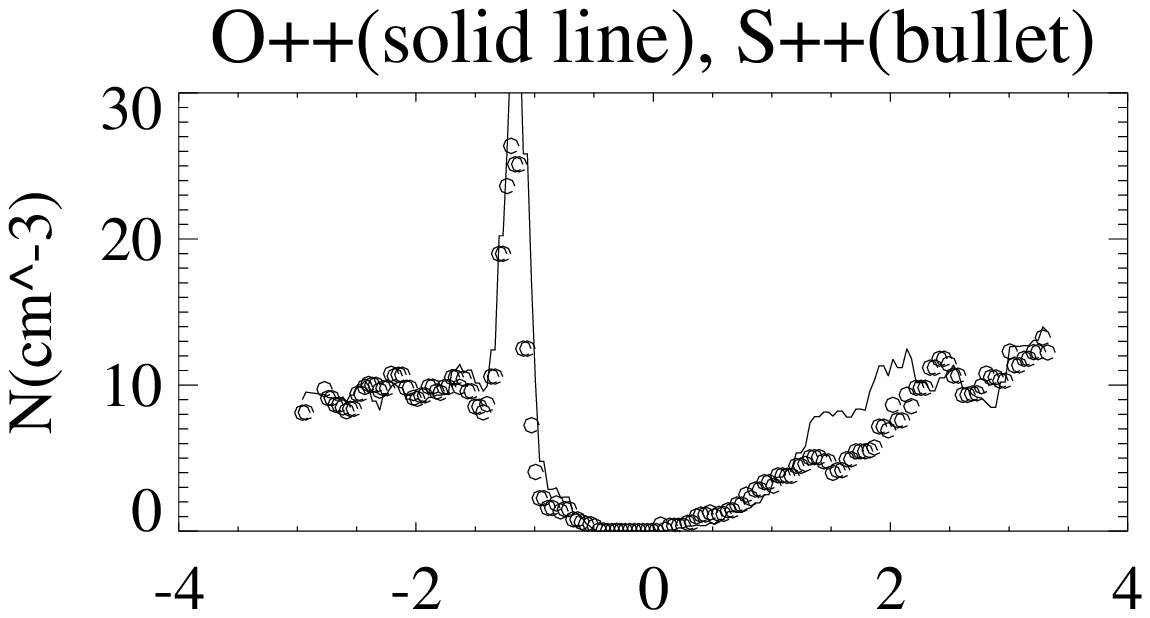}
\includegraphics{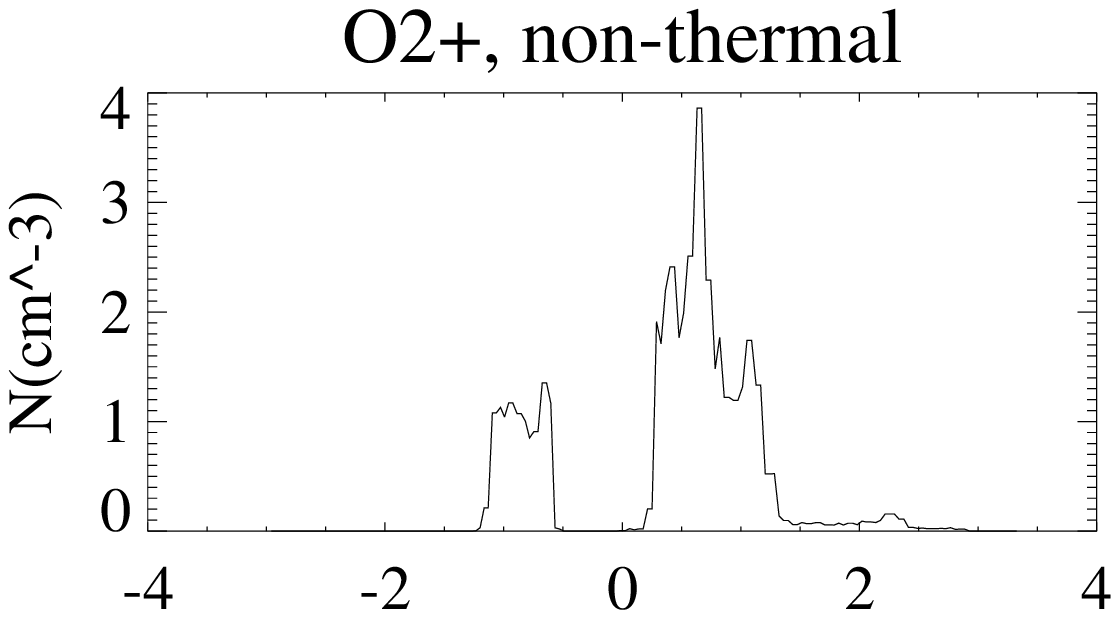}
\includegraphics{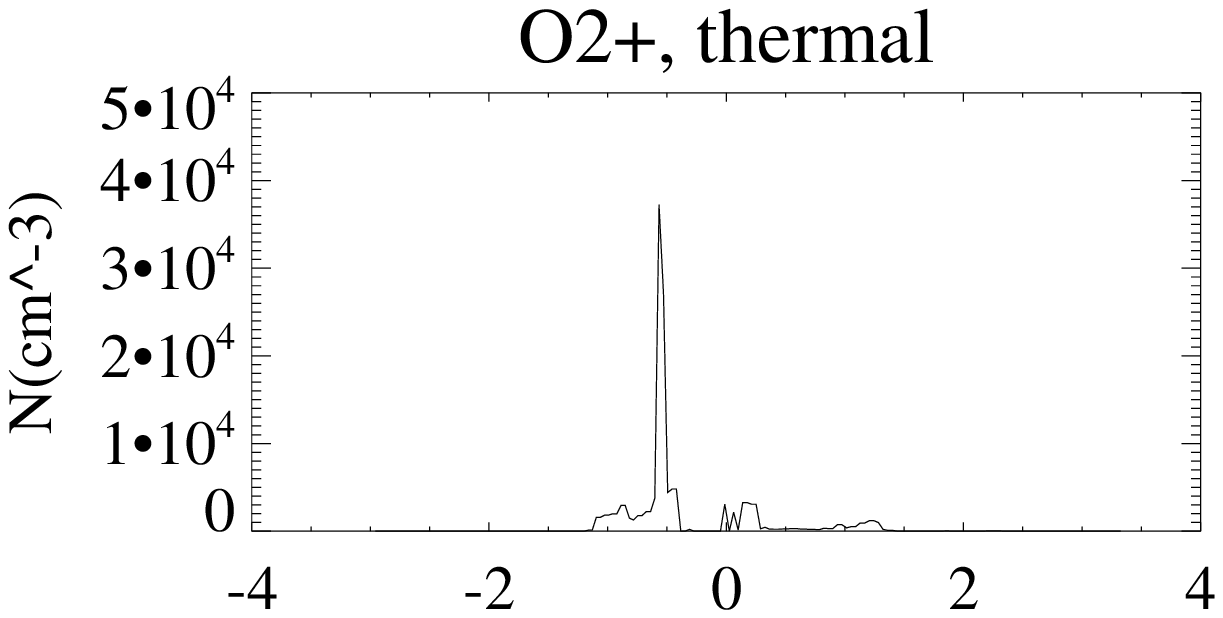}
\includegraphics{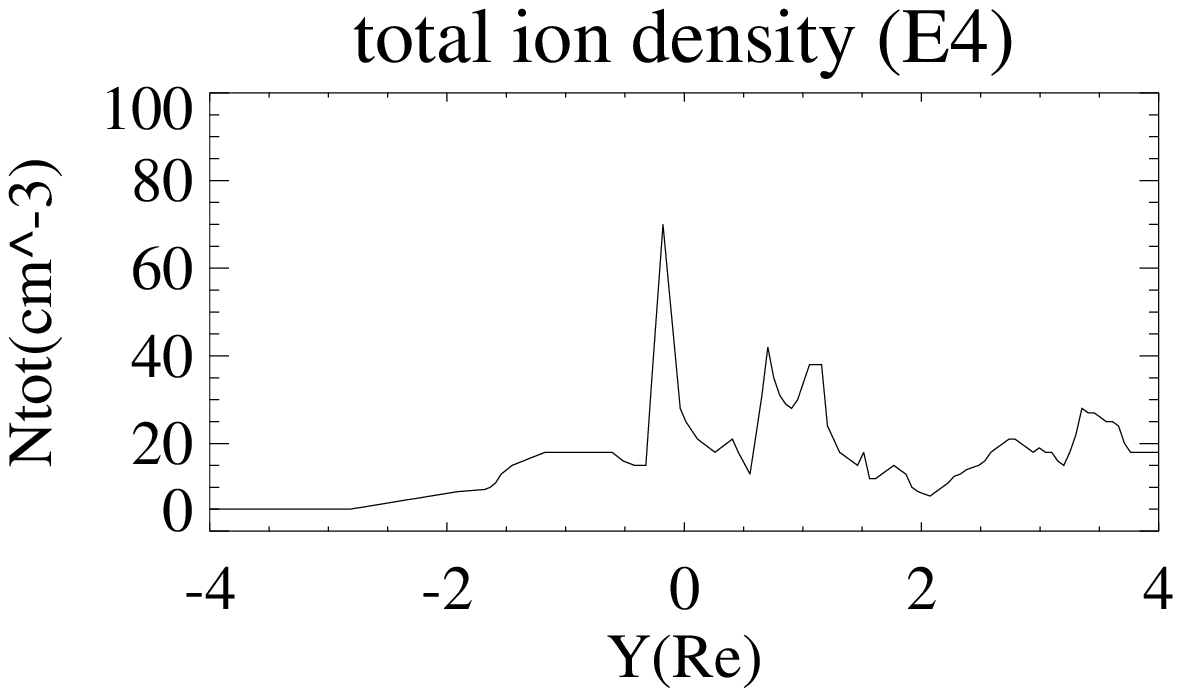}
\includegraphics{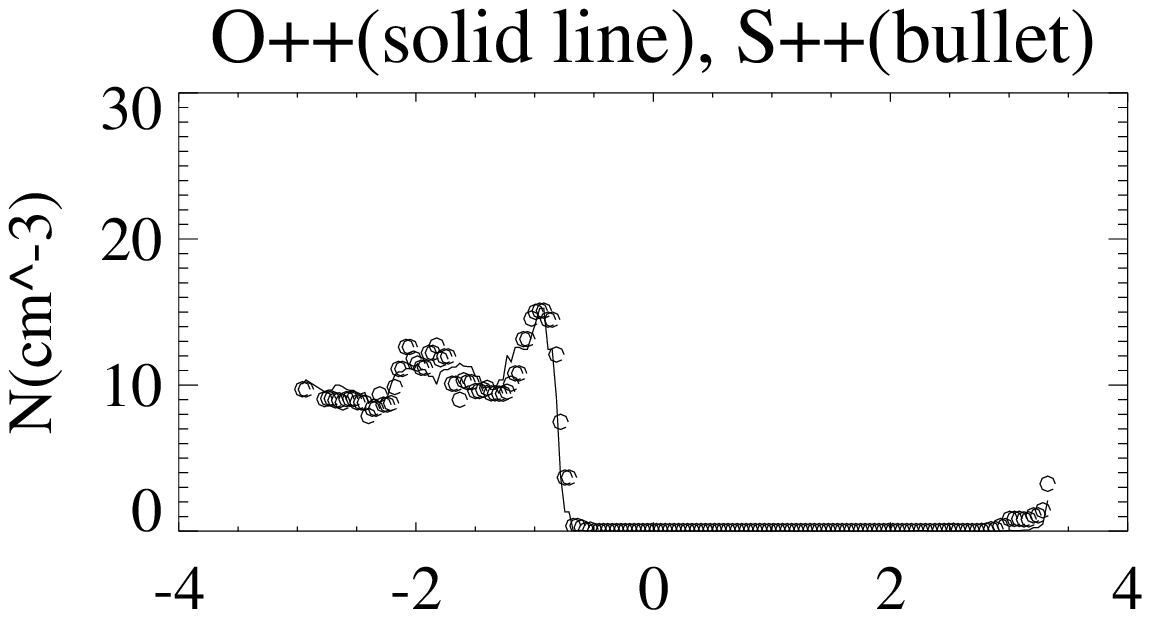}
\includegraphics{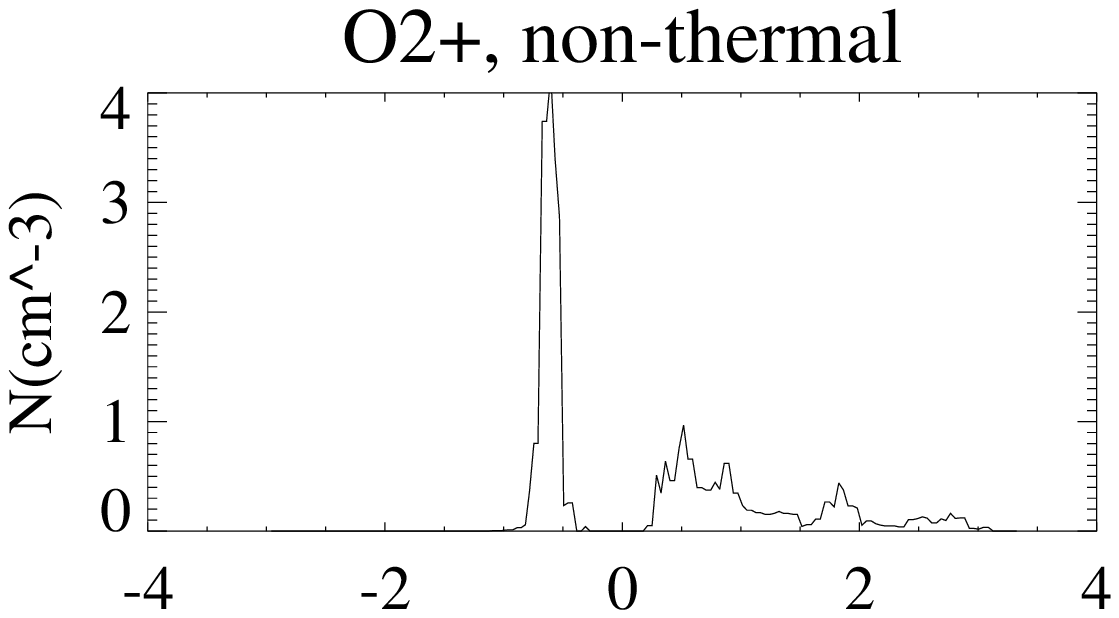}
\includegraphics{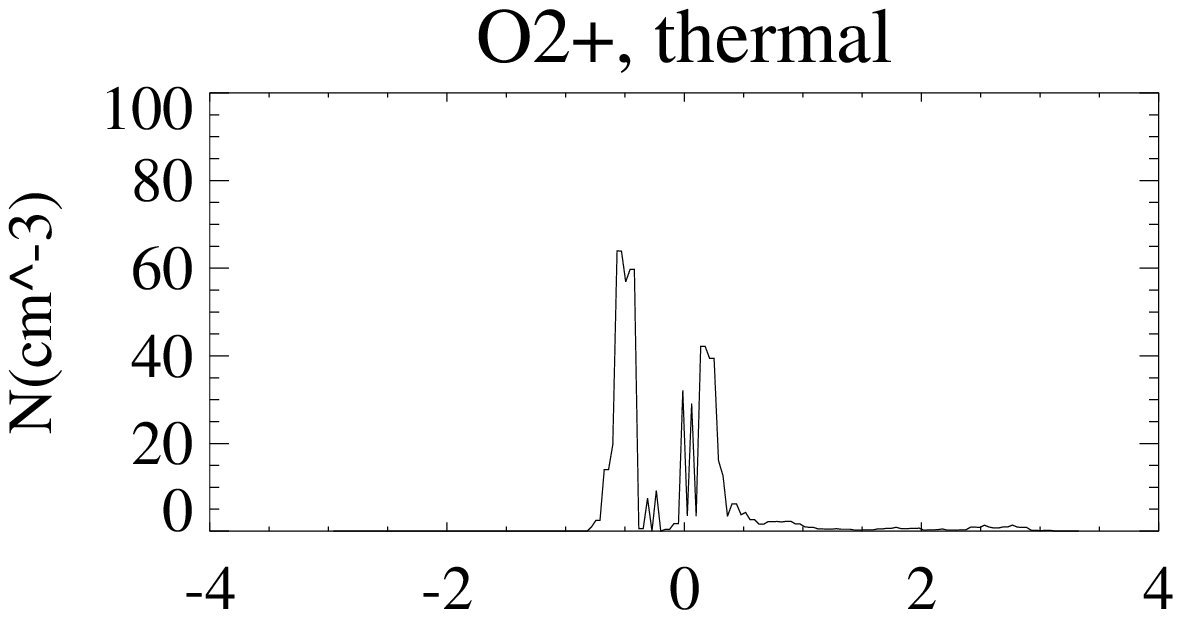}
\includegraphics{nicap17.eps}
\caption{1D cuts of the background $O^{++}$/$S^{++}$, and pickup non-thermal/thermal 
($O_2^+$) ion densities from simulation. 
$Y(Re)$ denotes a projection of satellite trajectory onto the $y$ axis.
Model I, case (a) (left) and case (b) (right). 
Bottom - E4 observation of the total ion density (Paterson et al. 1999).}
\label{fig:6}
\end{figure}

Figures~\ref{fig:6} shows 1-D density profiles of the background
and pickup ions along the E4 trajectory of the Galileo spacecraft.
One can see a strong plasma void in the center of the plasma wake.
There is also a sharp boundary with an overshoot in the density 
profiles on the side of the plasma wake in the 
Jupiter-direction, and a smooth boundary layer 
on the side in the anti-Jupiter direction, 
Fig.~\ref{fig:6} (top).
The density profile  for $O^{++}$ is similar the density 
profiles for the $S^{++}$ upstream ions.
Fig.~\ref{fig:6} (middle and bottom) also shows the density profiles
for the non-thermal (top) and thermal (bottom) $O_2^{+}$ pickup 
ions. One can see the split structure of the plasma tail.
The effect of splitting of the plasma tail was also observed in 
the hybrid modeling of weak comets 
(see, e.g., Lipatov, Sauer and Baumg\"atel, 1997; 
Lipatov, 2002).
The general feature of this plasma density is due to the effect
of the finite heavy gyroradius.
The total ion density profile observed in E4 pass is shown in 
Fig.~\ref{fig:6} (bottom).
The observed value of the density in these peaks is lower than 
in modeling and it may be explained by an overestimated 
density of $O_2^{+}$ pickup ions for case (a).
In the case (b), disagreement is not as strong,
an improvement of the atmosphere model is still required.

The modeling gives the following total fluxes for the $O_2^{+}$ 
pickup ions (case (a)): $1.4\times 10^{22}$\,mol/s (non-thermal) 
and $1.75 \times 10^{25}$\,mol/s  (thermal);
(case (b)): $0.8\times 10^{22}$\,mol/s (non-thermal) and 
$1.0 \times 10^{25}$\,mol/s (thermal)
across the back boundary $x=12 R_{E}$.

\begin{figure}
\vspace*{18.cm}
\includegraphics{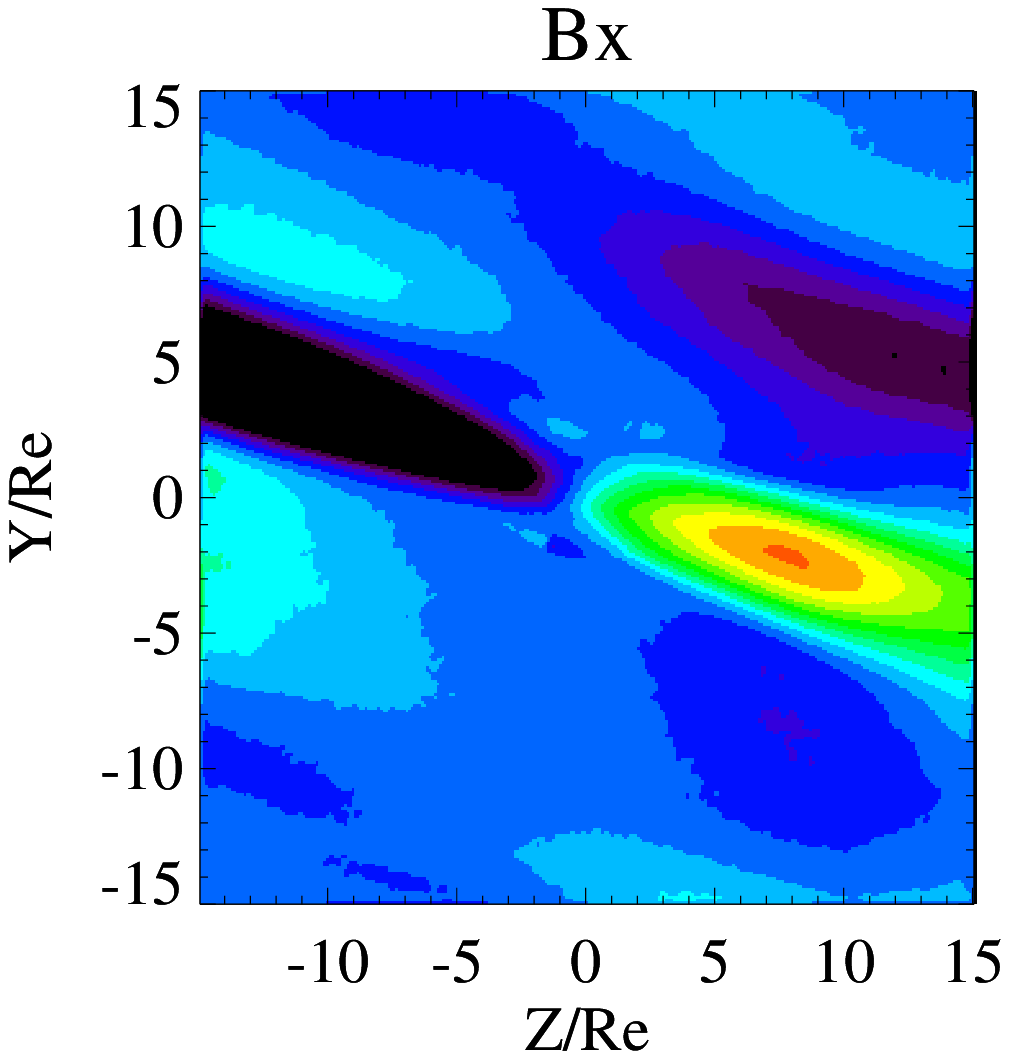}
\includegraphics{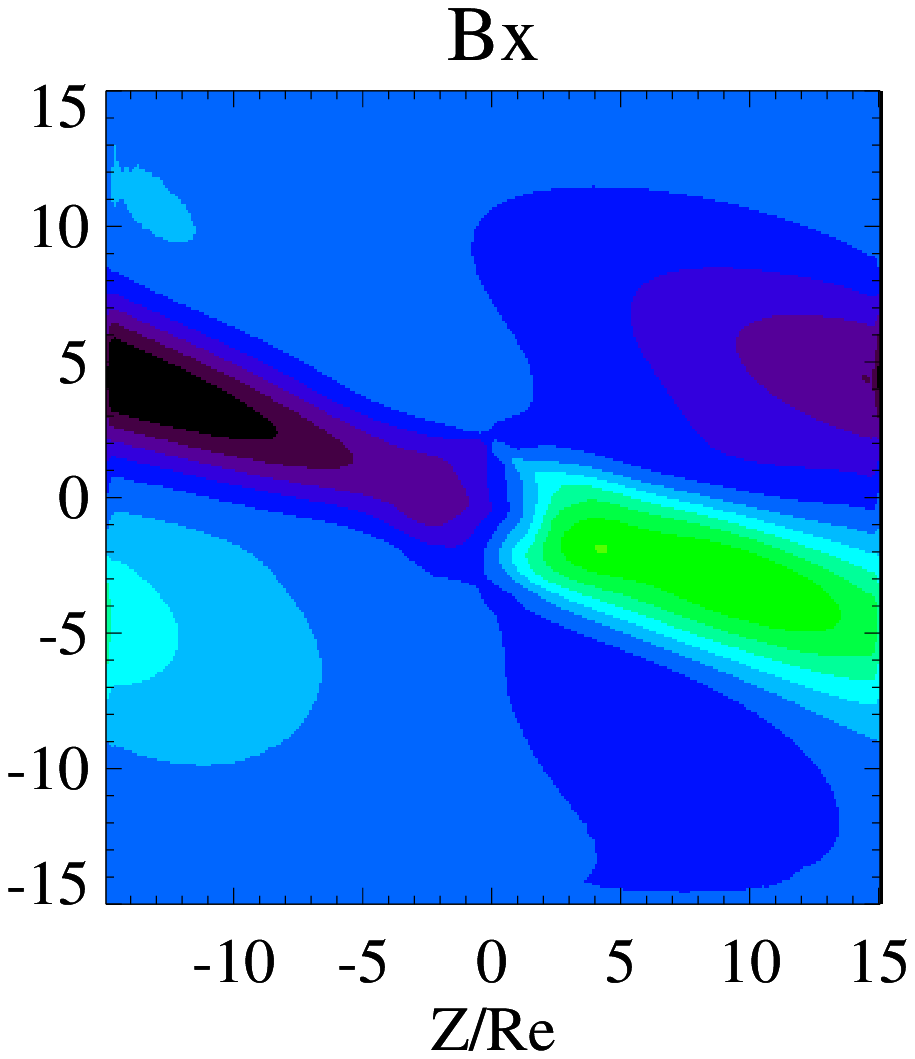}
\includegraphics{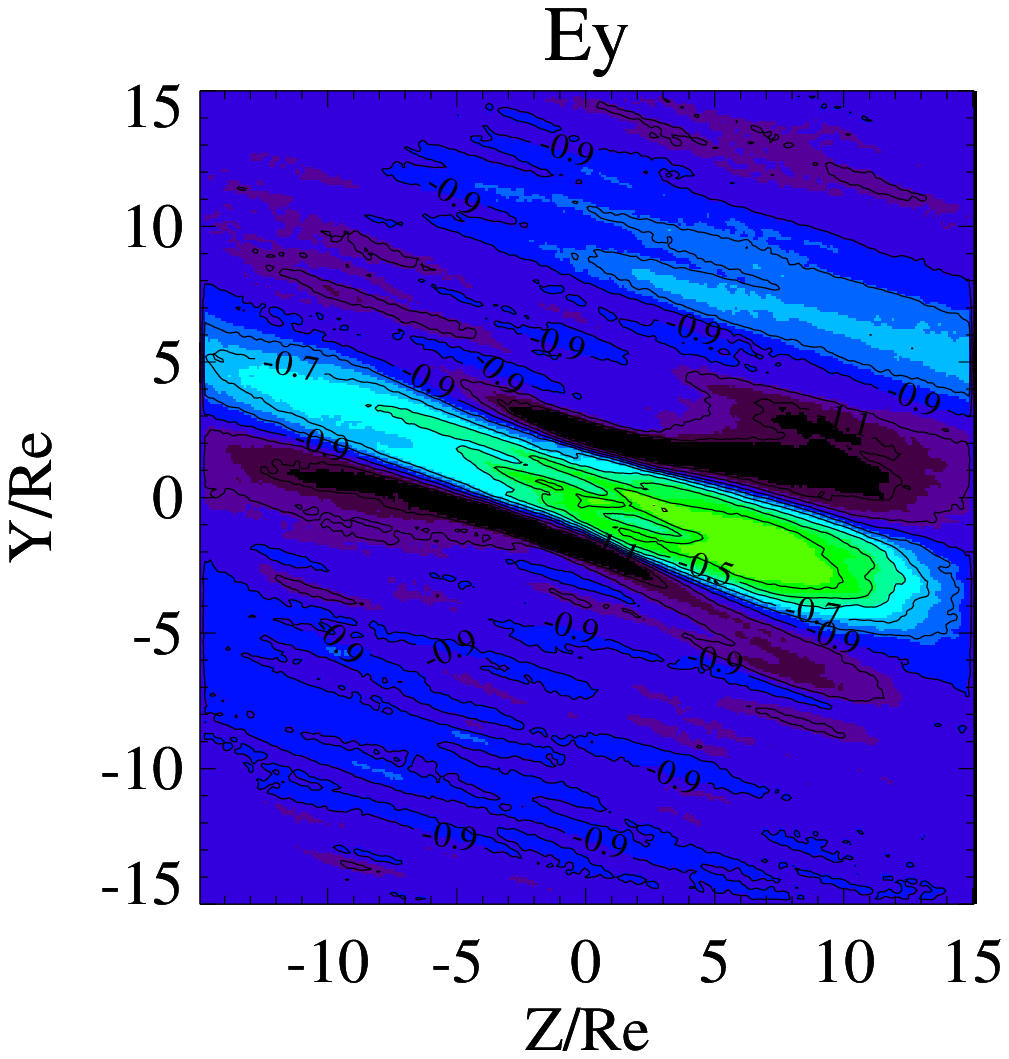}
\includegraphics{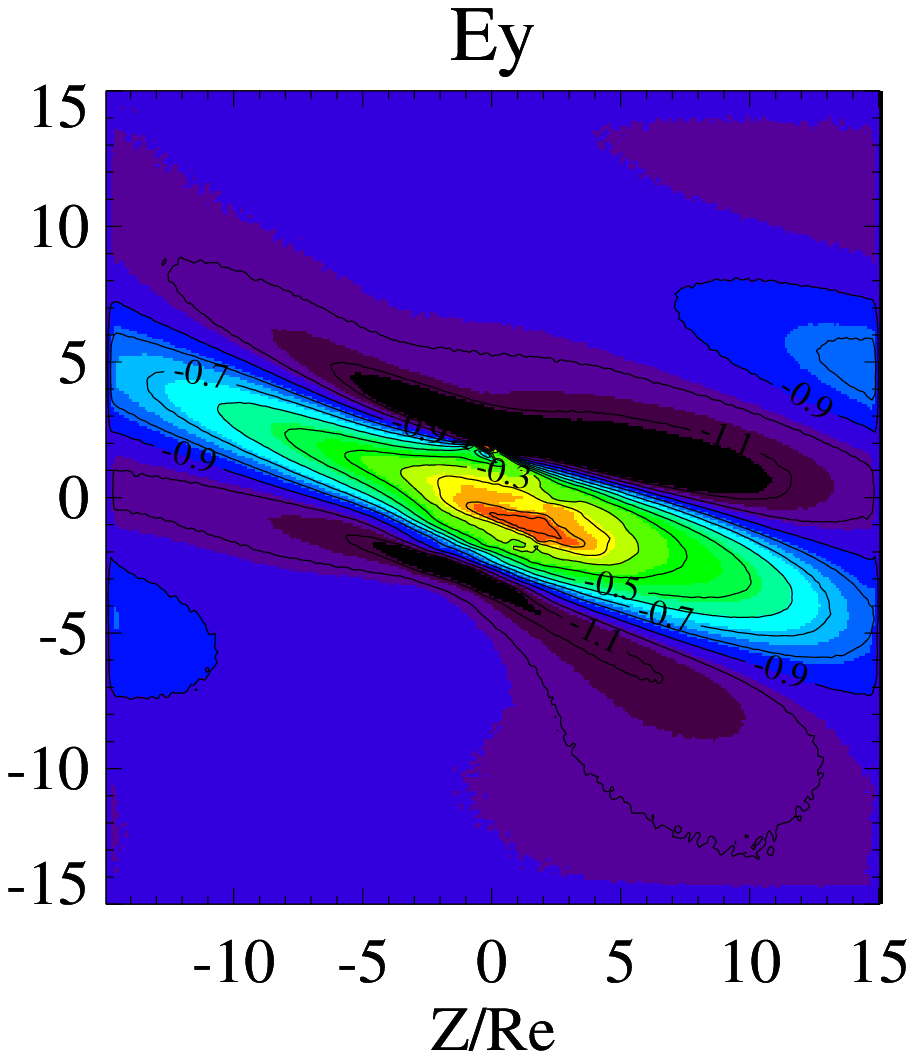}
\includegraphics{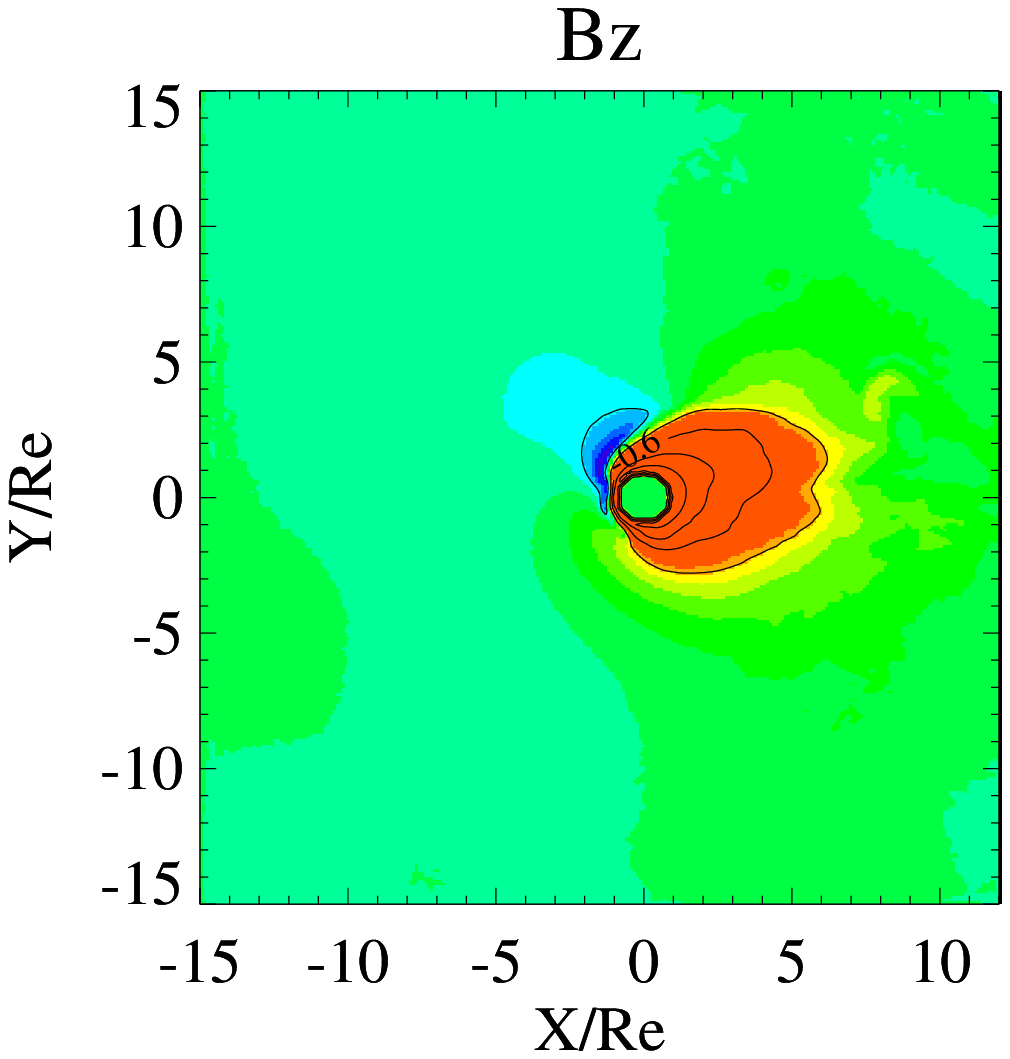}
\includegraphics{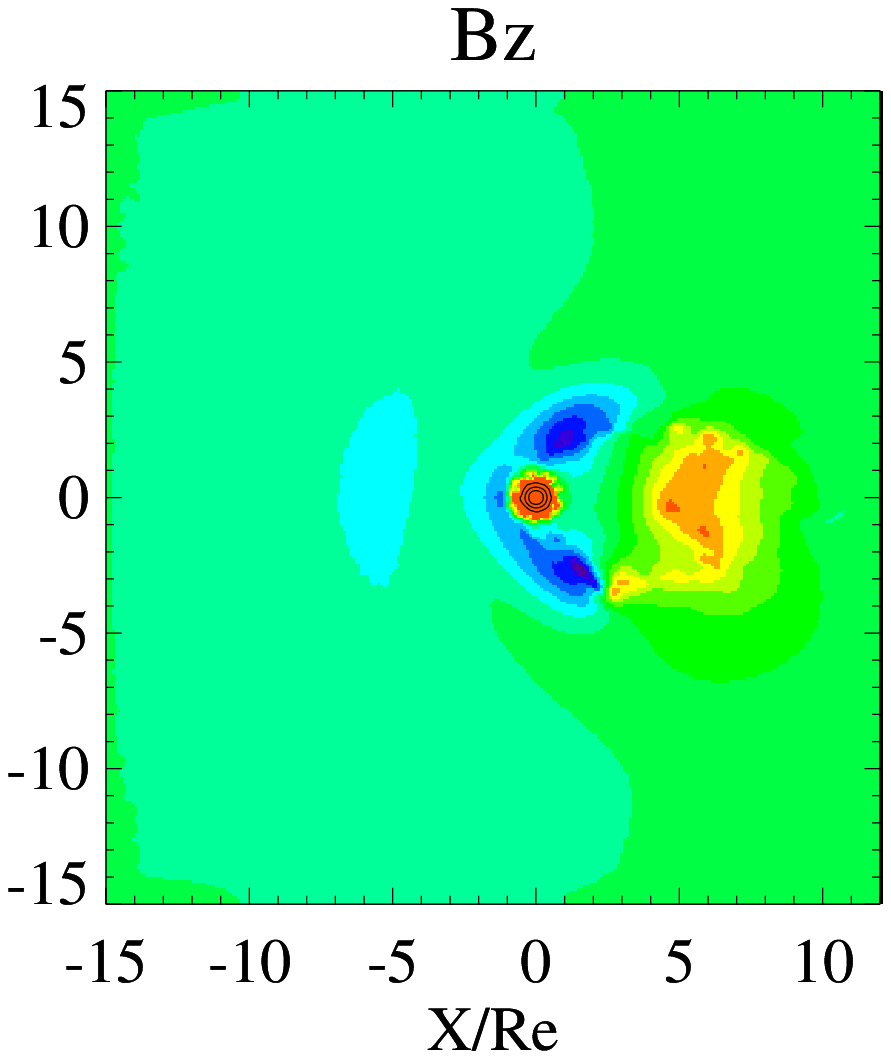}
\includegraphics{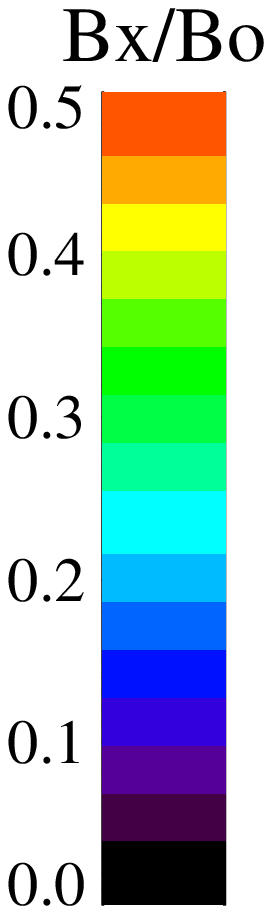}
\includegraphics{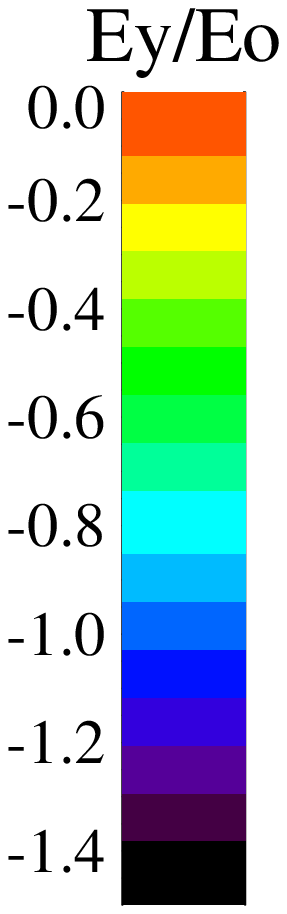}
\includegraphics{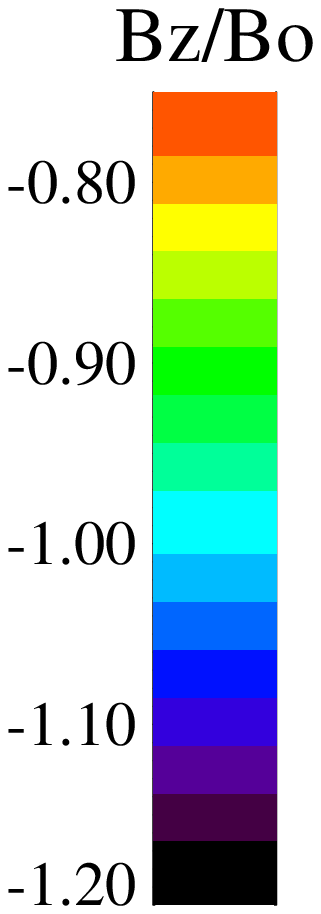}
\caption{2-D cuts of the $B_x$, $B_z$ magnetic 
and $E_y$ electric field profiles.
Model I, case (a) (left) and case (b) (right). 
$y-z$ cuts (top and middle) are located at $x/R_{E}=7$, 
and $y-x$ cuts (bottom) are
located at $z/R_{E}=0$.}
\label{fig:7}
\end{figure}

Let us consider a global distribution of the electric and magnetic field 
in Europa's environment. Figure \ref{fig:7} shows 
$B_x$, $B_z$ magnetic and $E_y$ electric field profiles for case (a)
(left) and case (b) (right). The
$y-z$ cuts (top and middle) are located at $x/R_{E}=7$, 
and $x-y$ cuts (bottom) are located at $y=0$. 
The figure demonstrates perturbations in the 
magnetic $B_x$ and electric $E_y$ field profiles, which are due to 
the  formation of an Alfv\'en wing. 
The increase in the magnetic field $B_z$ indicates the formation of an
asymmetrical magnetic barrier, Fig. \ref{fig:7} (bottom). 

The asymmetry of the modeling distributions in ${\bf B}$ appears 
to be caused by the 
finite gyroradius effects of incoming and pickup ions. 
A weak perturbation of the magnetic field was observed
near the ionosphere of Europa: 
compression of the upstream magnetic field and
decompression in the plasma wake. 

The modeling also shows the formation of an 
Alfv\'en wing in the direction of the main magnetic field. 
The formation of the Alfv\'en wing in a sub-Alfv\'enic flow near 
Europa is similar to a formation near Io, which was first studied  
analytically by Neubauer (1980). 
The pickup ions play an important role in the fine
structure of the Alfv\'en wing due to effects of mass loading. 
In particular, the scale of the front of the Alfv\'en wing must be 
determined by the gyroradius of pickup ions. 
Unfortunately, in our 3D hybrid kinetic simulation we cannot yet 
resolved these spatial scales. 

%The modeling also shows
%the difference between the modeled  and observed magnetic field 
%(Kivelson et al., 1999) along the E4 Galileo trajectory, 
%especially for the $B_x$ and $B_y$ magnetic field.
%We assume that these disagreements may be caused by 
%the non-realistic background magnetic field.

\subsection*{3.2\ \ Effects of inductive Europa's magnetic field} \vspace*{-3mm} 

In the first set of 
models (Sect. 3.1, model I, cases (a) and (b)), 
we used a homogeneous global magnetic field
as an initial condition. These models do not produce agreement
between the simulated and observed magnetic fields.

In the second set of modeling we take into account
the gradient of the global Jovian magnetic field for an 
initial magnetic field distribution.
In the paper by Kivelson, Khurana, Stevenson et al. (1999);
Kivelson et al. (1997); Kivelson et al. (2000), 
it has been shown that the $B_y$ component of the magnetospheric 
magnetic field has strong time variations at the position of Europa. 
In the MHD-fluid  approximation
the effects of such magnetic field variations are estimated in
Schilling, Neubauer and Saur (2007); Schilling, Neubauer and Saur (2008).
The initial plasma density and bulk velocity distribution in our
modeling were taken
from the E4 flyby data (Paterson et al., 1999).

We created the following model II, case (c) for simulation:
the density for thermal $O_2$ is the same as for model I, case (b), and 
the pickup electron temperature is lower than in model I, case (b).
The plasma density and bulk velocity distribution in our
modeling were taken
from the E4 flyby data (Paterson et al., 1999):
full magnetosphere corotation speed
$U_0=105$\,km/s;
upstream densities are
$\rho_{O^{++}}=10$\,cm$^{-3}$;
$\rho_{S^{++}}=10$\,cm$^{-3}$;
upstream ion and electron temperatures,
$T_{O^{++}}=100$\,eV;
$T_{S^{++}}=100$\,eV;
$T_{e,0}=20$\,eV.
The temperatures of electrons connected with non-thermal and thermal $O_2^+$ 
pickup ions are
$T_{e,non-thermal}=100$\,eV;
$T_{e,thermal}=100$\,eV.

In our hybrid kinetic modeling (model II) we use
a simple magnetic dipole model of the induced
oceanic magnetic field from the ten-hour corotation
variation of the background Jovian magnetic field at Europa 
(see paragraph ``Boundary Conditions", Sect. 2). 
And, finally, we fit the results of
modeling to the components of the measured magnetic field.

This is not yet a fully self-consistent approach but provides
a first approximation.
Also, the ocean may not be exactly a spherically symmetric
conducting shell and may ultimately require a higher-order
multipole model for the induced fields.

\begin{figure}
\vspace*{12.cm}
\includegraphics{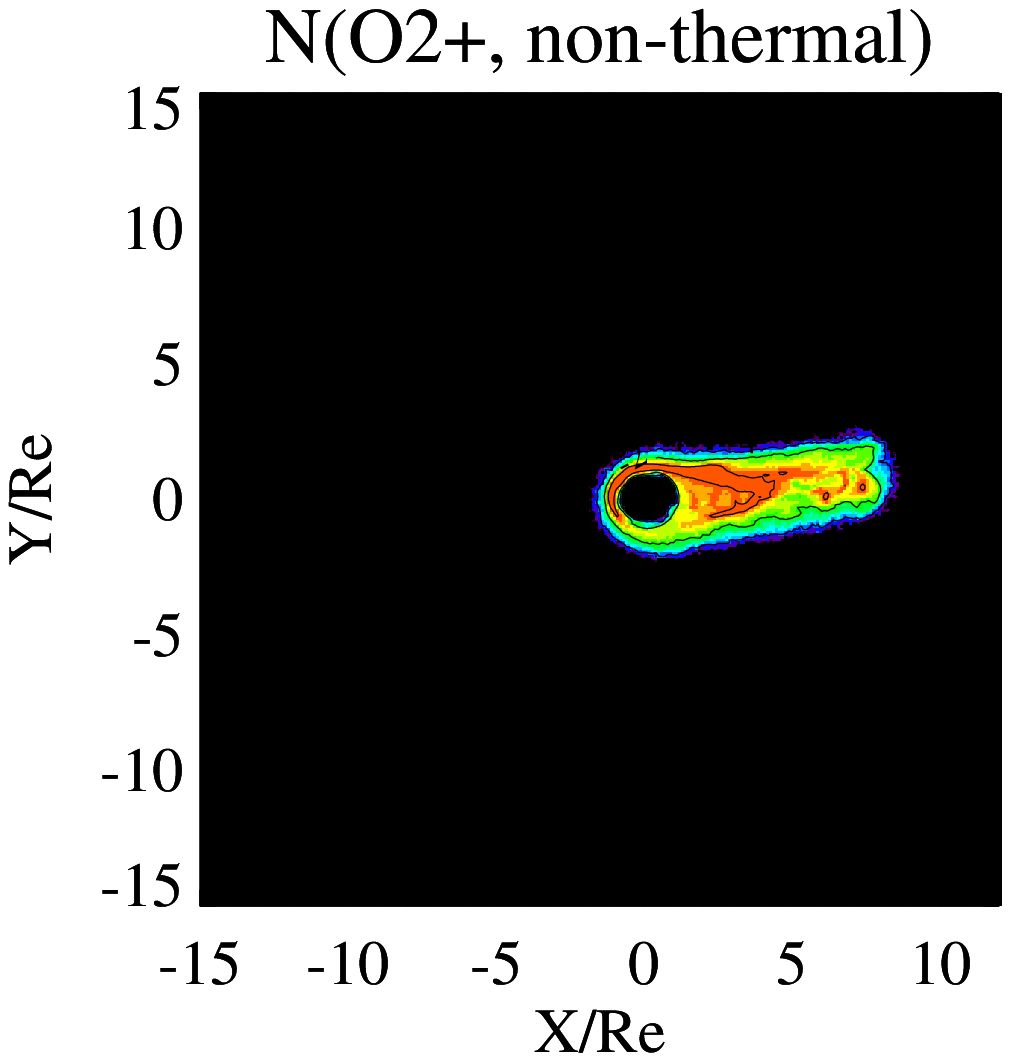}
\includegraphics{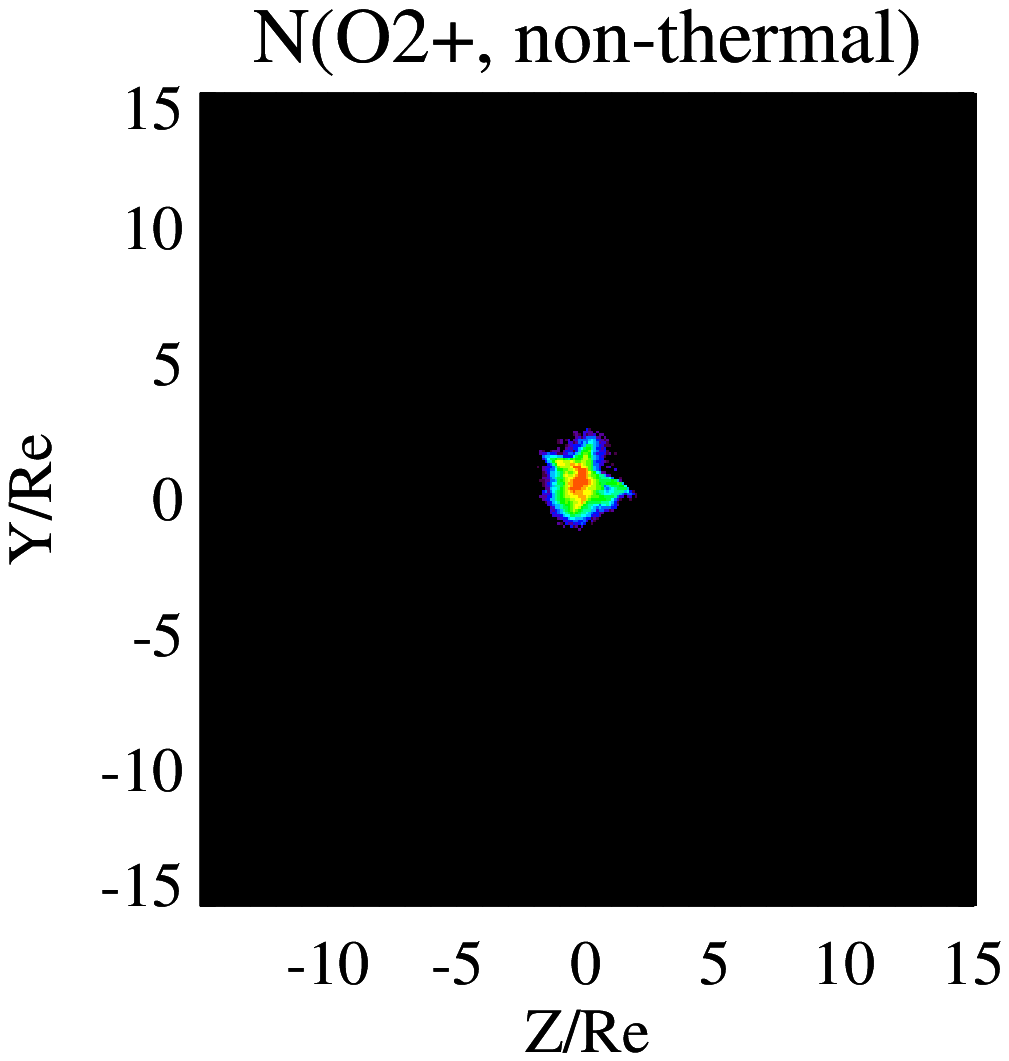}
\includegraphics{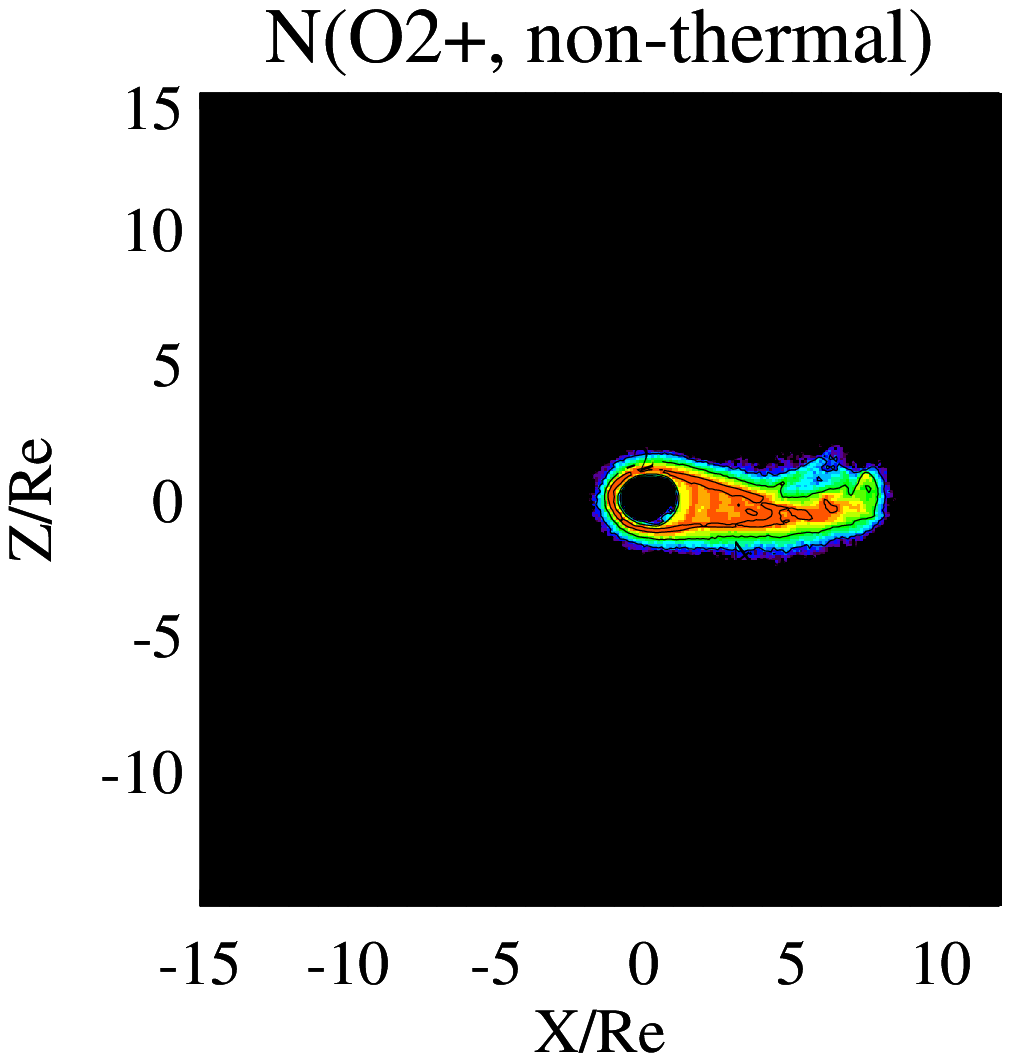}
\includegraphics{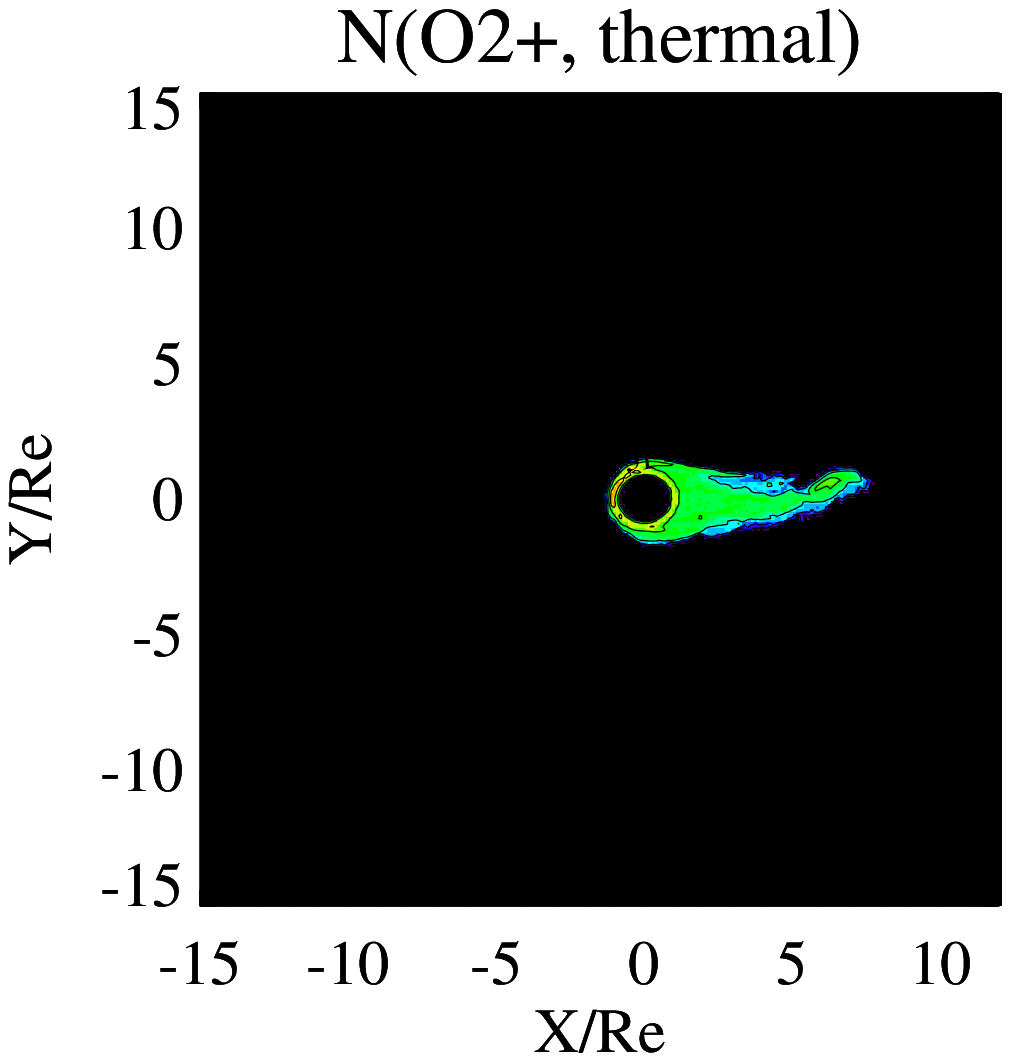}
\includegraphics{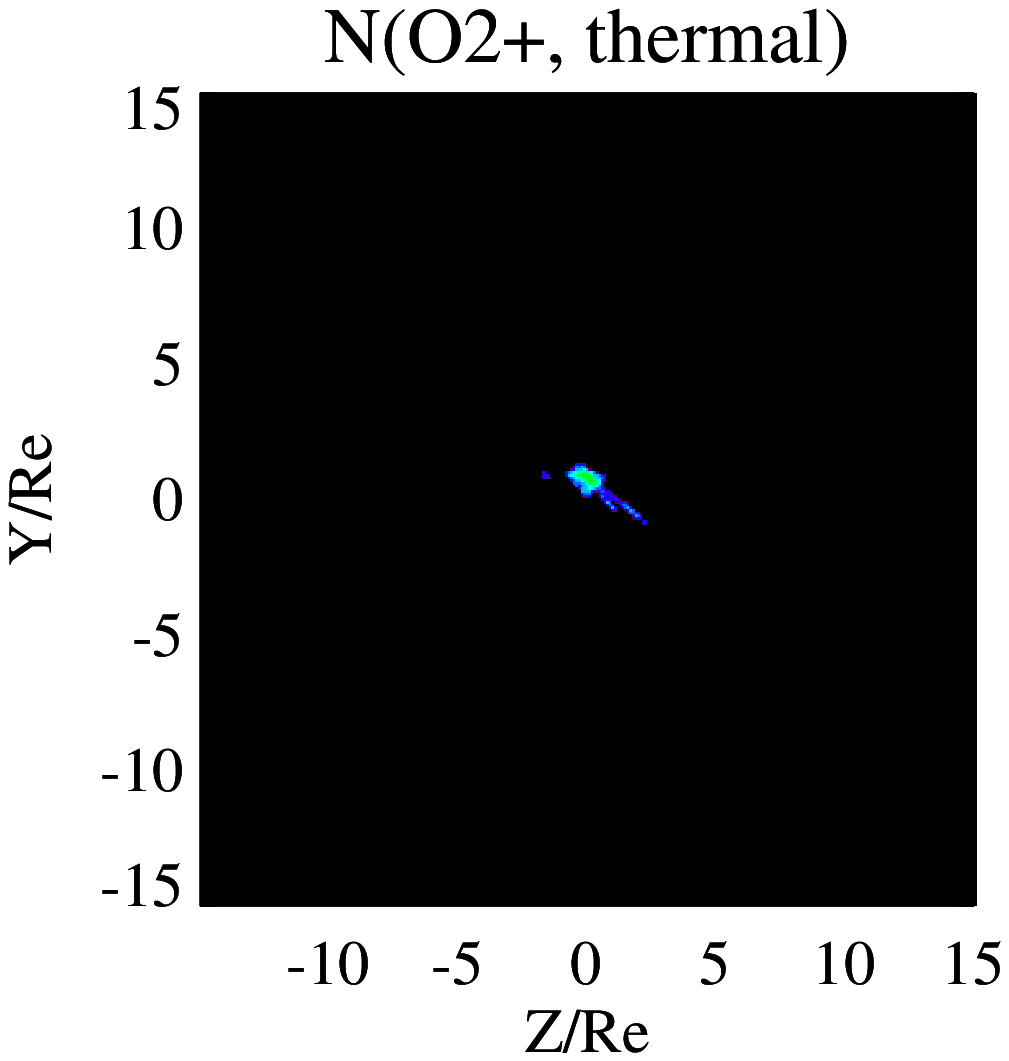}
\includegraphics{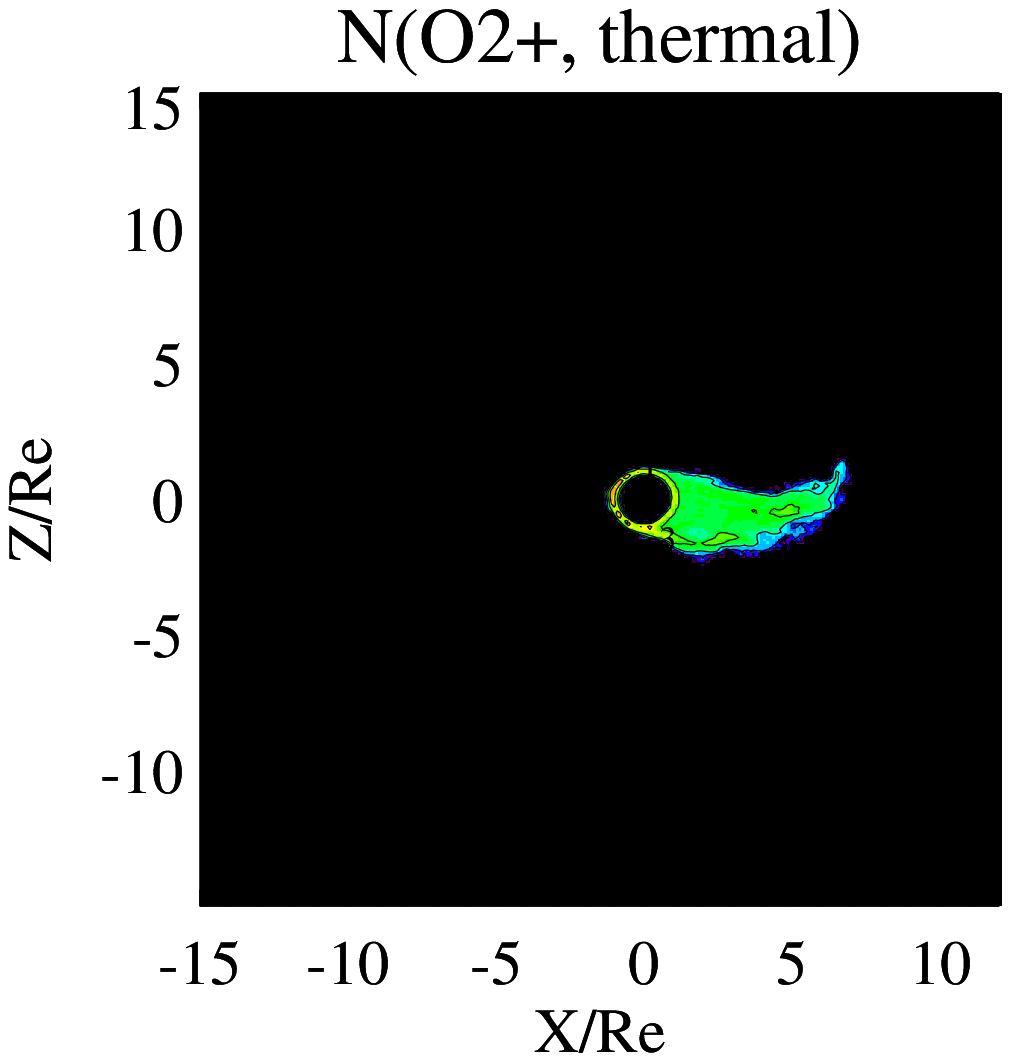}
\includegraphics{scalen3.eps}
\includegraphics{scalen4.eps}
\caption{2-D cuts of the pickup ion $O_2^+$ density profile.
Non-thermal $O_2^+$ (top), thermal $O_2^+$ (bottom). Model II, case (c).
$x-y$ cuts (left column) are located at $z=0$, 
$y-z$ cuts are located at $x/R_{E}=7$, and $x-z$ cuts (right column) are
located at $y=0$.}
\label{fig:8}
\end{figure}

Figure \ref{fig:8} demonstrates the 2D cuts for non-thermal
and thermal $O_2^+$ pickup ion densities. The figure does not show
any extension of the pickup ion profile in the $y$ and $z$ directions.
The plasma wake is narrower in the $y$ and $z$ directions compared to 
that produced by
model I, cases (a) and (b). 
The reason for this effect is the lower temperature of
electrons connected with pickup $O_2^{+}$ ions than in case (b), 
and a lower pickup ion production rate
near the surface of Europa than in case (a).

\begin{figure}
\vspace*{8.cm}
\includegraphics{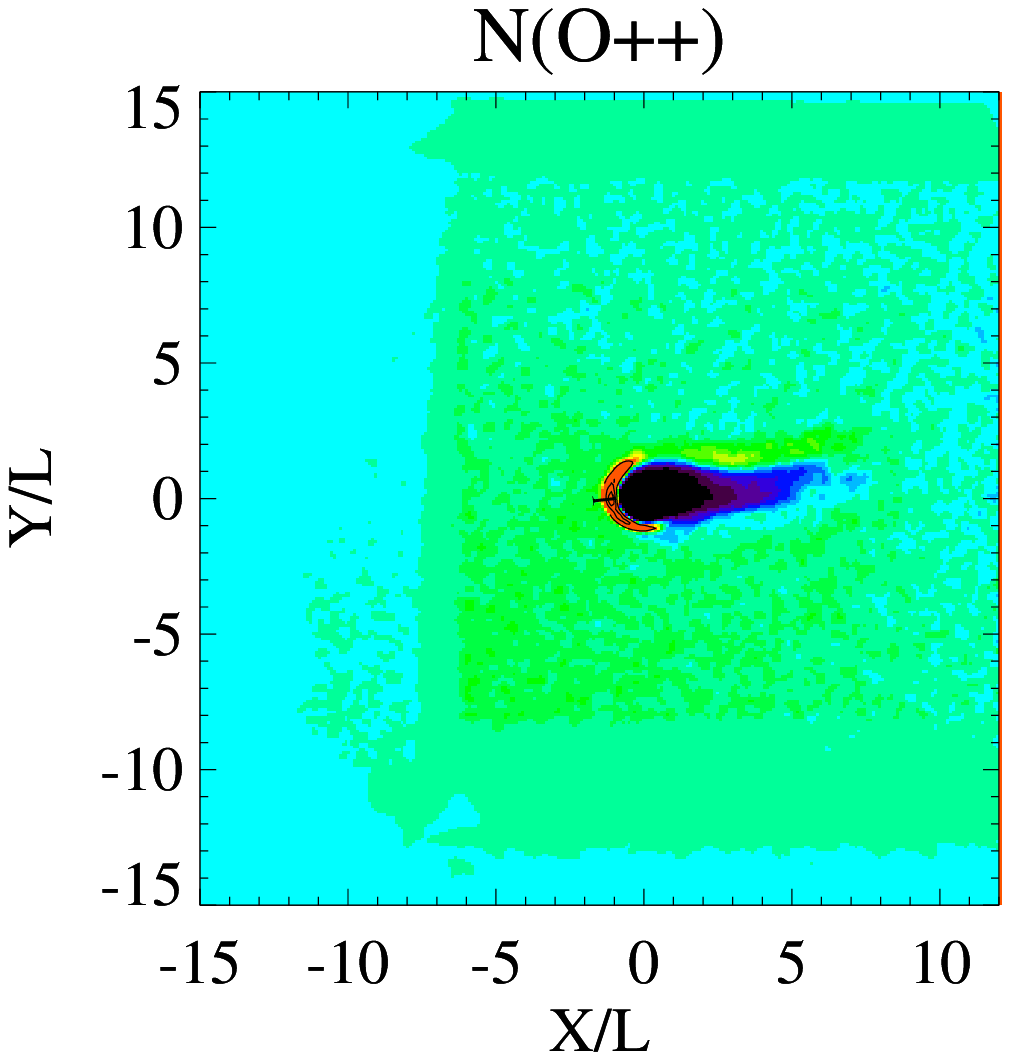}
\includegraphics{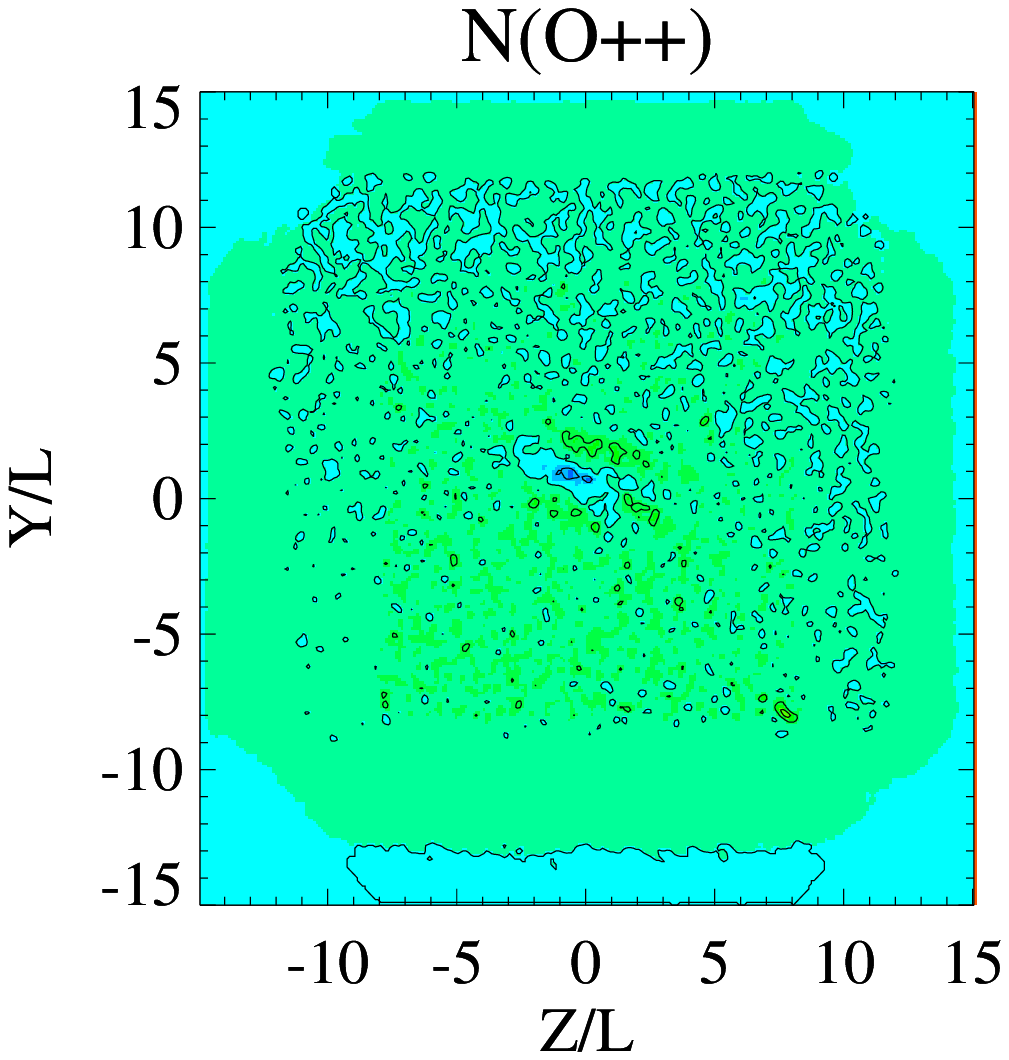}
\includegraphics{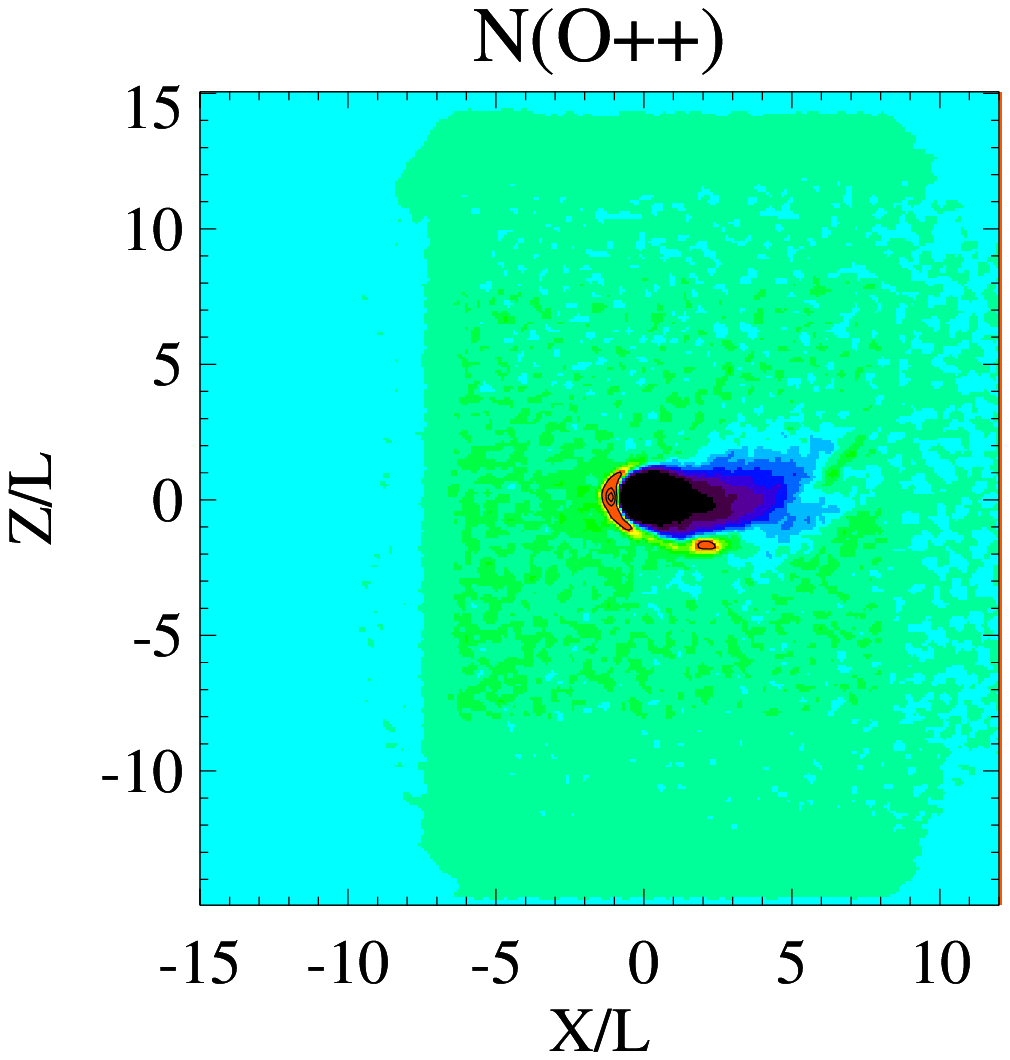}
\includegraphics{fig16.eps}
\caption{2-D cuts of the background $O^{++}$ ion density profiles.
Model II, case (c).
$x-y$ cuts (left column) are located at $z=0$, 
$y-z$ cuts are located at $x/R_{E}=7$, and $x-z$ cuts (right column) are
located at $y=0$.}
\label{fig:9}
\end{figure}

Figure \ref{fig:9} shows the 
distribution of the $O^{++}$ ion density in the 
$x$-$y$, $y$-$z$\,$(x=5\,R_{E})$ and $z$-$x$ planes. 
The narrow plasma wake may be explained 
by the cooler temperature of the electrons connected 
with $O_2^{+}$ pickup ions, resulting in
a smaller polarization electric field that is responsible 
for the expansion of Europa's ionosphere. 

One can also see an increase in the plasma
density near Europa due to the formation of a magnetic barrier 
(not shown here).
The density profile for $SO^{++}$ background ions is close 
to the density profile for $O^{++}$ ions as in model I, cases (a) and (b).

\begin{figure}
\vspace*{14.cm}
\includegraphics{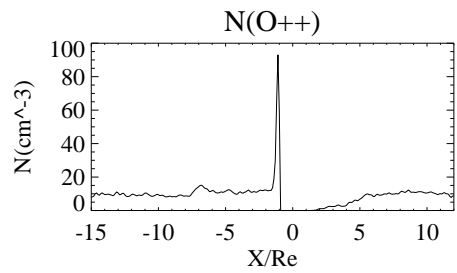}
\caption{1D cuts of the background $O^{++}$ ion density profile. 
The cut is located at $y=0$, $z=0$. 
Model II, case (c).}
\label{fig:10}
\end{figure}

Figure \ref{fig:10} shows a 1-D cut of the background $O^{++}$ density
along the $x$- axis ($y=0, z=0$). 
One can see jump in the background plasma density with
$N_{O^{++}, max}=90$cm$^{-3}$ (model II, case (c))
on the day-side of the ionosphere
and depletion in the plasma density in Europa's plasma wake.
Note that the jump in the plasma density
profile is stronger in model II, case (c) than is observed in model I, 
case (a).
The jump is located near the surface of Europa, as was observed 
in model I, cases (a) and (b).

\begin{figure}
\vspace*{14.cm}
\includegraphics{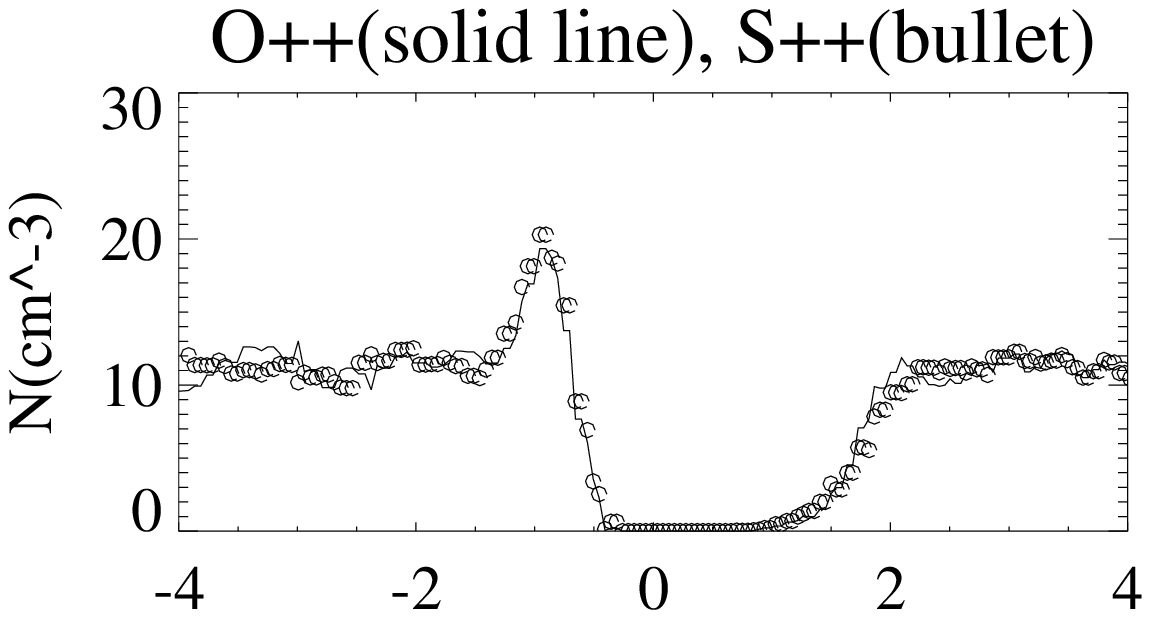}
\includegraphics{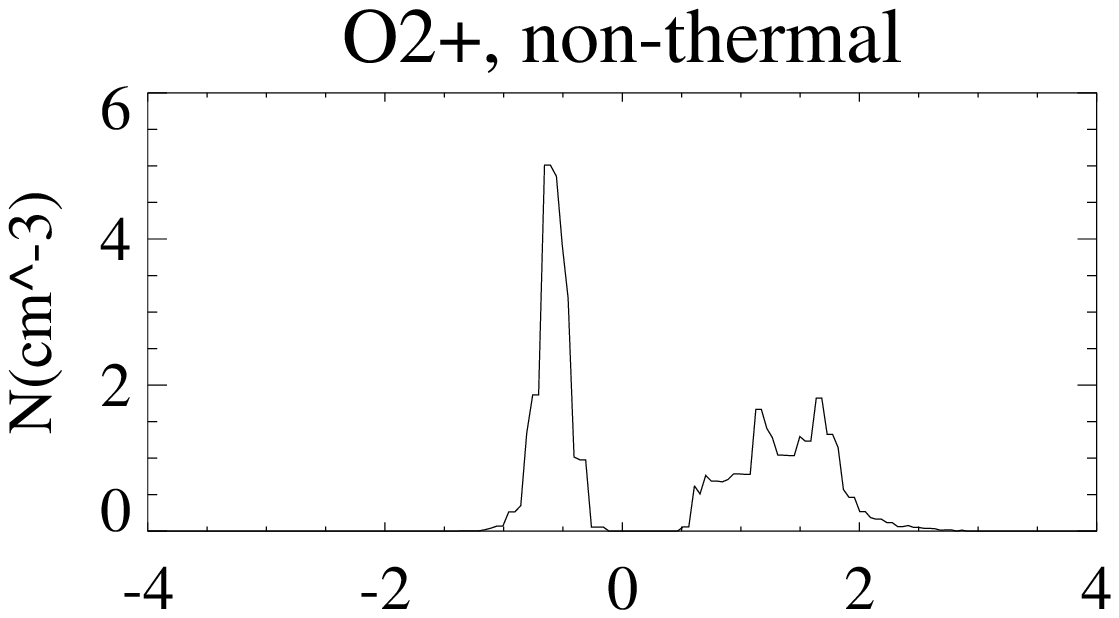}
\includegraphics{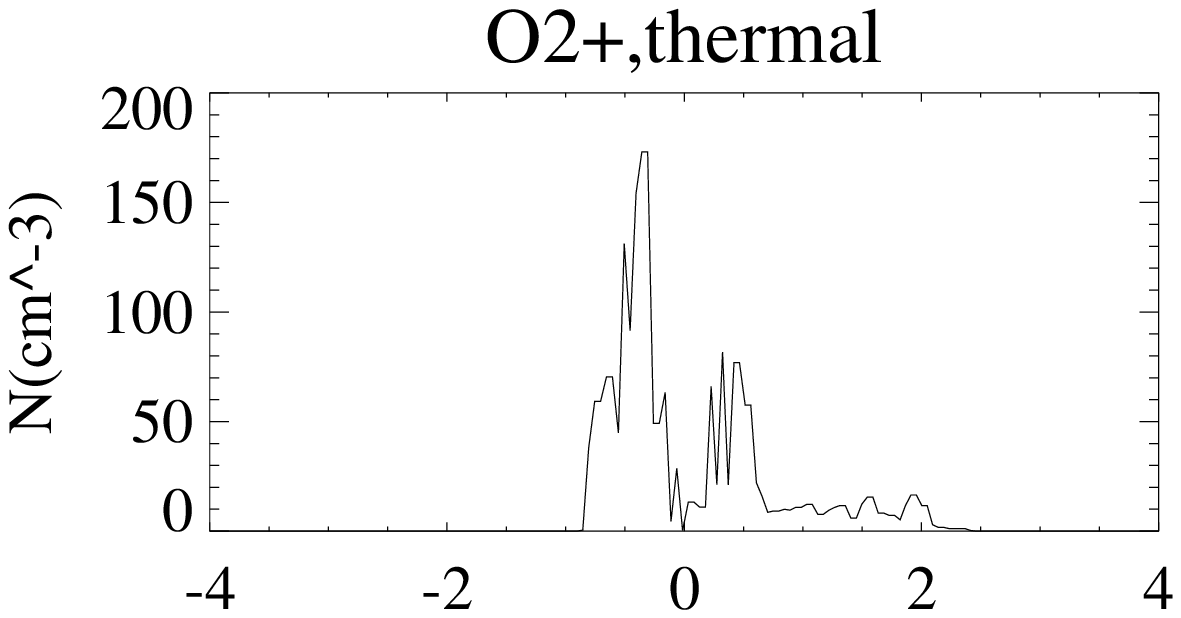}
\includegraphics{nicap17.eps}
\caption{
Background ($S^{++}$, $O^{++}$), non-thermal ($O_2^+$) and thermal
($O_2^+$) pickup ion densities from simulation. 
$Y(Re)$ denotes a projection of the spacecraft position onto the $y$ axis.
Model II, case (c).
Bottom - E4 observation of the total ion density (Paterson et al. 1999).}
\label{fig:11}
\end{figure}

Figures~\ref{fig:11} shows 1-D density profiles of the background
and pickup ions along the E4 trajectory of the Galileo spacecraft.
One can see a strong plasma void in the center of the plasma wake.
There is also a sharp boundary with an overshoot in the density 
profiles  on the left side of the plasma wake, and a smooth 
boundary layer on the right side, Fig.~\ref{fig:11} (top).
The density profile  for $S^{++}$ is similar the density 
profile for $O^{++}$ background ions.
Fig.~\ref{fig:10} (middle) shows the density profiles
for non-thermal and thermal $O_2{+}$ pickup 
ions.
The total ion density profile observed during the E4 pass is shown in 
Fig.~\ref{fig:11} (bottom).
Again, one can see two peaks in the total ion density profile. 
However, the observed value of the density in these peaks 
is lower than predicted by the model; 
this may be explained by an overestimated density 
of $O_2^{+}$ pickup ions for model II, case (c).

The modeling shows that the shape of Europa's global 
plasma environment
depends on a combination of the upstream plasma parameters 
and pickup ion and electron parameters.
For example, reducing in the temperature of electrons connected 
with pickup ions
results in a higher density of thermal $O_2^{+}$ pickup ions 
at the trajectory of the spacecraft (compare Fig. \ref{fig:6} 
(right) and Fig. \ref{fig:11}).
This effect is connected with the polarization electric field 
which is proportional to the gradient
of the electron pressure.
Reducing the temperature of the background upstream ions 
results in the widening of the plasma wake (compare Fig. \ref{fig:6} 
(left and right, top) and Fig. \ref{fig:11} (top)).
These effects were earlier demonstrated in the 3-D
hybrid simulation of Io's plasma environment (Lipatov and Combi, 2006).  
We have found the similarities between the
plasma environments of these objects.  
Indeed, Io and Europa have sufficiently thin exospheres 
and strong magnetic fields resulting in a small value 
of the ion gyroradius. 

\begin{figure}
\vspace*{14.cm}
\includegraphics{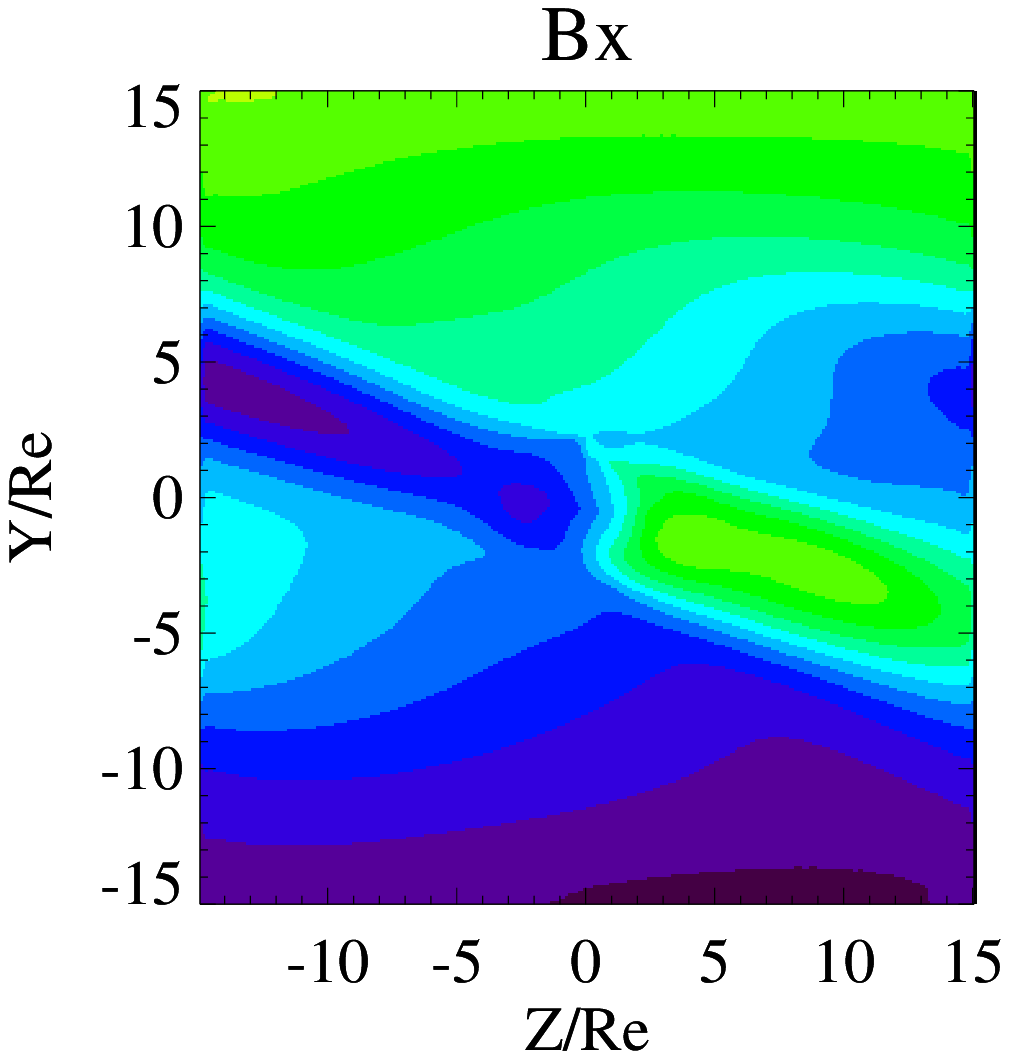}
\includegraphics{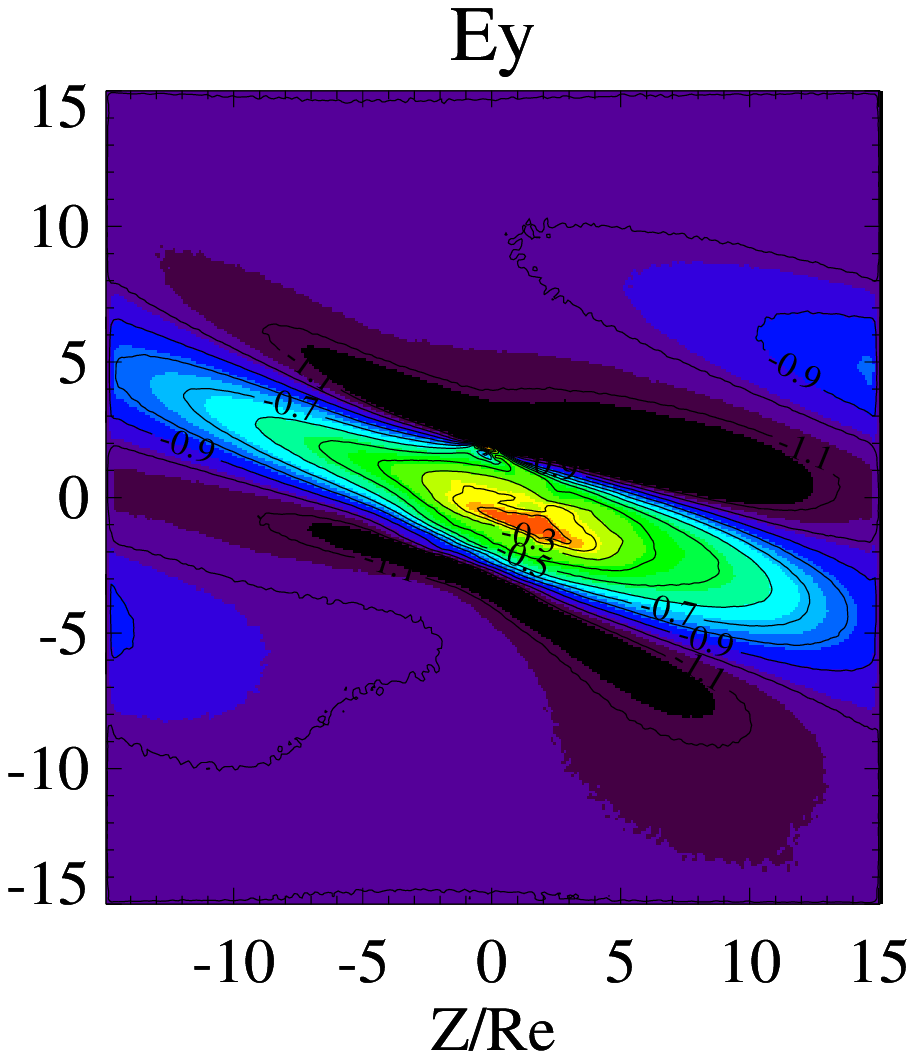}
\includegraphics{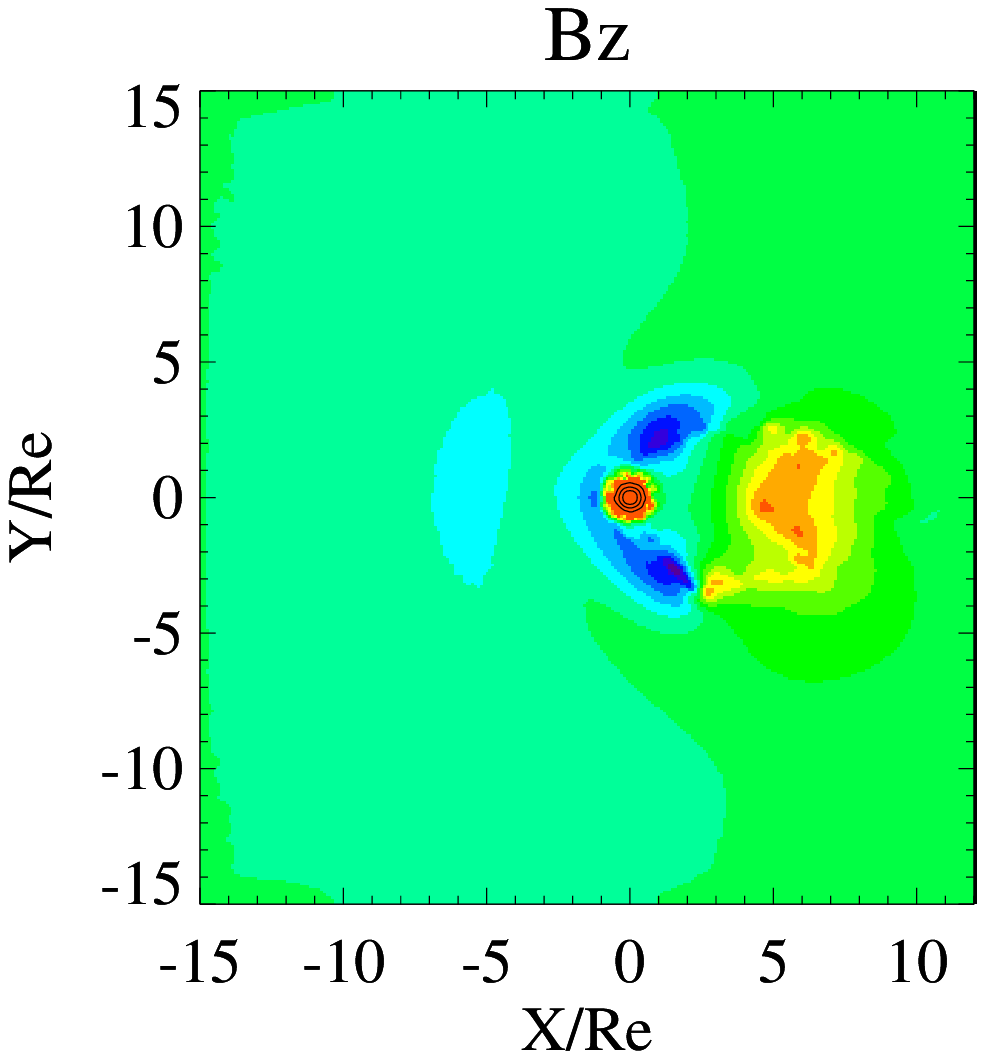}
\includegraphics{bx.eps}
\includegraphics{ey.eps}
\includegraphics{bz.eps}
\caption{Model II, case (c). 2-D cuts of the magnetic $B_x$ (top), 
electric $E_y$ (top) and 
magnetic $B_z$ (bottom) field profiles.  
$y-z$ cuts (top) are located at $x/R_{E}=7$, and $y-x$ cuts (bottom) are
located at $z/R_{E}=0$.}
\label{fig:12}
\end{figure}

Let us consider the global distribution for the electromagnetic 
field of model II, case (c).  
Figure \ref{fig:12} shows 2-D cuts for the magnetic
$B_x$, $B_z$ and electric $E_y$ field profiles. 
The distributions for the $B_z$, $E_y$ field shown in the figure
are close to the distributions for model I, case (b). 
However, there are significant
differences between the $B_x$ profiles for model I, case (a) and case (b), 
and model II, case (c). The differences between the $B_x$ profiles for
cases (a) and (b), Fig. \ref{fig:7} (top) are due to a much higher density 
of the thermal $O_2{+}$
pickup ions in the plasma wake, whereas
the differences between the $B_x$ profiles for cases (b) and (c) are 
due to the nonlinear interaction of the Alfv\'en wing with the inhomogeneous 
Jovian magnetic field in model II, case (c).

\begin{figure}
\vspace*{13.cm}
\includegraphics{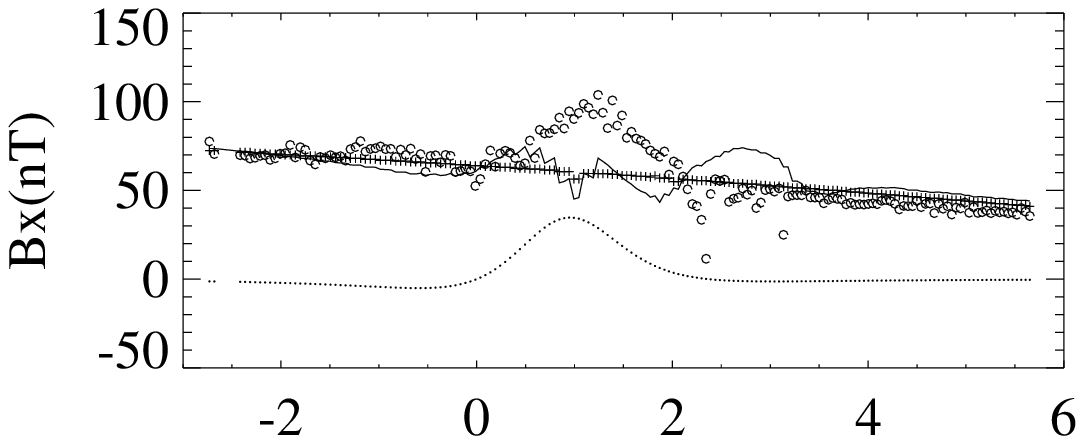}
\includegraphics{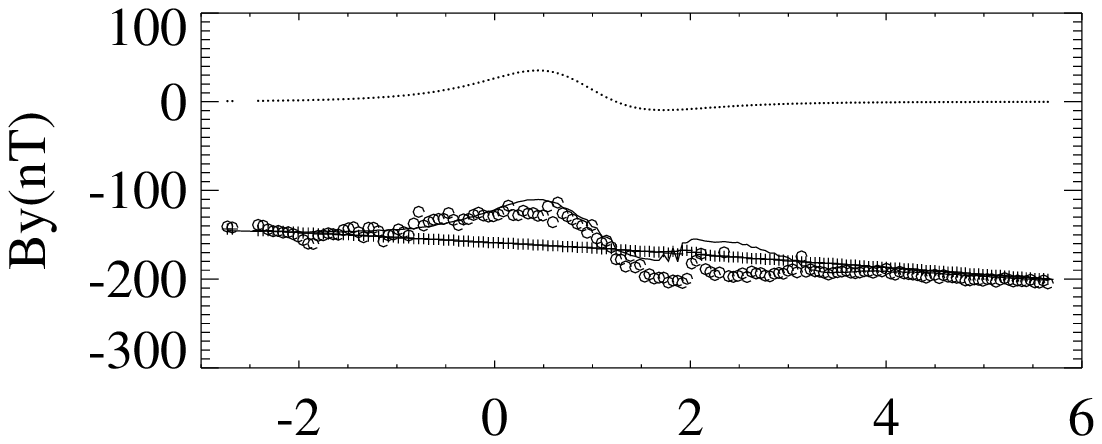}
\includegraphics{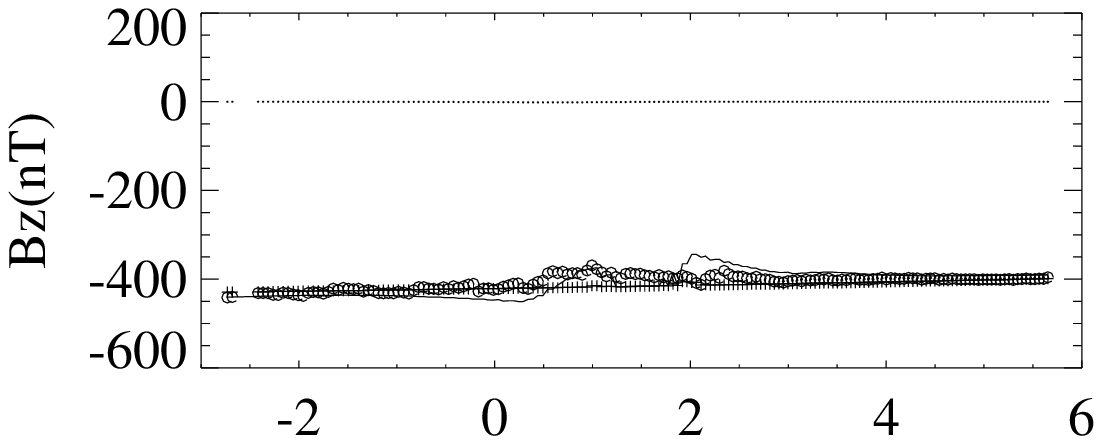}
\includegraphics{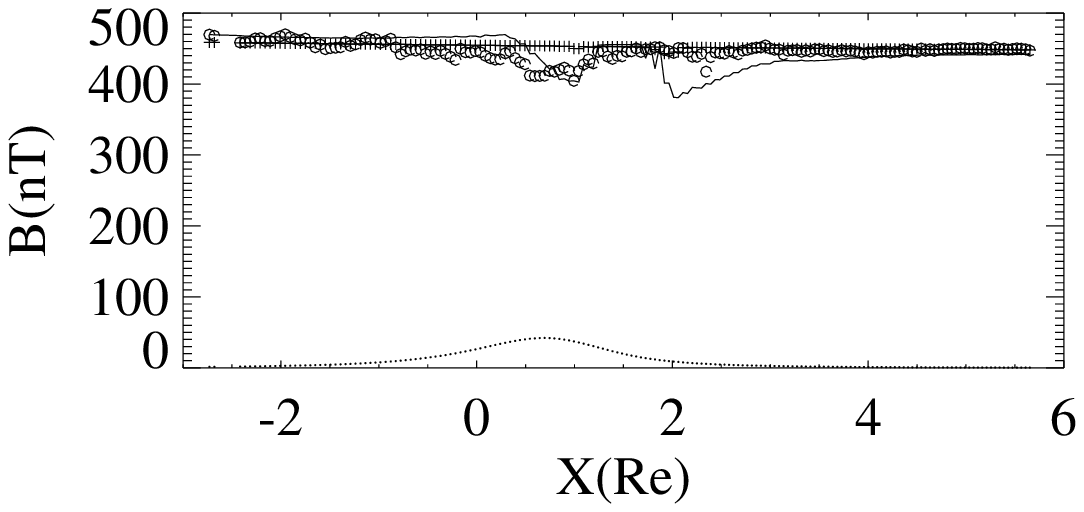}
\caption{
Magnetic field component profiles along the E4 trajectory
after fitting with inductive dipole magnetic field. 
Solid line - modeling,
(--) denotes dipole field, and 
(+) is the Jovian magnetic field at the position of Europa.
$\circ$ - Galileo's E4 flyby measurements (Kivelson et al. 1997).
$X(Re)$ denotes a projection of the spacecraft position onto the $x$-axis.}
\label{fig:13}
\end{figure}

Figure \ref{fig:13} shows the magnetic field components (solid line)
$B_x$, $B_y$ , $B_z$, and $|B|$
along the E4 trajectory of the  Galileo spacecraft. 
The magnetic
field components of the inductive magnetic dipole that simulates the 
effect of the nonstationarity of the Jovian magnetic field are shown
by a dotted line ($---$).
The circles ($\circ$) denote observational data
from Kivelson et al. (1997) and the
initial Jovian magnetospheric field at
the position of Europa (+++).
The simulation produces a satisfactory agreement with the observational
data for the $B_y$ magnetic field component, but not for the $B_x$ and
$B_z$ magnetic field components.
A multipole model for the oceanic magnetic
field may address this issue.
We will need to improve the model of the $O_2$ atmosphere, 
the resolution of
the ion trajectory, and the
gradient in the atmosphere/ionosphere density profiles
near the surface of Europa to obtain better agreement in
the $B_x$ and $B_z$ magnetic field components

\section*{4.\ \ Conclusions}

Hybrid modeling of Europa's plasma environment for the E4 
encounter with 3 ion species demonstrated several features:
\begin{itemize}
\item
The modeling shows a strong phase mixing in the plasma wake.
The plasma wake demonstrates the formation of time-dependent
structuring in the pickup ion tails (see, e.g., 
McKenzie, Sauer, Dubinin, 2001 for a weak comet case)
and the splitting of the pickup
ion tails. 
The splitting of the plasma wake has
the same nature as the splitting of the weak comet's plasma wake or 
the splitting of Titan's plasma wake.   
Such finite gyroradius effects were also observed
in 2.5 D hybrid and bi-fluid modeling of a weak comet
(see, e.g., Lipatov, Sauer, Baumg\"artel, 1997; Sauer et al., 1996; 
1997; Lipatov, 2002)
and in 3D hybrid modeling of Titan's plasma environment 
(Lipatov et al., 2011; 2012).
The further investigation of these fine structure needs 
an additional modeling
with much better resolution.
\item
The model shows a magnetic field barrier formation 
at the day-side portion 
of the ionosphere. The formation of an Alfv\'en wing in the plane 
of the external magnetic field was also observed. 
Note that the Alfv\'en wing was
earlier observed in a hybrid simulation of the plasma environment of 
Io and Europa by Lipatov and Combi (2006) and  by Lipatov et al. (2010). 
An MHD simulation of the plasma environment of Io and Europa also 
produces the formation
of an Alfv\'en wing (Saur et al.,  1999; 1998; Liu et al., 2000; 
Schilling et al., 2008).
\item
The ion and electron temperatures play an important role in plasma 
structure formation, and in creating the ion fluxes inside the 
ionosphere. 
These effects were observed earlier in a 3-D hybrid simulation of
Io's  plasma environment (Lipatov and Combi, 2006).
The hybrid model produces the correct pickup ion flux along 
the magnetic field, in contrast to
the MHD models which operate with pickup ions 
with a Maxwellian velocity distribution.
In the current paper we have presented
only three runs with different combinations of the upstream ion 
and pickup electron temperatures.
\item
The model's total ion density in the plasma wake
does not satisfactory match the observed density.
\item
The constant induced dipole moment (model II, case (c))  
improves a fit of the magnetic field $B_y$ component 
to the E4 trajectory. 
However, a fit of the magnetic field $B_x$ component 
is still not satisfactory due to
the imperfect model of the atmosphere/ionosphere 
and unsatisfactory numerical resolution of
the gyroradii on the grid cell.
\item
Use of an inhomogeneous background magnetic
field provides a good agreement between the
observed and simulated magnetic fields.
However, we still need to improve the resolution of the gradient
in the atmosphere density, the gyroradius of pickup
ions, and the fields in the internal non-conduction
ice shell and conduction ocean layers of Europa.
\end{itemize}

In our future computational models, we plan to include a 
nonstationary boundary condition for the magnetic field in order 
to take into account the spatially inhomogeneous 
and nonstationary background Jovian magnetic field. 
This model will also be appropriate for a
potentially nonspherical ocean shell.
We also plan
the use of a varying atmospheric density, a varying electron  
temperature (that plays key-role in the pickup ion dynamics), 
and sputtering processes (Johnson, 1990; Johnson et al., 1998) 
at the surface of Europa. 
We also plan to use a composite grid structure using  
the ``cubed sphere" technique (see, e.g. Koldoba et al, 2002) 
to improve the resolution of the a small
scales near the surface of Europa and to increase the size of the 
computational domain. 

The composite grid structure will allow us to
estimate the inductive magnetic field
from the ocean as a part of the total current closure that also includes
the external plasma currents.
This technique will allow us to study
wave-particle interaction effects in the far plasma wake, such as 
ion cyclotron waves
that have been observed in the Galileo flyby mission (see e.g.
Volwerk, Kivelson and Khurana, 2001; Kivelson, Khurana and Volwerk, 2009).
These models must include the induced magnetic field from
a putative subsurface ocean, and will also include particle trajectory
tracing for test particles, e.g. electrons and high-energy ions.

Note that the larger computational domain allows 
us to use the upstream parameters for the plasma 
and electromagnetic field 
instead of the use of the ``damping" boundary condition. 
However, in the outer region of the computational domain 
(large cell size) we have to use a drift-kinetic approach 
(see e.g. Lipatov et al., 2005)
for ion dynamics since we cannot approximate the 
ion trajectory there. 
We can also use a complex particle kinetic technique
(see e.g. Lipatov, 2012) which provides a flexible fluid/kinetic  description 
and may significantly save computational resources.

 \acknowledgments

A.S.L. was supported in part by the Project/Grant 00004129, and 00004549 
between the GPHI UMBC and NASA GSFC. J.F.C. was supported 
as Principal Investigator by the NASA Outer Planets 
Research Program. 
Computational resources were provided by the NASA Ames Advanced
Supercomputing Division (SGI - Columbia, Project SMD-09-1110).

\bibliographystyle{apsrev}

\begin{references} \vspace*{-3mm} 

Austin, J.V., Goldstein, D.B., 2000. Rarefied gas model of 
Io's sublimation-driven atmosphere. Icarus 148, 370-383.

\reference
Bagenal, F., 1994. Empirical model of the Io plasma torus: 
Voyager measurements. J. Geophys. Res. 99, 11043-11062.

\reference
Braginskii, S.L., 1965. Transport processes in a plasma.
In: Leontovich, M.A. (Ed.),
Reviews of Plasma Physics. Consultants Bureau, New York, pp. 205-240.

\reference
Cassidy, T.A., Johnson, R.E., McGrath, M.A.,  Wong, M.C.,
Cooper, J.F., 2007. The spatial morphology of Europa's 
near-surface $O_2$ atmosphere. Icarus 191, 755-764.

\reference
Combi, M.R., Kabin, K., Gombosi, T., De Zeeuw, D.L., 
Powell, K., 1998. Io's plasma environment during the Galileo 
flyby: Global three-dimensional MHD modeling with adaptive mesh 
refinement. J. Geophys. Res. 103, 9071-9081.

\reference
Combi, M.R., Gombosi, T.I., Kabin, K., 2002. Plasma Flow Past 
Cometary and Planetary Satellite Atmospheres. In: Mendillo, M., 
Nagy, A., Waite, J.H. (Eds.), 
Atmospheres in the Solar System: Comparative Aeronomy. Geophys. 
Monograph Series Vol. 130. AGU Washington, D.C., pp. 151-167.

\reference
Dols, V., Delamere, P.A., Bagenal, F., 2008.  
A multispecies chemistry model of Io's local interaction with the Plasma Torus, 
Journal of Geophysical Research (Space Physics), 113, 9208-+, 
doi:10.1029/2007JA012805.

\reference
Goertz, C.K., 1980. Io's interaction with the plasma torus.
J. Geophys. Res. 85, 2949-2956.

\reference
Herbert, F., 1985. ``Alfv\'en wing" models of the induced 
electrical current system at Io: A probe of the ionosphere of Io.
J. Geophys. Res. 90, 8241-8251.

\reference
Hewett, D.W., Langdon, A.B., 1987. Electromagnetic Direct
Implicit Plasma Simulation. J. Comput. Phys. 72(1), 121-155.

\reference
Huddleston, D.E., Strangeway, R.J., Warnecke, J., Russel, C.T., Kivelson, M.G.,
1997. Ion cyclotron waves in  the Io torus during the Galileo encounter:
Warm plasma dispersion analysis. Geophys. Res. Lett. 24 (17), 2143-2146.

\reference
Ip, W.-H., 1990. Neutral gas-plasma interaction: The case of the 
Io plasma torus. Adv. Space Res. 10(1), 15-18.

\reference
Jia, X., Walker, R.J., Kivelson, M.G., Khurana, K.K., Linker, 
J.A., 2008. Three-dimensional MHD simulation of Ganymede's 
magnetosphere. J. Geophys. Res. 113, A06212.

\reference
Johnson, R.E., 1990. Energetic Charge-Particle Interaction
with Atmospheres and Surfaces. Springer-Verlag, New York.

\reference
Johnson, R.E., Killen, R.M., Waite, J.H., Lewis, W.S., 1998. 
Europa's surface composition and sputter-produced ionosphere.
Geophys. Res. Lett. 25, 3257-3260. 

\reference
Johnson, R.E., Burger, M.H., Cassidy, T.A., Leblanc, F., Marcony, M.,
Smyth, W.H., 2009.
Composition and detection of Europa's sputter-induced atmosphere.
In: Pappalardo, R.T., McKinnon, W.B., Khurana, K.K. (Eds.), Europa. 
University of Arizona Press, Tucson, pp. 507-527.

\reference
Kabin, K., Combi, M.R., Gombosi, T.I., Nagy, A.F., DeZeeuw, D.L., 
Hansen, K.S., Powell, K.G., 1999. On Europa's magnetosphere 
interaction: A MHD simulation of the E4 flyby. J. Geophys. Res. 
104(A9), 19983-19992. 

\reference
Kabin, K., Combi, M.R., Gombosi, T.I., DeZeeuw, D.L., Hansen, 
K.S., Powell, K.G., 2001. Io's magnetospheric interaction: 
an MHD model with day-night asymmetry. Planetary and Space Sci. 
49, 337-344.

\reference
Kageyama, A., Sato, T., 2004. ``Yin-Yang grid": An overset 
grid in spherical geometry. Geochemistry, Geophysics and 
Geosystems 5(9), Q09005, doi:10.1029/2004GC000734.

\reference
Khurana, K.K., Kivelson, M.G., Stevenson, D.J., G. Schubert, 
Russell, C.T., Walker, R.J., Polanskey, C., 1998.
Induced magnetic fields as evidence for subsurface oceans in 
Europa's and Callisto. Nature 395, 777-780.

\reference
Khurana, K., Kivelson, M.G., Hand, K.P., Russell, C.T., 2009.
Electromagnetic induction from Europa's ocean and the deep interior.
In: Pappalardo, R.T., McKinnon, W.B., Khurana, K.K. (Eds.), Europa. 
University of Arizona Press, Tucson, pp. 571-586.

\reference
Kivelson, M.G., Khurana, K.K., Means, J.D., Russell, C.T., 
Snare, R.C., 1992.
The Galileo magnetic field investigation.
Space Sci. Rev. 60, 357.-383

\reference
Kivelson, M.G., Khurana, K.K., Joy, S., Russell, C.T., 
Southwood, D., Walker, R.J., Polanskey, C., 1997. 
Europa's magnetic signature: Report from Galileo's pass on 19 
December 1996. Science 276, 1239-1241.

\reference
Kivelson, M.G., Khurana, K.K., Stevenson, D.J., Benett, L., 
Joy, S., Russell, C.T., Walker, R.J., Polanskey, C., 1999.
Europa and Callisto: Induced or intrinsic fields in periodically
varying plasma environment. J. Geophys. Res. 104, 4609-4625.

\reference
Kivelson, M.G., Khurana, K.K., Russell, C.T., Volwerk, M.,
Walker, R.J., Zimmer, C., 2000.
Galileo magnetometer measurements strengthen the case for a 
subsurface ocean at Europa. Science 289, 1340-1343.

\reference
Kivelson, M.G., Khurana, K., Volwerk, M., 2009.
Europa's interaction with the Jovian magnetosphere.
In: Pappalardo, R.T., McKinnon, W.B., Khurana, K.K. (Eds.), Europa. 
University of Arizona Press, Tucson, pp. 545-570.

\reference
Koldoba, A.V., Romanova, M.M., Ustyugova, G.V., Lovelace, 
R.V.E., 2002. Three dimensional MHD simulation of accretion to 
an inclined rotator: The ``cubed sphere" method. Astrophys. J. 
576:L53-L56

\reference
Lellouch, E., 1996. Urey Prize Lecture. Io's Atmosphere: 
not yet understood. Icarus 124, 1-21. 

\reference
Linker, J.A., Khurana, K.K., Kivelson, M.G., Walker, R.J., 
1998. MHD simulation of Io's interaction with the plasma torus.
J. Geophys. Res. 103(E9), 19867-19877.

\reference
Lipatov A.S., K. Sauer and K. Baumg\"artel, 1997.
2.5-D hybrid code simulation of the
solar wind interaction with weak comets and related objects.
Adv. Space Res. 20(2), 279-282.

\reference
Lipatov, A. S., 2002. 
The Hybrid Multiscale Simulation Technology. An introduction
with application to astrophysical and laboratory plasmas.
Springer-Verlag, Berlin, Heidelberg and New York, pp. 413.

\reference
Lipatov, A.S., Motschmann, U., Bagdonat, T., 
Grie{\ss}meier, J.-M., 2005. The interaction of the stellar wind 
with an extrasolar planet -- 3D hybrid and drift-kinetic 
simulations. Planet. Space Sci. 53, 423-432.

\reference
Lipatov, A.S., Combi, M.R., 2006.
Effects of kinetic processes in shaping Io's global plasma
environment: A 3D hybrid model. ICARUS 180, 412-427.

\reference
Lipatov, A.S., Cooper, J.F., Paterson, W.R.,
Sittler Jr., E.C., Hartle, R.E., 2010.
Jovian plasma torus interaction with Europa:
3D Hybrid kinetic simulation. First results,
Planet. Space Sci. 58(13), 1681-1691.

\reference
Lipatov, A.S., Sittler Jr., E.C., Hartle, R.E., Cooper, J.F., 
Simpson, D.G., 2011.
Background and pickup ion velocity distribution dynamics in
Titan's plasma environment:
3D hybrid simulation and comparison with CAPS observations.
Adv. Space Res. 48, 1114-1125.

\reference
Lipatov, A.S., Sittler Jr., E.C., Hartle, R.E., Cooper, J.F., 
Simpson, D.G., 2012.
Saturn's magnetosphere interaction with Titan for T9 encounter:
3D hybrid modeling and comparison with CAPS observations.
Planet. Space Sci. 61, 66-78.

\reference
Lipatov, A.S., 2012.
Merging for Particle-Mesh Complex Particle Kinetic modeling of the
multiple plasma beams. J. Comput. Phys. 231, 3101-3118.

\reference
Liu, Y., Nagy, A.F., Kabin, K, Combi, M.R., DeZeew, D.R., 
Gombosi, T.I., Powell, K.G., 2000. Two species, 3D MHD 
simulation of Europa's interaction with Jupiter's magnetosphere. 
Geophys. Res. Lett. 27, 1791-1794.

\reference
Mankofsky, A., Sudan, R.N., Denavit, J., 1987. 
Hybrid Simulation of Ion Beams in Background Plasma. 
J. Comput. Phys. 70, 89-116.

\reference
Marconi, M.L., Dagum, L., Smyth, W.H., 1996. 
Hybrid fluid/kinetic approach to planetary atmospheres: 
An example of an intermediate-mass body. Astrophys. J. 469, 
393-401.

\reference
McGrath, M.A., Johnson, R.E., 1989.
Charge Exchange Cross Sections for the Io Plasma Torus.
J. Geophys. Res. 94(A3), 2677-2683.

\reference
McGrath, M.A., Hansen, C.J., Hendrix, A.R., 2009.
Observations of Europa's tenuous atmosphere.
In: Pappalardo, R.T., McKinnon, W.B., Khurana, K.K. (Eds.), Europa. 
University of Arizona Press, Tucson, pp. 485-505.

\reference
McKenzie, J.F., Sauer, K., Dubinin, E.M., 2001.
Stationary waves in a bi-ion plasma transverse to the magnetic field.
J. of Plasma Physics 65(3), 197-212.

\reference
Neubauer, F.M., 1980. Nonlinear standing Alfv\'en wave current 
system at Io - Theory. J. Geophys. Res. 85, 1171-1178.

\reference
Neubauer, F.M., 1998. The sub-Alfv\'enic interaction of the
Galilean satellites with the Jovian magnetosphere. 
J. Geophys. Res. 103, 19834-19866.

\reference
Neubauer, F.M., 1999. Alfv\'en wing and electromagnetic 
induction in the interiors: Europa and Callisto.
J. Geophys. Res. 104, 28671-28684.

\reference
Paranicas, C., Cooper, J.F., Garrett, H.B., Johnson, R.E., Sturner, S.J., 
2009. Europa's Radiation Environment and Its Effects on the Surface.
In: Pappalardo, R.T., McKinnon, W.B., Khurana, K.K. (Eds.), Europa. 
University of Arizona Press, Tucson, pp. 529-544.

\reference
Paterson, W.R., Frank, L.A., Ackerson, K.L., 1999. Galileo 
plasma observation at Europa: Ion energy spectra and moments.
J. Geophys. Res. 104(A10), 22779-22791.

\reference
Sauer, K., Bogdanov, A., Baumg\"artel, K., Dubinin, E., 1996.
Plasma environment of comet Wirtanen during its low-activity stage.
Planet. Space Sci. 44 (7), 715-729.

\reference
Sauer, K., Lipatov, A.S., Baumg\"artel, K., Dubinin E., 1997.
Solar Wind-Pluto Interaction Revised. Adv. Space Res.
20(2), 295.

\reference
Saur, J., Strobel, D.F., Neubauer, F.M., 1998. Interaction of
the Jovian magnetosphere with Europa: Constrains on the
neutral atmosphere. J. Geophys. Res. 103, E9, 19947-19962.

\reference
Saur, J., Neubauer, F.M., Strobel, D.F., Summers, M.E., 1999.
Three-dimensional plasma simulation of Io's interaction with the 
Io plasma torus: Asymmetric plasma flow.    
J. Geophys. Res. 104, 25105-25126.

\reference
Schilling, N., Khurana, K.K., Kivelson, M.G., 2004.
Limits on an intrinsic moment in Europa.
J. Geophys. Res. 109, E05006.

\reference
Schilling, N., Neubauer, F.M., Saur, J., 2007.
Time-varying interaction of Europa with the jovian magnetosphere:
Constrains on the conductivity of Europa's subsurface ocean. Icarus 192, 41-55.

\reference
Schilling, N., Neubauer, F.M., Saur, J., 2008.
Influence of the internally induced magnetic field on the plasma
interaction of Europa.
J. Geophys. Res. 113, A03203, doi:10.1029/2007JA012842.

\reference
Schreier, R., Eviatar, A., Vasyli\"unas, V.M., Richardson, J.D., 1993. 
Modeling the Europa plasma torus. J. Geophys. Res.
98, 21231-21243.

\reference
Shematovich, V.I., Johnson, R.E., Cooper, J.F., Wong, M.C., 2005.
Surface-bounded atmosphere of Europa. Icarus 173, 480-498.

\reference
Sittler, Jr., E.C., Strobel, D.F. 1987. Io plasma torus electrons:
Voyager 1. J. Geophys. Res. 92, 5741-5762.

\reference
Smyth, W.H., Marconi, M.L., 2006. 
Europa's atmosphere, gas tori, and magnetospheric
implications, Icarus 181, 510-526.

\reference
Southwood, D.J., Kivelson, M.G., Walker, R.J., Slavin, J.A., 
1980. Io and its plasma environment. J. Geophys. Res. 85, 
5959-5968. 

\reference
Southwood, D.J., Dunlop, M.W., 1984. Mass pickup in 
sub-Alfv\'enic plasma flow: A case study for Io. Planet. Space 
Sci. 32, 1079-1089.

\reference
Tikhonov, A.N., Samarskii, A.A., 1963.
Equations of Mathematical Physics. Mac Millan, New York,
pp. 765

\reference
Umeda, T., Omura, Y., Matsumoto, H., 2001. An improved masking
method for absorbing boundaries in electromagnetic particle
simulations. Comp. Phys. Comm. (137), 286-299.

\reference
Van'yan, L.L., Lipatov, A.S., 1972. Three-dimensional
hydromagnetic disturbances generated by a magnetic dipole in an
anisotropic plasma. Geomagn. and Aeronomy 18(5), 316-318. 

\reference
Volwerk, M., Kivelson, M.G., Khurana, K.K., 2001. 
Wave activity in Europa's wake: Implications for ion pickup. 
J. Geophys. Res. 106(A11), 26033-26048.

\reference
Wolf-Gladrow, D.A., Neubauer, F.M., Lussem, M., 1987.
Io's interaction with the plasma torus: A self-consistent model.
J. Geophys. Res. 92, 9949-9961. 

 \end{references}

%\begin{thebibliography}{}

%\end{thebibliography}

%\end{article}

\newpage
\setlength{\oddsidemargin}{0.2in}

\noindent

\newpage
\noindent

\newpage
\noindent

\newpage
\noindent

\newpage
\noindent

\newpage
\noindent

\newpage
\noindent

\newpage
\noindent

\newpage
\noindent

\newpage
\noindent

\newpage
\noindent

\newpage
\noindent

\newpage

\end{document}